\documentclass[aps,prd,nopacs,nofootinbib,notitlepage]{revtex4-1}

\usepackage{amsfonts}
\usepackage{amsmath}
\usepackage{mathtools}
\usepackage{xcolor}

\usepackage{bm}
\usepackage{graphicx}
\usepackage{fancybox}
\usepackage{comment}
\usepackage{multirow}
\usepackage{tipa}
\definecolor{refs}{RGB}{245,156,74}
\usepackage[colorlinks=true,hyperfootnotes=true,citecolor=cyan]{hyperref}
\usepackage{enumitem}
\usepackage{comment}
\usepackage{subcaption}
\usepackage{esint}
\usepackage{bbold}
\usepackage{csquotes}
\usepackage{slashed}

\usepackage{tikz}
\usetikzlibrary{arrows.meta, positioning, shapes.geometric}

\newcommand{\cev}[1]{\reflectbox{\ensuremath{\vec{\reflectbox{\ensuremath{#1}}}}}}

\newcommand{\be}{\begin{equation}}
\newcommand{\ee}{\end{equation}}
\newcommand{\ba}{\begin{eqnarray}}
\newcommand{\ea}{\end{eqnarray}}
\newcommand{\bs}{\begin{subequations}}
\newcommand{\es}{\end{subequations}}

\newcommand{\bbe}{\boldsymbol{\mathrm{e}}}

\newcommand{\ie}{\text{\textschwa}}
\newcommand{\bbie}{\boldsymbol{\textbf{\textschwa}}}

\newcommand{\bxi}{\boldsymbol{\xi}}

\newcommand{\bA}{\boldsymbol{A}}

\newcommand{\bF}{\boldsymbol{F}}

\newcommand{\bk}{\boldsymbol{k}}

\newcommand{\bL}{\boldsymbol{L}}
\newcommand{\bM}{\boldsymbol{M}}
\newcommand{\bO}{\boldsymbol{O}}

\newcommand{\bT}{\boldsymbol{T}}
\newcommand{\bt}{\boldsymbol{t}}

\newcommand{\bX}{\boldsymbol{X}}

\newcommand{\diff}{\textrm{d}}

\newcommand{\bdiff}{\boldsymbol{\mathrm{d}}}
\newcommand{\bDiff}{\boldsymbol{\mathrm{D}}}

\newcommand{\lp}{\left(}
\newcommand{\rp}{\right)}
\newcommand{\lb}{\left[}
\newcommand{\rb}{\right]}

\newcommand{\nn}{\nonumber}

\newcommand{\+}{ \prescript{+}{}}
\newcommand{\m}{ \prescript{-}{}}
\newcommand{\g}{ \prescript{g_\pm}{}}
\newcommand{\x}{ \prescript{\pm}{}}
\newcommand{\viff}{\bdiff\tilde{\varphi}}

\DeclareMathSymbol{\mrq}{\mathord}{operators}{`'}

\begin{document}

\title{A $Spin(4)$ gauge theory of space, time, gravitation, matter and dark matter}

\date{\today}

\author{Tomi Koivisto}
\email{tomi.koivisto@ut.ee}
\address{Laboratory of Theoretical Physics, Institute of Physics, University of Tartu, W. Ostwaldi 1, 50411 Tartu, Estonia}
\address{National Institute of Chemical Physics and Biophysics, R\"avala pst. 10, 10143 Tallinn, Estonia}
\author{Lucy Zheng}
\email{luxi.zheng@ut.ee}
\address{Laboratory of Theoretical Physics, Institute of Physics, University of Tartu, W. Ostwaldi 1, 50411 Tartu, Estonia}
\author{Tom Zlosnik}
\address{Institute of Theoretical Physics and Astrophysics,
University of Gda\'{n}sk,
80-308 Gda\'{n}sk, Poland}

\begin{abstract}

A gauge-theoretic framework for spacetime and gravitation is proposed,    
in which a {\it Cartan khronon} field breaks the symmetry between space and time, enabling the emergence of temporality within a fundamentally
Euclidean setting. Based on a $Spin(4)$ gauge structure, the theory provides a real-valued formulation of chiral spacetime, wherein the effects typically attributed to dark matter may instead be accounted for by the dynamics of gravitation. 
New results are presented that are relevant to a broad range of phenomena, including cosmology, large-scale structure, gravitational waves, black holes, and potential signatures accessible to laboratory experiments.
%By avoiding the pitfalls of complexification and reinterpreting chiral spacetime geometry through a real, dynamical, and gauge-theoretic lens, the {\it Cartan khronon} theory offers a fresh and compelling framework for revisiting the foundations of spacetime and gravitation.

\end{abstract}

\maketitle 

\tableofcontents

\section{Introduction}
\label{intro}

The Euclidean path integral is a powerful and widely used tool in modern
theoretical physics, particularly in the study of quantum field theories. 
The formulation and  
development of gauge theories, such as quantum electrodynamics and quantum
chromodynamics have relied heavily on this method. 
%A more rigorous and mathematically well-defined formulation of the path integral for the fields in the standard model of particle physics, indispensable in the study of non-perturbative phenomena, is often only possible in Euclidean space. 
A Euclidean formulation of the path integral is often the most effective setting for studying non-perturbative phenomena in the standard model, offering a level of control that is generally unavailable in Lorentzian signature.
The bridge between the results obtained in Euclidean space
and the observables in Minkowski spacetime (when available) is established via a Wick
rotation, an analytic continuation that relates Schwinger functions in the
Euclidean regime to Wightman functions in the Lorentzian regime \cite{Osterwalder:1973dx,Osterwalder:1974tc}. 

One may also recall that the quantum fields describing particles in the standard model are most naturally associated with unitary representations not of the non-compact Lorentz group directly, but of the compact Euclidean rotation group, via the so-called unitarian trick. This mathematical structure further underscores the prominent role of Euclidean space in the formulation of quantum field theory.

%We may also recall that the quantum fields that describe particles in the standard model are most
%naturally associated with unitary representations not of the non-compact 
%Lorentz group directly, but of the compact Euclidean rotation group, via the so-called unitarian trick. This mathematical structure further underscores the central role that Euclidean space plays in the formulation of quantum field theory.

%In this light, it is natural to entertain the possibility that the Euclidean formulation of physics is not merely a convenient mathematical tool, but reflects a more fundamental description of reality.
In this light, it is natural to entertain the possibility that the Euclidean formulation of physics is not merely a convenient mathematical tool, but may provide an alternative, and perhaps equally fundamental, description of reality.
An expression of this profound perspective appears in a 1978 lecture by Stephen Hawking, reprinted 
in \cite{doi:10.1142/1301}:
\begin{displayquote}
In fact, one could take the attitude that quantum theory and indeed the
whole of physics is really defined in the Euclidean region and that it is
simply a consequence of our perception that we interpret it in the Lorentzian regime.
\end{displayquote}
Yet, integrating gravitation into this Euclidean framework introduces significant challenges. Unlike the gauge theories of particle physics, which are formulated on fixed spacetime backgrounds, gravity governs the dynamics of spacetime itself. The principle of general covariance in general relativity implies that there is no preferred time coordinate, complicating the straightforward application of Euclidean techniques to gravitational theories. Hawking immediately acknowledged this difficulty \cite{doi:10.1142/1301}:
\begin{displayquote}
I feel that one should adopt a similar Euclidean approach in quantum gravity
and supergravity. Of course one cannot simply replace the time coordinates by
imaginary quantities because there is no preferred set of time coordinates in
general relativity.
\end{displayquote}
The ``problem of time" is a generic structural issue in quantum gravity \cite{Isham:1992ms,Anderson:2012vk}, not unique to the Euclidean approach. However, the Euclidean context makes it particularly acute, as the formalism lacks any time direction, distinguished or otherwise, from the outset. While the Euclidean path integral remains a powerful tool, its application to gravity forces a confrontation with the fact that time in general relativity is not a background parameter but rather a subtle concept whose operational definition remains challenging and tied to the choice of coordinates.

 It is natural to seek a pregeometric foundation for a Euclidean description of spacetime, particularly in formulations where metric geometry is not fundamental but composed from a primordial spinor \cite{Wetterich:2021ywr,Wetterich:2021hru}. In this framework, the spacetime manifold and its geometric properties, such as distance and angle, emerge from bilinear combinations of a fundamental spinor field, without presupposing a metric structure \cite{Akama:1978pg,Wetterich:2003wr}. Within such an approach, even the signature of the effective metric need not be taken as a prior assumption but can itself be regarded as an emergent property, arising dynamically from the underlying spinor degrees of freedom. 
We will show that this can be consistently realised in a new gauge-theoretic formulation, where a spinor bilinear defines a solution-dependent temporal direction within an otherwise Euclidean four-dimensional structure \cite{Zlosnik:2018qvg}. 
%We will show that this can be consistently realised in the new kind of gauge theory that addresses the longstanding problem of time by allowing the spinor bilinear to spontaneously select a direction in an abstract 4-dimensional space, thereby giving rise to the arrow of time as an emergent feature . 
The proposal is that in such a theory, the distinction between Euclidean and Lorentzian signatures can be understood as a matter of perception arising from how the spinorial substrate is interpreted. 

%The new Lorentz gauge theory appears to be a rather unique setting in which such an approach is possible, or at the very least, it stands out as minimal in terms of assumptions. 
The new Lorentz gauge theory appears to provide a comparatively economical setting wherein to pursue such an approach. 
Conventional gauge theories of gravity often require a proliferation of auxiliary fields\footnote{However, related formulations of fermion–gravity couplings exist in which no tetrad or solder form is postulated from the outset. An approach grounded on spin-base invariance, which goes back to the early works of Schrödinger and Bargmann in 1932, has been developed more recently \cite{Gies:2013noa,Gies:2015cka,Lippoldt:2016ayw}, see also \cite{Weldon:2000fr,10.1063/5.0081140}. In these formulations the gravitational background is encoded in a spacetime metric via the Clifford algebra $\gamma_{(\mu}\gamma_{\nu)}=-g_{\mu\nu}\mathbb{1}$ and the (explicit) introduction of a solder form is avoided. The affine connection entering the fermionic covariant derivative is not introduced as a gravitational gauge field, though Palatini-type extensions with an independent connection can be considered, allowing for torsion \cite{Gies:2015cka}.}. 
The venerable and active framework of Poincaré gauge theory gravity  \cite{Kibble:1961ba,Shaposhnikov:2025znm} can also be formulated in terms of Yang-Mills-like gauge fields supplemented by a Higgs-like scalar field \cite{Grignani:1991nj,Koivisto:2022uvd}. 
This gauge structure can be seen to descend from a contraction limit of anti–de Sitter or de Sitter gravity frameworks which have the virtue of being based on semi-simple, and in fact simple symmetry groups \cite{Westman:2014yca,Addazi:2024rzo}. 
On top of the usual spin connection, this requires 5+4$\times$4 extra field components for the symmetry breaking field + the translation/extra-dimensional rotation gauge fields. Moreover, to describe the chiral aspect of spacetime, complexification is required in the conventional frameworks, and thereby the total number of real field components is further doubled. 
In contrast, the Lorentz gauge theory introduces, in addition to the spin connection, only the 4 real components of the {\it Cartan khronon} scalar field $\phi^I$ and grounds spacetime and gravity on the semi-simple and compact gauge group\footnote{For a recent different take on compact gravity structure group, see \cite{Partanen:2023dkt}.}. The khronon $\phi^I$ with the $SO(4)$ index $I$ can be regarded as a bilinear formed from the primordial spinor and the symmetry group taken as the double-cover $Spin(4)$.

%Despite its technical elegance, 
%An application for the formulation lies in its robust and unambiguous treatment of the flow of time, achieved with a minimal field content
This offers a minimal field content for gravity with an in-built clock field\footnote{Degrees of freedom corresponding classically to a pressureless perfect fluid have long been proposed as relational clocks in attempts to address the problem of time in quantum gravity \cite{Brown:1994py}, see in particular \cite{Husain:2011tk}. The dust is introduced as an additional matter component in order to define a physical time variable, in contrast to the present framework wherein we will see that  a (generalised) dust-like degree of freedom arises as an internal prediction of the $Spin(4)$ gauge theory in its geometric phase.}. There is no in-built dimension, but scales can always be introduced to a theory. As the units used to measure time and energy are fundamentally arbitrary, their operational relation to the underlying fields - such as the dimensionless $\phi^I$ - can be, in a precise technical sense, ``imaginary''. While the $\phi^I$ itself is a real scalar, assigning physical units to its evolution involves a transformation familiar from Wick rotation. This does not imply that the coordinates themselves are imaginary, but that the relation between unit conventions and field dynamics is encoded in an abstract, complex-valued structure. 
The recipe is as follows:
\begin{itemize}
\item Let the field break the symmetry as $\phi^I = \phi\delta^I_4$.
\item Identify the synchronous Euclidean time coordinate with the khronon\footnote{Let $U$ be such that $\diff\phi_{\mid U} \neq 0$, and appeal to the submersion theorem using
$\phi_{\mid U}$ to define $\tau$ as part of a coordinate system $(\tau,x^i)$ on $U$. Note that different synchronous coordinates in a patch $V$ are consistently matched within an overlap $U\cup V \neq 0$.} $\tau = \sqrt{\kappa}\phi$.  
\item Set the connection on-shell.   
\item Introduce the synchronous Lorentzian time $t \leftarrow -i\tau$, so that $\phi = m_P t$.  
\end{itemize}
The synchronous gauge is defined by the vanishing of lapse and shift. 
The essential point of the second step, however, is not a particular coordinate choice but the preferred foliation selected by the khronon field. Whenever $\diff \phi \neq 0$, the level sets of $\phi$ define this foliation in a coordinate-independent way. One may then introduce synchronous coordinates adapted to it, and in that adapted description $\phi$ is taken to be affine in the synchronous time parameter $\tau$. 
Thus, the identification above should not be viewed as a gauge-invariant relation between two arbitrary scalar functions, but as the canonical parametrisation of the foliation defined by $\phi$. In non-adapted coordinates, the same content is expressed as
$\phi = \sqrt{\kappa}\tau(\tau',x',y',z')$. The invariant content of the prescription is therefore the restriction to solution sectors for which the khronon defines such a foliation, not the explicit coordinate form of the relation.
%The identification in the second step can not be viewed as gauge-invariant per se, but rather as a well-defined gauge fixing whose physical content is gauge-independent because the synchronous time is uniquely determined. As the key secret indgredient of our recipe, it is worth reiterating that the physical content is captured by the {\it restriction to solutions for which the field $\phi$ is linear in $\tau$}. This classification is coordinate-independent, and one is free to describe the class of solutions in arbitrary coordinates - though, of course, the relation $\phi = \sqrt{\kappa}\tau'$ only holds for the unique synchronous time coordinate $\tau' = \tau$ and generalises to $\phi = \sqrt{\kappa}\tau(\tau',x',y',z')$ otherwise. 
%Note that the identifications in steps 2 and 4 are both manifestly gauge-invariant.  
The $\leftarrow$ denotes, not an equality but {\it substitution}. In particular, both $\tau$ and $t$ are real, and similarly, both $m_P$ and $\kappa^{-\frac{1}{2}}$ are real units of energy (in our convention the speed of light $c \equiv 1$, and as the notation suggests, $m_P$ will be identified with the Planck mass). 
At this stage, the introduction of the scales does not constitute additional physical input. They merely fix the units used to interpret the otherwise dimensionless khronon field 
(and the associated frame field as will be seen in section \ref{bartels}).
We could generalise the final step of the recipe by allowing rotations by an arbitrary angle in the complex plane, but we omit this as it is unnecessary for the purposes of the present article. Moreover, the choice of $\phi$ being linear in time(s) in the second and the fourth steps was made solely for clarity; the formulation remains manifestly coordinate-invariant throughout, and we will indeed implement the recipe with $\phi(x)$ taken as a more general function of the coordinates in the course of this work.
%We could generalise the last step of the recipe by considering rotation by an arbitrary angle in the complex plane, but omit this as unnecessary for the purposes of the present article. Also, it is only for clarity we chose the $\phi$ to be linear in time(s) in the last two steps above; our formulation is manifestly coordinate-invariant throughout and indeed we will implement the recipe also with $\phi(x)$ as a generic function of coordinates. 

A Wick rotation should not be confused with the unitarian trick. The former is a (possibly local) analytic continuation relating Euclidean and Lorentzian descriptions once a physical notion of time has been identified, whereas the latter is a global group-theoretic device motivated by the absence of non-trivial unitary representations of the Lorentz group. In the present framework, neither construction is required as a fundamental input. The theory is defined directly with a compact $Spin(4)$ gauge group, 
and the khronon field selects a canonical Euclidean time direction through the foliation defined by its level sets. In sectors where $\bdiff\phi\neq0$, the involution $\phi \rightarrow -\phi$ then induces a natural reflection of the foliation and thereby a corresponding reflection on all fields.
This suggests that the framework may provide a natural setting in which Osterwalder–Schrader reflection positivity could be formulated and assessed intrinsically, without relying on a background metric or on a background-specific Wick rotation.
 This should be contrasted with the standard Osterwalder–Schrader setting on flat $\mathbb{R}^4$, where global analytic continuation between Minkowski and Euclidean metrics relies on special properties of the underlying manifold \cite{Osterwalder:1973dx,Osterwalder:1974tc}. 
 %Here, no such global complexification is assumed. Instead, the dynamical emergence of time provides a natural setting in which Osterwalder–Schrader reflection positivity can be meaningfully formulated and assessed, without relying on background-specific Wick rotations. 
 It is also useful to clarify that we do not construct a dynamical Landau-Higgs-type symmetry-breaking mechanism, in which a potential drives the system from a symmetric to a symmetry-reduced vacuum. Rather, the $Spin(4)$ gauge theory admits qualitatively distinct classes of classical solutions. These include fully symmetric configurations, such as $\phi^I=0$, as well as solutions for which $\phi_I\phi^I\neq 0$, admitting a geometric interpretation and, when appropriate, a Lorentzian one. In this sense, the emergence of a preferred temporal direction should be understood as a reduction of symmetry at the level of solutions\footnote{Whether such solution sectors are dynamically or topologically connected -- for example via anomaly-induced effects or renormalisation-group (stochastic) flows of the effective couplings -- is an open question. In a cosmological setting it was proposed that the fully symmetric configuration with $\phi^I = 0$ could simply correspond to an initial condition for the universe \cite{Koivisto:2023epd}. Addressing possible transitions between sectors more rigorously in a general setting requires a well-defined action principle for the fundamental Spin$(4)$ variables, which is laid out in this article as a first step. A possible scenario involving dynamically varying couplings is briefly mentioned in Sec.~\ref{dymaxion}, but a detailed analysis is left for future investigation.}.

%\footnote{Whether such solution sectors are dynamically or topologically connected - for example via anomaly-induced effects or renormalisation-group (stochastic) flows of the effective couplings - is an open question. Addressing this requires a well-defined action principle for the fundamental Spin$(4)$ variables, which is laid out here as a first step.
%A possible scenario involving dynamically varying couplings is briefly mentioned in section \ref{dymaxion}, but a detailed analysis is left for future investigation.}, rather than as a dynamical phase transition in the conventional field-theoretic sense.

For concreteness, consider the example of a spatially flat radiation-dominated Friedmann-Lema{\^i}tre-Robertson-Walker (FLRW) universe, characterised by a scale factor that grows like the square root of the time coordinate. The square of the dimensionless scale factor can then be taken to be $\phi$, choosing an appropriate normalisation. So the line element in this case reads $\diff s^2 = \diff\tau\diff\tau + \phi\diff x_i\diff x^i = -\diff t\diff t + \phi\diff x_i\diff x^i$ 
in terms of the Euclidean time $\tau$ and the Lorentzian time $t$, respectively. The line element is manifestly real in both cases. If we insisted on using the dimensionless field $\phi$ as the time coordinate, we would write the line element as $\diff s^2 = \kappa\diff\phi\diff\phi + \phi\diff x_i\diff x^i = -m_P^{-2}\diff \phi\diff \phi + \phi\diff x_i\diff x^i$. This illustrates that the Wick-rotated picture can be interpreted as the Lorentzian observers living in imaginary space dimension or, equivalently, as measuring distances in imaginary energy units from the Euclidean perspective (or mutatis mutandis from the opposite perspective). A yet third expression of the same element in the two ``frames''
is $\diff s^2 = \diff\tau\diff\tau + \kappa^{-\frac{1}{2}}\tau\diff x_i\diff x^i = -\diff t\diff t + m_P t\diff x_i\diff x^i$.  Both ``frames'' are dynamically equivalent descriptions of the same field configurations, and as long as one doesn't confuse the units in one ``frame'' with the units in the other, both descriptions are real. Whereas the radiation-dominated example serves to illustrate the simple idea, across this article we'll construct much more nontrivial examples from first principles, including an exact rotating solution and the perturbed FLRW universe without any symmetries, and we'll see in all these examples that the physics can be consistently described in both the real Euclidean and the real Lorentzian ``frames''. 
These results support the conjecture that the geometric phase can be reached consistently, such that the physically relevant solutions of general relativity arise within the underlying $Spin(4)$ theory\footnote{Not all Euclidean solutions may be relevant. Counter-examples may be provided by well-known gravitational instantons without Lorentzian counterparts. (However, the `recipe' may avoid some of the known obstructions to consistent Wick rotations by the treatment of scales, in particular those that may appear in the metric components.)}.
%Therefore, we will confidently conclude that the geometric phase can be reached consistently, such that all physically relevant solutions of general relativity arise within the underlying $Spin(4)$ theory.

%Therefore, we will be confident to conclude that the above recipe is generally applicable within the class of solutions admitting a geometric phase.  

Significant progress and many interesting results have been obtained within the Euclidean quantum gravity programme, see e.g. \cite{doi:10.1142/1301,Hebecker:2018ofv,Lehners:2023yrj}. However, these advances often rely on complexified coordinates and/or metrics, coordinate-dependent (or tetrad-frame-dependent) constructions, or other auxiliary prescriptions introduced to recover the desired aspects of Lorentzian dynamics \cite{Visser:2017atf,Samuel:2015oea,Baldazzi:2018mtl,Singh:2020hpv,Kontsevich:2021dmb,Witten:2021nzp,Valtancoli:2023wab,BeltranJimenez:2024ufa,Banerjee:2024tap}. Such strategies, while often effective within specific contexts, reflect the fact that a fully coherent and compelling formulation of Euclidean quantum gravity remains elusive at the foundational level. One of the approaches considered is ``the real way'' wherein one introduces an extended action parameterised by two real numbers providing a handle to interpolate between Euclidean and Lorentzian solutions \cite{BarberoG:1995tgc}. Ref.\cite{Ashtekar:1995qw} defines a mapping $W$ on phase space spanned by $E,K$ such that\footnote{More explicitly, the triad and its conjugate would transform as $E^{a}_{i}\rightarrow i E^{a}_{i}$, $K_{a}^{i}\rightarrow -i K_{a}^{i}$. This can be interpreted as flipping the sign of the `space space' part of the metric in order to change signature.} 
for functions $f(E,K)$ on phase space we have $W(f(E,K)) = f(iE,-iK)$. A dynamical approach has been considered in the literature as well, wherein the gravitational metric is Euclidean but has a disformal relation, via a vector field or a gradient of a scalar field, to the Lorentzian spacetime metric which couples to matter \cite{Iliev:1998su,article,Girelli:2008qp,Mukohyama:2013ew,Svidzinsky:2015xbl,Nash:2023zza,Nash:2023zza,Nash:2025xhe,Feng:2025xsi}. In this approach, an extra field distinguishes the direction of time allowing a mapping from the postulated Riemannian to a pseudo-Riemannian metric. Our pregeometric theory, however, does not postulate the metric but constructs it from the Lorentz-covariant derivatives of the khronon field $\phi^I$ as a composite field. Perhaps closest to our approach is the Wick rotation in tangent space implemented upon the zeroth component of tetrad frame \cite{Samuel:2015oea} thus mapping real Lorentz metrics to real Euclidean metrics; in this proposal the time coordinate ambiguity is translated into Lorentz frame ambiguity\footnote{This is reminiscent of the situation in the context of the conjugate problem, ``the problem of energy'' in general-relativistic physics: though it was realised a long time ago that in a (teleparallel) tetrad formulation one obtains coordinate-invariant expressions for energy-momenta, the issue of frame-noncovariance was resolved only recently, see \cite{Gomes:2022vrc,Gomes:2023hyk} for some discussions and many references.}.  

In what follows, we develop the theoretical structure and physical implications of a $Spin(4)$ gauge theory. After introducing the basic mathematical ingredients, particularly the chiral decomposition and the construction of the covariant field strengths, we deduce the action principle that governs the dynamics of the Cartan khronon and the gauge connection. 
This sets the stage for a systematic exploration of the theory's vacuum structure, cosmological solutions, and spherically symmetric configurations.
To demonstrate the theory in a concrete and tractable setting, we focus in section \ref{righthand} on the case of right-handed gravity, which serves as a proving ground for key conceptual features, including the emergence of time and the recovery of standard gravitational dynamics. Sections \ref{cosmology} and \ref{largescale} are devoted to cosmology, where we analyse the $Spin(4)$ theory in full generality, constructing the perturbation formalism systematically up to linear order. Section \ref{spherical} turns to the spherically symmetric case, laying the groundwork for further exploration
of exact inhomogeneous configurations. Matter couplings and spinor structures are addressed in \ref{mattercouplings}. As the examples will show, the proposed framework enables a consistent real-valued description of spacetime, gravitational dynamics, and dark matter phenomena. We close in section \ref{perspectives} with an outlook on near-term observational handles that could falsify or corroborate the theory.  For a full overview of the material, we refer the reader to the table of contents above. Readers who prefer a visual roadmap may wish to glance at the schematic diagram in section \ref{perspectives}, which summarises the structure of the theory and the implementation of the above recipe.

\section{Formulation of the theory}
\label{formulation}

In this section, we first introduce the relevant mathematical formalism that allows one to split the $Spin(4)$ gauge connection 
into two $SU(2)$ connections, based on the reducibility of the algebra $\mathfrak{so}(4)=\mathfrak{so}(3)\oplus\mathfrak{so}(3)$.  
Then we consider the possible action principles that can be constructed from the khronon field $\phi^I$ using the
gauge-covariant derivative.

\subsection{The reducibility of the Lorentz group}
\label{reducibility}

A generic object $X$ in the adjoint representation of the Euclidean Lorentz algebra $\mathfrak{so}(4)$ is specified by its components $X^{IJ}=-X^{JI}$, where the indices $I,J,\dots$ run from 1 to 4. The object can be split into its self-dual and anti-self-dual components by using the appropriate projectors. These projectors are
\be \label{projectors}
P_\pm^{IJ}{}_{KL} \equiv \frac{1}{2}\lp \delta^{[I}_K\delta^{J]}_L \pm \frac{1}{2}\epsilon^{IJ}{}_{KL}\rp \equiv \frac{1}{2}\lp 1 \pm \star\rp^{IJ}{}_{KL}\,,
\ee
and we denote $\x X^{IJ} \equiv P_\pm^{IJ}{}_{KL}X^{KL}$. It follows that $\star \x X^{IJ} = \pm \x X^{IJ}$, which is why they are called self-dual and anti-self-dual, respectively. One can easily check that the $P_\pm^{IJ}{}_{KL}$ indeed are projectors, i.e. they are complete, $X=\+ X + \m X$, orthogonal, $\prescript{+-}{}X = \prescript{-+}{}X=0$ and idempotent, $\prescript{++}{}X=\+X$, $\prescript{--}{}X=\m X$. In the definitions
(\ref{projectors}), we have referred to the two $SO(4)$-invariant structures, the Kronecker metric $\delta^I_J$ and the totally antisymmetric Levi-Civita symbol $\epsilon^{IJ}{}_{KL}$, for which our conventions are such that $\delta^I_J = \text{diag}(1,1,1,1)^I{}_J$ and $\epsilon^{12}{}_{34}=1$. From now on, throughout this article, we may nonchalantly lower and raise $SO(4)$ indices as this is always understood wrt ($\equiv$ with respect to) the Kronecker metric $\delta_{IJ}$. 

The generators $o_{IJ}$ of the algebra 
\be \label{algebra1}
[o_{IJ},o_{KL}] =4\delta_{[L[I}o_{J]K]} = 2\lp\delta_{L[I}o_{J]K} - \delta_{K[I}o_{J]L}\rp = \delta_{LI}o_{JK} - \delta_{LJ}o_{IK} - \delta_{KI}o_{JL} +\delta_{KJ}o_{IL}\,, 
\ee
could be represented with the antisymmetric matrices $(o_{IJ})^K{}_L = 2\delta^K_{[I}\delta_{J]L}$; or as well like $o_{IJ} = 2x_{[I}\partial_{J]}$; in \ref{gammas} we will recall the appropriate spinor representation; and in the appendix \ref{quaternion} we consider a quaternionic matrix representation.   

There's no canonical split into rotations and boosts, the {\it Cartan khronon} will be needed for that. At this point, we'll just choose to denote rotations around the 4th axis in a different way, and 
introduce the small Latin letters for the other three indices, so that $i,j,\dots$ run from 1 to 3. Let us also introduce 3-dimensional totally antisymmetric Levi-Civita symbol $\epsilon_{ijk} \equiv \epsilon_{ijk4}$, and allow ourselves to freely raise and lower the 3-dimensional indices with the 3-dimensional Kronecker metric $\delta_{ij}$. 
Then we define
\bs
\ba
o^i  & \equiv & \frac{1}{2}\epsilon^{4ijk}o_{jk} = -\frac{1}{2}\epsilon^{ijk}o_{jk}\,, \\ 
b^i & \equiv & -o_4{}^i\,,
\ea
\es  
so that we can rewrite the algebra (\ref{algebra1}) as
\be \label{algebra2}
[o^i,o^j]=\epsilon^{ijk}o_k\,, \quad [o^i,b^j] = \epsilon^{ijk}b_k\,, \quad [b^i,b^j] = \epsilon^{ijk}o_k\,.  
\ee
Then as usual we define the linear combinations of the generators,  
\bs
\ba
l^i & \equiv & \frac{1}{2}\lp o^i + b^i\rp\,, \\ r^i & \equiv & \frac{1}{2}\lp o^i - b^i\rp\,, 
\ea
\es
so that we can yet rewrite (\ref{algebra2}) as
\be \label{algebra3}
[l^i,l^j]=\epsilon^{ijk}l_k\,, \quad [r^i,r^j] = \epsilon^{ijk}r_k\,, \quad [l^i,r^j] = 0\,.  
\ee
Now going back to (\ref{projectors}), we can verify that
\bs
\label{rlgen}
\ba
\+ o^{4i} & = & \frac{1}{2}\lp o^{4i} - \frac{1}{2}\epsilon^i{}_{jk}o^{jk}\rp = \frac{1}{2}\lp -b^i + o^i  \rp = r^i\,, \\ 
\m o^{4i} & = & \frac{1}{2}\lp o^{4i} + \frac{1}{2}\epsilon^i{}_{jk}o^{jk}\rp = \frac{1}{2}\lp -b^i - o^i\rp  = -l^i\,. 
\ea
\es
Since parity reverses boosts (electric/polar/odd) but not rotations (magnetic/axial/even), it is clear that the parity operator
\be
P_{\leftrightarrow}^{IJ}{}_{KL} = \text{diag}(-1,-1,-1,1)^I_K \text{diag}(-1,-1,-1,1)^J_L\,,
\ee
flips $l^i \leftrightarrow r^i$. It then seems that $P_\leftrightarrow P_\pm = - P_\mp$, but it does hold that $P_\leftrightarrow \star P_\pm = \star P_\mp$. Note that there inevitably exist parity-duality-odd objects $X$ and parity-duality-even objects $\star X$. Which is which may depend on specific conventions, but either choice can be implemented consistently. 
 
%I think this is consistent since one considers $P_\leftrightarrow [\x o^{4i}] = (P_\leftrightarrow P_\pm ) (P_\leftrightarrow o^{4i}) = (-P_\mp)(-o^{4i}) = \prescript{\mp}{}o^{4i}$.  
To be explicit, we can decompose any element $X \in \mathfrak{so}(4)$ as
\bs
\ba
X & = & \frac{1}{2}X^{IJ}o_{IJ} = \frac{1}{2}\lp \+ X^{IJ}\+ o_{IJ} + \m {X}^{IJ}\m o_{IJ}\rp =  \+ X^{4i}\+ o_{4i} + \frac{1}{2}\+ X^{ij}\+ o_{ij}
+ \m X^{4i}\m o_{4i} + \frac{1}{2}\m X^{ij}\m o_{ij} \nn \\
& = & \lp \+ X^{4k} - \frac{1}{2}\+X^{ij}\epsilon^k{}_{ij}\rp\+ o_{4k} +  \lp \m X^{4k} + \frac{1}{2}\m X^{ij}\epsilon^k{}_{ij}\rp\m o_{4k} \equiv  \+ X^i r_i + \m X^i l_i\,,
\ea
where
\be \label{LRcomponents}
\x X^i = \pm 2\x X^{4i} = -\epsilon^i{}_{jk}\x X^{jk} =  \pm X^{4i} - \frac{1}{2}\epsilon^i{}_{jk}X^{jk}\,. 
\ee
\es
These are our conventions for representing the $X$ equivalently in the adjoint representation of $\mathfrak{so}(3)+\mathfrak{so}(3)$  (which in this case is equivalent to the sum of two fundamental representations). We may refer to this representation as the left/right basis. 

Had we chosen instead the $SO_{\mathbb{C}}(1,3)$ structure group, the differences would begin from (\ref{projectors}), which would be modified like $\star \rightarrow -i\star$, the imaginary unit $i$ being required for consistency of the projections in the Lorentzian regime. In this article we stick to the all-plusses signature and to a strictly real formulation; for a pedagogical review of formulations with different signatures, see \cite{Krasnov:2020lku}.

\subsection{The gauge field and field strength}

The gauge field, also known as (the pullback of) the connection, plays a main role in gauge theory. The connection is a 1-form, and we will denote as $\bA$. For clarity, we use bold symbols for $n$-forms when $n>0$. Valued in the adjoint representation of the $\mathfrak{so}(4)$ algebra, $\bA = {\tiny{\frac{1}{2}}}\bA^{IJ}o_{IJ}$, and it is conventional to refer to $\bA$ as the spin connection (though conventionally, it takes values in the $\mathfrak{so}(1,3)$ algebra). According to the development in subsection \ref{reducibility} above, the components of the spin connection in the left/right basis are related to its components in the anti/self-dual basis as 
\bs
\ba
\x \bA^1 & = & \pm \bA^{41} - \bA^{23} = \pm 2\x \bA^{41} = - 2\x \bA^{23} \,, \\
\x \bA^2 & = & \pm \bA^{42} + \bA^{13} = \pm 2\x \bA^{42} = + 2\x \bA^{13}\,, \\
\x \bA^3 & = & \pm \bA^{43} - \bA^{12} = \pm 2\x \bA^{43} =  - 2\x \bA^{12}\,.
\ea
\es 
The connection defines the gauge-covariant derivative $\bDiff = \bdiff + \bA$, wherein the $\bdiff$ is the usual exterior derivative. 
For example, consider the $SO(4)$-covariant derivative of a particular component of an anti/self-dual scalar (meaning a 0-form) element $\x X^{IJ}$,  
\ba
2\bDiff\x X^{41} & = & \bDiff \lp \pm X^{41} - X^{23} \rp  = \pm \lp \bdiff\bX^{41} + \bA^4{}_2 X^{21} + \bA^4{}_3 X^{31} + \bA^1{}_2 X^{42} + \bA^1{}_3X^{43}\rp \nn \\
& - & \lp \bdiff X^{23} + \bA^2{}_1 X^{13} + \bA^2{}_4 X^{43} + \bA^3{}_1 X^{21} + \bA^3{}_4 X^{24}\rp \nn \\
& = & \bdiff\lp \pm X^{41} - X^{23}\rp  + \x \bA^2\lp X^{21} \pm X^{43}\rp + \x \bA^3\lp X^{31} \pm X^{24}\rp \nn \\ 
& =  & \bdiff\x X^1 + \x\bA^2 \x X^3 - \x\bA^3 \x X^2\,.
\ea
This demonstrates that for any $\bX^{IJ}$ it holds that
\be
\pm 2\bDiff\x \bX^{4i} = \x \bDiff\x \bX^i = \bdiff\x \bX^i + \epsilon^i{}_{jk}\x \bA^j\wedge\x \bX^k\,. 
\ee
We don't need different notations for the covariant derivative acting on different objects: the connection coefficients are determined by the covariance group of the respective object. So, for example
\bs
\ba
\bDiff \x \bX^i & = & \bdiff \x\bX^i + \epsilon^i{}_{jk}\x\bA^j\wedge\x\bX^k\,, \\
%\bDiff \bX^a & = & \bdiff \bX^a + \epsilon^a{}_{bc}\bA^b\wedge\bX^c\,, \\
%\bDiff \bX^{\bar{a}} & = & \bdiff \bX^{\bar{a}} + \epsilon^{\bar{a}}{}_{\bar{b}\bar{c}}\bA^{\bar{b}}\wedge\bX^{\bar{c}}\,, \\
\bDiff \bX^I & = & \bdiff \bX^I + \bA^I{}_J\wedge\bX^J\,, 
\ea
\es  
and this generalises to arbitrary tensors with mixed indices as well. The appropriate derivatives of the connections are called the gauge field strengths, or, in more geometrical language, the curvatures\footnote{With the connections, one has to recall that they're not covariant, and therefore the covariant derivatives of their components do not make sense, but the components of their covariant derivatives are well-defined. If taking this into account, we can state $\bF \equiv \bDiff\bA$ and $\bDiff\bF=0$. The $\equiv$ symbol in (\ref{sosocurv}) and (\ref{toinencurv}) are appropriate from a geometrical perspective, so we have included it for clarity, though when allowing also an algebraic perspective wherein multiplication is understood as commutation, i.e. $\bDiff\bA = \bdiff\bA + [\bA,\bA]$, the equality of the 2nd $=$ 3rd expressions follows from (\ref{algebra1}). Thus, instead of (\ref{sosocurv}) we better understand $\x \bF^{i} =  (\bDiff\x\bA)^i = \bdiff\x\bA^i + \frac{1}{2}\epsilon^i{}_{jk}\x\bA^j\wedge\x\bA^k $ and instead of (\ref{toinencurv}), $\bF^{IJ} \equiv (\bDiff\bA)^{IJ} \equiv \bdiff\bA^{IJ} + \bA^I{}_K\wedge\bA^{KJ}$. However, since we don't need the elegance of the algebraic perspective here, it is sufficient to deal with the additional definitions like (\ref{sosocurv}) and (\ref{toinencurv}).}. The components of the curvatures are given as
\bs
\ba
\x \bF^{i} & \equiv & \bdiff\x\bA^i + \frac{1}{2}\epsilon^i{}_{jk}\x\bA^j\wedge\x\bA^k = \pm 2\x\bF^{4i}\,,  \label{sosocurv}
%\bF^{a} & = & (\bDiff\bA)^a = \bdiff\bA^a + \epsilon^a{}_{bc}\bA^b\wedge\bA^c = 2\+\bF^{4a}\,, \\
%\bF^{\bar{a}} & = &  (\bDiff\bA)^{\bar{a}} = \bdiff \bA^{\bar{a}} + \epsilon^{\bar{a}}{}_{\bar{b}\bar{c}}\bA^{\bar{b}}\wedge\bA^{\bar{c}}
%= -2\m \bF^{4\bar{a}}\,,
\ea
where, as usual
\be
\bF^{IJ} \equiv \bdiff\bA^{IJ} + \bA^I{}_K\wedge\bA^{KJ}\,, \label{toinencurv}
\ee
\es 
and the covariant derivative of any curvature is zero, e.g. $\bDiff\bF^{IJ}=\bDiff\x\bF^i=0$. 
%$\bDiff\bDiff\bDiff=0$ \cmt{this is taken to apply when $\bDiff\bDiff\bDiff=0$ is applied to certain fields? E.g. $\bDiff\bDiff\bDiff V^{I}$ might be non-zero for some vector $V^{I}$}. 
This is due to the Jacobi identity satisfied by Lie algebras, and its consequences in geometry are known as Bianchi identities. 

\subsection{The action principle}

As mentioned in the introduction, the fundamental field of the theory can be regarded as the (Euclidean) spinor $\psi$. However, for the purposes of the current article we formulate the theory directly in terms of $\phi^I =\bar{\psi}\gamma^I \psi$ (the spinor notations are clarified in section \ref{mattercouplings}), so that the {\it Cartan khronon} field $\phi^I$ is a dimensionless scalar in the fundamental representation of the $Spin(4)$ torsor. By torsor, we simply mean the group without an identity element, and the physical 
relevance of this distinction is that the field $\phi^I$ lives in a space without an origin. %\cmt{This seems important! It means that in this framework we cannot, for example, interpret $\phi_{I}\phi^{I}$ as a clock field?} 
Since only differences in $\phi^I$ are physical, the action should not change under constant shifts of the field. In particular, we will interpret the khronon as a clock field which measures the lapse of time in terms of relative durations between events rather than giving an absolute reading associated with a given event. Thus, we require invariance of the action principle under global $T(4)$ translations, defined as $\phi^I \rightarrow \phi^I + \xi^I$, wherein $\bDiff \xi^I = \bdiff \xi^I + \bA^I{}_J\xi^J=0$. 

This restriction taken into account, the building blocks for the action principle are the 1-form $\bDiff\phi^I$, and the two independent 2-forms $\x\bF^{IJ}$. %The action should further involve only even powers of the 1-form $\bDiff\phi^I$. 
Of course, the action principle should be $SO(4)$-invariant as well, and in 4 spacetime dimensions, the action should be an integral over a 4-form. In the end, we are then allowed only a 6-parameter action. This action turns out to contain only even powers of the 1-form $\bDiff\phi^I$, and therefore has no danger of breaking potential reflection symmetries in the matter sector, as it realises the Z$_2$ symmetry\footnote{The clock field in the models of Mukohyama {\it et al} also implement both a shift symmetry and a reflection symmetry \cite{Mukohyama:2013ew,Feng:2025xsi}.} under $\phi^I \rightarrow -\phi^I$. 

Let us consider this in detail. Firstly, without the khronon field we may only write down the two independent 4-form invariants,
\bs
\label{actions}
\ba \label{topo1}
I_{(0)} & =  & g_{G}\int \epsilon_{IJKL}\bF^{IJ}\wedge\bF^{KL} + g_P\int \bF^{IJ}\wedge\bF_{IJ}  \nn \\
& = &  \lp g_P + 2g_G\rp \int \+\bF^{IJ}\wedge\+\bF_{IJ}  + \lp g_P - 2g_G\rp \int \m\bF^{IJ}\wedge\m\bF_{IJ} \nn \\
& = &   \oint \lp g_G\epsilon_{IJKL} + g_P\eta_{IK}\eta_{JL}\rp\lp \bA^{IJ}\wedge\bF^{KL} - \frac{1}{3}\bA^{IJ}\wedge\bA^{K}{}_M\wedge\bA^{ML}\rp\,.
\ea
In the second line we have expressed the action in terms of the anti/self-dual curvatures, and in the third line shown explicitly how the action reduces to a boundary term (under the assumption that the associated de Rham cohomology class vanishes).
Thus, the 4-form integrand in $I_{(0)}$ is a total derivative of a 3-form. The $g_G$-term is known as the Gauss-Bonnet term and the $g_P$-term is known as the Pontryagin term. 
%The corresponding terms for the anti/selfdual curvatures are by construction be proportional. 
Note that we cannot write these terms in the left/right basis, considering 4-forms like $\sim \int \x \bF^i\wedge\x\bF_i$, because no canonical split is available.  

We saw that only topological field theories may be constructed without the building block $\bDiff\phi^I$. With that at hand, we can build the 4-form integrand,
\be
I_{(4)} = -\frac{\lambda}{4!}\int \epsilon_{IJKL}\bDiff\phi^I\wedge\bDiff\phi^J\wedge\bDiff\phi^K\wedge\bDiff\phi^L = -\lambda\int \star 1\,. 
\ee
This will turn out to correspond to a cosmological constant term. To have dynamics for the spin connection, we need to consider yet the two quadratic terms contained in
\ba \label{I2}
I_{(2)} & = & \frac{1}{2}\int \epsilon_{IJKL}\bDiff\phi^I\wedge\bDiff\phi^J\lp g_+ \+\bF^{KL} +  g_- \m\bF^{KL}\rp  =  \int \bDiff\phi^I\wedge\bDiff\phi^J\wedge\lp g_+ \+\bF_{IJ} -  g_- \m\bF_{IJ}\rp\,.  
\ea
This parameterisation, introducing the two coupling constants $g_\pm$ for the anti/self-dual field strengths $\x\bF^{IJ}$, respectively, was introduced in \cite{Nikjoo:2023flm}. Besides the topological invariants (\ref{topo1}), there is yet the torsional Nieh-Yan term
\be
I_{(NH)} = \tilde{g}\int\lp \bF_{IJ}\wedge\bF^{IK}\phi^J\phi_K - \bF_{IJ}\wedge\bDiff\phi^I\wedge\bDiff\phi^J\rp = \tilde{g}\oint \bDiff\phi_I\wedge\bF^{IJ}\phi_J\,.  
\ee
\es
It is not difficult to see that (\ref{actions}) completely exhausts all possible actions which are compatible with the local $SO(4)$ symmetry and the global $T(4)$ invariance. 

If we did not insist on the latter, the action
densities in (\ref{actions}) could be multiplied by arbitrary polynomials of the scalar singlet $\phi^I\phi_I$. We also note that if the global shift invariance was allowed to be respected only up to 
a boundary term, we could also consider the actions
\bs
\label{actions2}
\be
\tilde{I}_{(2)} = -\int\phi^I\phi^J\bF^{K}{}_J\wedge\lp \tilde{g}_+\+\bF_{IK} - \tilde{g}_-\m\bF_{IK}\rp \overset{\text{b}}{=} I_{(2)}\,, \label{topo2}
\ee
and 
\be \label{action2cc}
\tilde{I}_{(4)} = \frac{\tilde{\lambda}}{8}\int\epsilon_{IJKL}\phi^I \phi^M\bF^J{}_M\wedge\bDiff\phi^K\wedge\bDiff\phi^L  \overset{\text{b}}{=} I_{(4)}\,,
\ee
\es
where $\overset{\text{b}}{=}$ means equivalence up to a boundary term (and setting $\tilde{g}_\pm=g_\pm$, $\tilde{\lambda}=\lambda$). For the special choice $\tilde{g}_-=-\tilde{g}_+=\tilde{g}$ the boundary term implied in (\ref{topo1}) involves the Nieh-Yan invariant (\ref{topo1}). For this special choice, the global shift invariance is reinstated into the action (\ref{topo2}), but in general one may consider linear combinations which split the torsional invariant into its anti/self-dual halves in a way which is shift-invariant only up to a boundary term.     
If we broke the shift invariance with functions of $\phi^I\phi_I$, this equivalence would also be broken, and such generalisations of (\ref{actions2}) would thus introduce yet new classes of extended theories. However, in this article, we insist on the exact shift invariance. Note that this invariance may not be thought of as a symmetry in the usual sense. It is a transformation that leaves the action invariant (in the cases (\ref{actions2}) only up to boundaries) even when the equations of motion don't apply, but the constraint $\bDiff\xi^{I}=0$ means that the parameters $\xi^{I}$ may ultimately depend non-locally on the connection field $\bA$, and this can manifest in the equations of motion not being invariant under this transformation \cite{Glavan:2024svx}. Further, a constant $\xi^I$ may exist only in some special cases.

To recapitulate, the most general action for the $Spin(4)$ gauge theory is dictated by the invariance principle and includes the six dimensionless parameters $g_G$, $g_P$, $\tilde{g}$, $g_\pm$ and $\lambda$. The action is a functional of the two fields, $\bA$ and $\phi$. We also include the possible source terms for these 
fields, an energy-momentum current $\bt_I$ and a spin current $\bO_{IJ}=-\bO_{JI}$ respectively. At this point, these currents are introduced as a parametrisation for the material sources. In Section \ref{EandScurrents} their explicit form will be derived from the Lagrangian for the matter fields.
The action is   
\be \label{TheAction}
I = I_{(0)} + I_{(NH)} + I_{(2)} + I_{(4)} - \int\bDiff\phi^I\wedge\bt_I - \int\bA^{IJ}\wedge\bO_{IJ}\,, 
\ee
where the  $I_{(0)}$, $I_{(2)}$ and $I_{(4)}$ are as in (\ref{actions}). At the classical level, the dynamics of the theory only depend on the coupling constants $g_\pm$ and $\lambda$. The action (\ref{TheAction}) can be called {\it pregeometric} because it admits solutions that do not describe a metric manifold. The completely symmetric solution $\phi^I=0$ is the basic example. More generally, the theory can have solutions with non-vanishing $\bDiff\phi^I$ from which one still cannot compose an invertible coframe field and thus lacks a non-degenerate metric. A phase of the theory which admits an invertible tetrad could thus be characterised as a `geometric phase' of the theory. In such a phase, one can recover a standard metric spacetime manifold in terms of a soldered bundle. We describe this in detail in the next section. 

\section{Right-handed gravity}
\label{righthand}

In this section, we illustrate the workings of the pregeometric theory in terms of a simplified model. We set $g_G=g_P=\tilde{g}=g_-=\lambda=0$ and let $g_+=1$, and also ignore the possible material spin currents described by $\bO_{IJ}$. Thus, we study the theory
\be \label{action+}
I_+ = \int \bDiff\phi^I\wedge\bDiff\phi^J\wedge\+\bF_{IJ} - \int\bDiff\phi^I\wedge\bt_I\,.
\ee
We will first investigate the most basic examples of solutions, the vacuum solutions reducing to the Minkowski and the case of flat FLRW (Friedmann-Lema\^{i}tre-Robertson-Walker) cosmology. These cases have been solved in the previous literature on Lorentz gauge theory, but the important subtlety we introduce in this section is their derivation via Wick rotation from the Euclidean space. We then proceed to study more involved spacetimes and present the explicit Kerr solution in the Lorentz gauge theory.

\subsection{Spacetime structure}

The EoM (equations of motion) derived by varying $I_+$ wrt the two fields, the khronon and the connection, are respectively
\bs
\label{EoM1}
\ba
\bDiff\lp 2\+\bF_{IJ}\wedge\bDiff\phi^J - \bt_I\rp & = & 0 \,, \label{khronon1} \\
\bDiff\+{}\lp \bDiff\phi^I\wedge\bDiff\phi^J\rp & = & 2\phi^{[I}\+\bF^{J]}{_K \wedge \bDiff\phi^K} - \phi^{[I}\bt^{J]}\,. 
\ea
\es
The khronon field equation (\ref{khronon1}) states that a certain 3-form is a (covariantly) closed form. We call this 3-form $\bM_I$ and use it also in the connection EoM. 
Thus, we can rewrite the EoM (\ref{EoM1}) as
\bs
\label{EoM2}
\ba
2\+\bF_{IJ}\wedge\bDiff\phi^J & = & \bt_I + \bM_I\,, \\
\bDiff\+{}\lp \bDiff\phi^I\wedge\bDiff\phi^J\rp & = & \phi^{[I}\bM^{J]}\,, \label{EoM2B}\\
\bDiff\bM_I & = & 0\,.  
\ea
\es
Until here, the equations look identical regardless of whether the gauge group $SO_\mathbb{C}(1,3)$ or $SO(4)$ is considered. When breaking the symmetry to $SO(3)$ by imposing that $\phi^I = \phi\delta^I_0$,
some signs become different, and most importantly, the imaginary unit which is necessary in the $SO_\mathbb{C}(1,3)$ formulation, is now absent. 

Using the formalism introduced in \ref{reducibility}, we can express the system (\ref{EoM2}) in the symmetry-broken phase in the form
%Fix the gauge $\phi^A = \delta^A_4\tau$. Then we can write
\bs
\label{efepre}
\ba
\+\bF_i\wedge\bDiff\phi^i & = & \bt_4 + \bM_4\,, \label{efe1pre} \\
\+\bF_i\wedge\bdiff\phi - \frac{1}{2}\epsilon_{ijk}\+\bF^j\wedge\bDiff\phi^k & = & -\bt_i -\bM_i\,, \label{efe2pre} \\
\lp \+\bF_i-\m\bF_i\rp\wedge\bdiff\phi + \epsilon_{ijk}\lp\+\bF^j-\m\bF^j\rp\wedge\bDiff\phi^k & = & 2\bM_i = 0\,, \label{efe3pre}  \\
\bDiff\bM_I & = & 0\,.
\ea
\es
The two equations in (\ref{efe2pre},\ref{efe3pre}) correspond to the ($4i$) and the ($ij$) components of the equation (\ref{EoM2B}), and their combination sets $\bM^i=0$ to zero. Thus, the 3-form $\bM^I$ is aligned with $\phi^I$ in the absence of spin sources in the right-handed gravity model wherein $g_-=0$. We return to this below.

\subsubsection{The Bartels frame}
\label{bartels}

At this point, it is convenient to take a further step toward the conventional spacetime structure. We have a distinguished direction for time measured by $\phi$, the appearance of the three 1-forms $\bDiff\phi^i$ suggests their identification with the Bartels frame\footnote{Martin Bartels (1769-1836) can be regarded as a key figure behind the discoveries of non-Euclidean geometries \cite{keyfigure}. He also developed a method of moving frames and, for example, derived the Frenet-Serret formulas \cite{LUMISTE199746}. The $\bbe^i$ that we call the Bartels frame can be obtained as a spatial pullback of the 4-dimensional $\bbe^I$ which is generally known as the Cartan('s moving) frame, after Eli{\'e} Cartan (1869-1951) who greatly developed further the existing geometrical methods \cite{akivis2011elie}.}, a spatial triad of frames that may span a 4-dimensional spatial metric $\sim \bDiff\phi^i\otimes\bDiff\phi_i$ on the 3-dimensional hypersurfaces orthogonal to the flow of time. The conventional (co)frame fields\footnote{On terminology: in this article we call the 1-forms $\bbe^I$ frame fields. Sometimes tetrads, which would be the inverse vectors of those 1-forms, are called the frame and the quartet $\bbe^I$ is then called the coframe or the cotetrad.}, however, are dimensionless, whereas $\bDiff\phi^i = -\bA^{4i}\phi$ has the dimension of energy (= the dimension of one per length or one per time unit). Let us simply introduce an arbitrary mass unit, let it be given by $\kappa^{-1/2}$, in terms of which to define the dimensionless Bartels frame. In terms of this mass unit, we can locally identify the khronon field with a coordinate $\tau$ in the conventional units of one per energy. 
%In the following, we shall denote derivatives wrt $\tau$ with a dot, since in the end $\tau$ can be identified with the Euclidean time. 
Thus,  
\bs \label{deftetrads}
\ba
\bdiff\phi & \equiv & \kappa^{-\frac{1}{2}}\bdiff\tau\,, \\  
\bbe^i & \equiv & \sqrt{\kappa}\bDiff\phi^i =  -\frac{\tau}{2}\lp \+\bA^i - \m\bA^i\rp\,. 
\ea
\es       
It follows that the torsion of the Bartels frame is a gauge-invariant measure of the chiral asymmetry, 
\be
\bT^i \equiv \bDiff\bbe^i = -\frac{\tau}{2}\lp \+\bF^i - \m\bF^i\rp\,.
\ee
In the manifestly $SO(4)$-covariant form, the above definitions read 
\bs
\ba
\bbe^I & \equiv & \sqrt{\kappa}\bDiff\phi^I\,, \\
\bT^I & \equiv & \bDiff\bbe^I = \sqrt{\kappa}\bF^I{}_J\phi^J\,. 
\ea
\es
The torsion is always orthogonal to the khronon, since $\phi_I \bT^I =\sqrt{\kappa}\phi^I\phi^J\bF_{IJ}=0$. 
Note that these identities are kinematical and independent of the dynamics governed by the action chosen for the theory. We note that the torsion appears, for example, in the cosmological constant term (\ref{action2cc}), reflecting the generalised structure of spacetime in the present framework.

Let us remark in passing that one could also adopt an alternative prescription for dimensional assignments. In particular, the introduction of the length scale $\sqrt{\kappa}$ (as well as the energy scale $m_P$) could in principle be avoided by subscribing to a formulation of physics using unconventional but ``natural'' dimensionalities \cite{Volovik:2020rjz}. 
In such a description, the Bartels frame $\bDiff\phi^i$
 would instead carry the dimension of energy, while the metric would have the dimension of energy squared. Correspondingly, time intervals and spatial lengths would be dimensionless. From our perspective, the ``natural'' dimensionless variables belong to the fundamental pregeometric sector of the theory, while the dimensionful variables that we employ correspond to the physically intuitive observables emerging in the geometric phase of the theory. The assignment of physical dimensions is essential for the emergence of a Lorentzian spacetime interpretation in our framework.

\subsubsection{The Minkowski solution}

Now we may return to the dynamics of the minimal right-handed gravity model. Though $\bM_i=0$, the conservation equations
$\bDiff\bM^i = \bA^{i4}\wedge\bM_4 = -\kappa^{-1/2}\bbe^i\wedge\bM_4 = 0$ nevertheless yield nontrivial  constraints, and in particular restrict the possibly nonzero component $\bM^4$ to be given by a single scalar function $\hat{\rho}$ such that
(see section 3.1 of Ref.\cite{Koivisto:2023epd} for a detailed derivation)
\be
\bM_4 = -\sqrt{\kappa}\hat{\rho}\star\bdiff\tau\,, \quad \text{where} \quad \star\bdiff\tau = -\frac{1}{6}\epsilon_{ijk}\bbe^i\wedge\bbe^j\wedge\bbe^k\,.   
\ee 
The formula for $\star\bdiff\tau$ follows directly from the identifications above when $\star$ is understood as the $SO(4)$ Hodge operator that was already used in (\ref{projectors}). Now we may put the system of equations (\ref{efepre}) in the form 
\bs
\label{efe}
\ba
\+\bF_i\wedge\bbe^i & = & \sqrt{\kappa}\bt_4 - \kappa\hat{\rho}\star\bdiff\tau\,, \label{efe1} \\
\+\bF_i\wedge\bdiff\tau - \epsilon_{ijk}\+\bF^j\wedge\bbe^k & = & -\sqrt{\kappa}\bt_i\,, \label{efe2} \\
\bT_i\wedge\bdiff\tau + \epsilon_{ijk}\bT^j\wedge\bbe^k & = & 0\,, \label{efe3}  \\
\bdiff\lp \hat{\rho}\star\bdiff\tau\rp & = & 0\,. \label{efe4}
\ea
\es
The conservation equation (\ref{efe4}) shows that $\hat{\rho}$ behaves like the energy density of an ideal dust. The connection EoM (\ref{efe3}) sets the torsion of the right-handed connection to vanish, and then (\ref{efe1},\ref{efe2}) are expected to reduce to the effective Einstein equations in the presence of an ideal dust source. 

%\subsection{Minkowski solution}

%Now the factors of $i$ are absent, some signs are flipped and the number 0 is changed to the number 4 when comparing (\ref{efe}) with the analogous equations for SO$_\mathbb{C}$(1,3). 
To begin with, it is pertinent to check how we recover the Minkowski spacetime. Consider the antiself-dual vacuum solution for the connection, 
\bs
\label{guys}
\ba
\+ \bA^i & = & 0 \quad \Rightarrow \quad \+\bF^i = 0\,, \label{plusguy} \\
\m \bA^i & = & \frac{2}{\tau}\bdiff x^i \quad \Rightarrow \quad \m\bF^i = \frac{2}{\tau^2}\lp - \bdiff \tau\wedge\bdiff x^i + \epsilon^i{}_{jk}\bdiff x^j\wedge\bdiff x^k\rp\,. \label{minusguy}
\ea
\es 
%\cmt{Am I correct that this solution covers a `wedge' of Minkowski spacetime?}
Due to (\ref{plusguy}), the two first equations (\ref{efe1},\ref{efe2}) are trivially satisfied. The torsion is now $\bT^i = \frac{1}{2} \m \bF^i\tau$, which satisfies (\ref{efe3}), and the Bartels frame $\bbe^i = \bdiff x^i$ describes the Euclidean space. To make contact with the standard pseudo-Riemannian formalism in terms of
a metric tensor, we have to reconsider $\bbe^4$. The strange part now is that we do not identify $\bbe^4$ with the differential of a time coordinate like $\bdiff t$, but implement the Wick rotation  $\bbe^4 = \bdiff\tau \rightarrow i\bdiff t$. Then we can identify the invariant $\bbe^I\otimes\bbe_I$ with the usual Minkowski metric. The imaginary unit only enters into the operational relation between a dynamical field and the time-like component of an observer's frame. All fields and all coordinates are real both before and after the Wick rotation\footnote{Though, from the pseudo-Riemannian viewpoint one could in hindsight argue that the connection field had complex components. However, since we first solve for the connection and Wick-rotate the thus obtained second order field equations, the imaginary unit does not enter into the equations at any stage. (A possible interpretation is that, prior to the collapse of the wave function, the connection need not satisfy on-shell conditions. We leave this line of thought aside in the present work.)}. In practical terms, this is equivalent to setting $\bdiff t \equiv \bbe^0$, where $\bbe^0$ is a frame component in a tangent space with the Minkowski metric $\eta_{AB}$ with the indices $A$, $B$ running from 0 to 3. Constructing the robust differential-geometric formalism for such a mapping between different
frame bundles could be interesting. However, towards the aim of this article, we rather focus on demonstrating the practical workings of the theory with more nontrivial physical examples.  

\subsubsection{Contrasting with SO(1,3)}

In the $SO_{\mathbb{C}}(1,3)$ gauge theory, the above Minkowski solution is not unique. One can also device the simple solution $(\phi^1,\phi^2,\phi^3,\phi^4) = \kappa^{-\frac{1}{2}}(x,y,z,\tau)$, identifying the khronon components directly with the spacetime coordinates \cite{Koivisto:2022uvd}. This solution implies the vanishing of the connection coefficients $\bA^{IJ}=0$ and thus has also vanishing curvature and torsion; therefore, it cannot be gauge-equivalent to the above solution (\ref{guys}). The simple solution $(\phi^1,\phi^2,\phi^3,\phi^4) = \kappa^{-\frac{1}{2}}(x,y,z,\tau)$ exists in the SO(4) formulation, but it does not represent a spacetime, meaning that it cannot be Wick-rotated into a Minkowski space since the solution is not compatible with the first step of our recipe: it is not realised physically. Problems with the solution in the Lorentz-signature version of the theory, with possible implications to an alternative formulation of a Lorentz gauge theory \cite{Wiesendanger:2018dzw,Wiesendanger:2020lwa,Wiesendanger:2024dnp}, were discussed in Ref.\cite{Koivisto:2024asr}. 

Another, related point of comparison also arises from the investigations of Ref.\cite{Koivisto:2024asr} (see section V there), where it was discovered that the $SO_{\mathbb{C}}(1,3)$ theory admits an additional third dynamical phase with no counterpart in the underlying Euclidean formulation. Although mathematically consistent, this phase led to unphysical consequences, most notably the appearance of caustic singularities even in static, spherically symmetric configurations. Such behavior, reminiscent of pathologies in ``mimetic matter'' models \cite{Gorji:2020ten}, resulted in e.g. black hole solutions that failed to be asymptotically flat, indicating a breakdown of physical viability. These anomalies illustrate the dangers of enlarging the phase space by complexification without a firm foundational principle. %and highlight the importance of a controlled, geometrically grounded formulation. 

The above examples point to a more general issue. Once the phase space is enlarged by
complexification, the question is not only which solutions exist, but also which
real slice should be regarded as physically relevant. This motivates recalling the
coordinate-independent geometric notion of Wick rotation that can be given in terms of real analytic
slices of a common complexification \cite{Pessers_2016}. Namely, one may say that a real
analytic Euclidean metric $g_{\rm E}$ and a real analytic Lorentzian metric $g_{\rm L}$ on a domain
$U$ are Wick-related if there exists a complexification $(U_{\mathbb C},g_{\mathbb C})$ together with
totally real embeddings $\iota_{\rm E},\iota_{\rm L}:U\hookrightarrow U_{\mathbb C}$ such that
\begin{equation}
g_{\rm E}=\iota_{\rm E}^*g_{\mathbb C},\qquad
g_{\rm L}=\iota_{\rm L}^*g_{\mathbb C}.
\end{equation}
In this sense, the formal replacement $t \leftarrow i\tau$ acquires invariant meaning only when
both real metrics arise as slices of the same complexified geometry. However, this viewpoint also makes clear that complexification alone does not in general select
a unique continuation. The ambiguity is already present in the flat case: the complexification of
$(\mathbb{R}^4,\delta)$ admits real slices of different signatures, so besides the Euclidean and
Lorentzian slices one may also obtain a split-signature slice. Thus, the existence of a common
complexification does not by itself determine which real slice should be regarded as physical.

Our `recipe' is more restrictive, and therefore avoids this ambiguity by construction. Rather
than allowing an arbitrarily chosen real slice, we flip the sign only of the distinguished proper-time
direction selected by the field $\phi^I$. Writing locally $g_{\rm E}=\diff\tau^2+h$,
with $h$ positive definite on the hypersurfaces $\phi=\mathrm{const}$, we define
$g_{\rm L}=-\diff t^2+h$. Hence, the continuation in our framework is specifically a proper-time sign flip, not an arbitrary continuation to some other real slice. In particular, a split-signature metric would require flipping also a spatial direction, and is therefore excluded by the prescription itself.

 To summarise: in contrast to the $SO_{\mathbb{C}}(1,3)$ formulation, the present Euclidean-based theory offers a more controlled and geometrically grounded framework, insofar as it appears to avoid the particular pathologies encountered in the above examples. Therefore, it could provide a more reliable platform for the path integral approach, where only physically admissible configurations are summed over. The extent to which the structure of the theory naturally enforces consistency remains to be explored at both the classical and quantum levels.

\subsection{Expanding spacetime}
\label{flrw}

The case of the Minkowski solution is rather special, since therein the Bartels frame reduces to the constant delta function. It is
crucial to check whether the prescription works in the general case with a dynamic Bartels frame. For this purpose, we will consider the cosmological Ansatz,
\bs
\label{FLRW}
\be
%\Big\{ 
 \begin{matrix} \bA^{4i} & = & A(\tau)\bdiff x^i \\ \bA^{ij} & = &  B(\tau)\epsilon^{ij}{}_k\bdiff x^k   \end{matrix} \quad \Rightarrow \quad 
 \begin{matrix} \+\bA^{i} & = \lp A - B\rp\bdiff x^i \\  \m\bA^{i} & = -\lp A + B\rp\bdiff x^i   \end{matrix}  %\quad \Rightarrow \quad 
 % \begin{matrix} \+\bF^{i} & = -\lp \dot{A} + \dot{B}\rp\bdiff t\wedge\bdiff x^i  + \frac{1}{2}\lp A+B\rp^2\epsilon^i{}_{jk}\bdiff x^j\wedge\bdiff x^k
  %\\  \-\bF^{i} & = \lp \dot{A} - \dot{B}\rp\bdiff t\wedge\bdiff x^i  \end{matrix}  
 \ee
\es    
This Ansatz is homogeneous and isotropic, thus allowing only the two functions of $\tau$ (or equivalently, functions of $\phi$). For simplicity, we have chosen the form corresponding to flat FLRW cosmology. We will denote derivatives wrt to $\tau$ with a dot.  
The curvature of the connection (\ref{FLRW}) is 
\bs
\ba
\+\bF^{i} & = & \lp \dot{A} - \dot{B}\rp\bdiff \tau\wedge\bdiff x^i  + \frac{1}{2}\lp A - B\rp^2\epsilon^i{}_{jk}\bdiff x^j\wedge\bdiff x^k\,, \\
\m\bF^{i} & = & -\lp \dot{A} + \dot{B}\rp\bdiff \tau\wedge\bdiff x^i  + \frac{1}{2}\lp A + B\rp^2\epsilon^i{}_{jk}\bdiff x^j\wedge\bdiff x^k\,,
\ea
\es
which, in the symmetry-broken phase $\phi^I = \delta^I_4\kappa^{-1/2}\tau$, corresponds to the torsion
\be \label{FRWtorsion}
\bT^i = -\dot{A}\tau\bdiff \tau\wedge\bdiff x^i + AB\tau\epsilon^i{}_{jk}\bdiff x^j\wedge\bdiff x^k\,.  
\ee
Plugging into (\ref{efe}), these three 3-form equations plus one 4-form equation become, respectively\footnote{With slight abuse of notation, we have here exploited the $SO(4)$ Hodge symbol for the dual of the spacetime coordinate differentials. Covariant forms would be the $SO(4)$ duals of $SO(4)$-valued objects, like $\star \bbe^i$, or the spacetime duals of coordinate-indexed objects, like $\ast \bdiff x^i$, where $\star$ is wrt the Kronecker metric and $\ast$ would be wrt the spacetime metric.}, 
\bs \label{frw}
\ba
3A\tau\lp A - B\rp^2\star\bdiff x^4 & = & \sqrt{\kappa}\bt^4 + \kappa\hat{\rho}(A\tau)^3\star\bdiff x^4\,, \label{frw1} \\ 
\lb \lp \dot{A} - \dot{B}\rp A\tau + \lp A - B\rp^2\rb\star\bdiff x^i & = & -\sqrt{\kappa}\bt^i\,, \label{frw2} \\
A\tau\lp \dot{A}\tau + B\rp \star\bdiff x^i & = & 0\,, \label{frw3} \\
\lb \dot{\hat{\rho}}A\tau + 3\hat{\rho} \lp \dot{A}\tau + A\rp\rb \star 1 & = & 0\,. \label{frw4} 
\ea
\es
From (\ref{frw3}) we obtain $B=-\dot{A}\tau$. We can identify the scale factor as $a \equiv -A\tau$, since the cosmological Bartels frame is then $\bbe^i = a\bdiff x^i$. Now, taking the standard form of the source terms $\bt^4 = \sqrt{\kappa}\rho a^3\star\bdiff x^4$ and $\bt^i = -\sqrt{\kappa} p a^2\dot{\tau}\star\bdiff x^i$, the two equations (\ref{frw1},\ref{frw2}) become 
\be \label{Eflrw}
3\lp \dot{a}/a \rp^2 = \kappa\lp \rho + \hat{\rho}\rp\,, \quad 2\ddot{a}/a + \lp \dot{a}/a \rp^2  = -\kappa p\,. 
\ee
Fixing the scale $1/\kappa = -m_P^2$ for the standard gravitational coupling, the Wick-rotation $\tau \rightarrow it$ yields us the standard two Friedmann equations,
\be
3 H^2 = m_P^{-2}\lp \rho + \hat{\rho}\rp\,, \quad \text{and} \quad 2\partial_t H + 3H^2 = -m_P^{-2} p\,, \quad \text{where} \quad H \equiv \partial_t\log{a}\,,
\ee
and the final equation (\ref{frw4}) gives the expected conservation law for ideal dust in FLRW background, $\partial_t\hat{\rho} + 3H\hat{\rho} = 0$.

This worked only too well. Still, like it was in the Minkowski example, it remains unclear why the Wick rotation is needed. Up to interpretations of the variables, the equations (\ref{Eflrw}) in terms of the Euclidean time $\tau$ are exactly the usual FLRW equations for an expanding universe, allowing the same physical interpretation as usual. So, effectively we have a ``geometrodynamical'' set-up with a stack of Bartels frame configurations $\bbe^i$ ordered according to an external parameter, and everything looks the same whether the parameter $\tau$ or $t$ is considered.    

The necessity of the Wick rotation in more generic spacetimes becomes apparent when considering perturbations around the FLRW background. We will demonstrate this with tensor fluctuations. The generic spin-2 fluctuation of the connection can be parameterised with two symmetric, transverse-traceless 3$\times$3 tensors: an even-parity $h_{ij}$ and an independent, odd-parity $\tilde{h}_{ij}$, which appear in the perturbed connection components as
\bs
\label{GWansatz}
\ba
\bA^{4i} &  =  & A\lp \delta^i_j + h^i{}_j\rp \bdiff x^j\,, \\
\bA^{ij} & = & B\lp \epsilon^{ij}{}_k + \epsilon^{ijl}\tilde{h}_{lk}\rp\bdiff x^k\,. 
\ea
\es
Plugging into the connection EoM (\ref{efe3}) and linearising, we obtain the three 3-form equations
\be 
\lb  2\lp \tau \dot{A} + B \rp\delta^i_j  - \tau A \dot{h}_{ij} + B\lp h^i{}_j - \tilde{h}^i{}_j\rp +  \epsilon_{(i}{}^{kl}h_{j)k,l}\rb\star\bbe^i = 0\,,     
\ee
which, plugging in the background solution, yields the solution for the odd tensor perturbation in terms of the even tensor perturbation,
\be \label{hsolution}
\tilde{h}_{ij} = h_{ij} + \frac{1}{B}\lp \epsilon_{(i}{}^{kl}h_{j)k,l} + a \dot{h}_{ij}\rp\,. 
\ee
Tensor perturbations decouple from the energy equation (\ref{efe1}), but plugging (\ref{hsolution}) into the momentum equations gives (\ref{efe2}), in the absence of tensor perturbations in the matter source $\bt_i$,
\bs
\label{GWEL}
\be \label{GWE}
\lb \ddot{h}_{ij} + 3(\dot{a}/a) \dot{h}_{ij} + a^{-2}\nabla^2 h_{ij} \rb \star\bbe^j =0\,.
\ee
The 3-gradient is defined as $\nabla^2 \equiv \delta^{ij}\partial_i\partial_j$. 
In terms of the Euclidean time $\tau$, this does not describe waves but rather an exponential instability. Thus it is clear that we
should perform the Wick rotation to the standard Lorentzian time $t$ to arrive at the standard equation describing the propagation
of gravitational waves in FLRW background,
\be \label{GWL}
\partial_t^2 h_{ij} + 3H\partial_t h_{ij} - a^{-2}\nabla^2 h_{ij} = 0\,. 
\ee
\es
We will generalise these computations in section \ref{cosmology} to scalar and vector perturbations in the most general 6-parameter 
theory (\ref{TheAction}). 

Here it is useful to phrase the continuation slightly more systematically. For simplicity,
neglect the expansion and set $a=1$. Then a real Fourier mode of (\ref{GWL}) may be written as
\bs
\begin{equation}
h^{\rm L}_{ij}(t,\mathbf{k})
=
Q_{ij}(\mathbf{k})\cos(kt)
+
\frac{\Pi_{ij}(\mathbf{k})}{m_P}\sin(kt)\,,
\end{equation}
where $Q_{ij}$ is a dimensionless configuration amplitude, while $\Pi_{ij}$ carries one power
of energy and parametrises the momentum datum of the mode. Under the Wick rotation
$t\to -i\tau$, together with the continuation of scales $\kappa^{-1/2}= i m_P$, the same mode is
represented in the Euclidean description as
\begin{equation}
h^{\rm E}_{ij}(\tau,\mathbf{k})
=
Q_{ij}(\mathbf{k})\cosh(k\tau)
+
\frac{\Pi_{ij}(\mathbf{k})}{\kappa^{-1/2}}\sinh(k\tau)\,.
\end{equation}
\es
Thus the dimensionless configuration parameter $Q_{ij}$ is unchanged, whereas the time-odd
branch is continued through the corresponding Wick rotation of its momentum scale. Equivalently,
if one absorbs the scales into the coefficients, then the coefficient of the odd branch Wick-rotates,
while the dimensionless configuration parameter does not. In this sense, both branches admit a real description in their respective pictures once the continuation of the relevant scale is taken into account. 

This illustrates one familiar feature and one new feature of our framework. The familiar one is
that oscillatory Lorentzian evolution may correspond to hyperbolic Euclidean evolution. The new
ingredient is that, in our prescription, the continuation acts not only on the time variable but also
on the physical scales. In particular, the odd branch is associated with momentum data, and its
natural scale Wick-rotates together with the time variable, $\kappa^{-1/2}= i m_P$. Thus the
dimensionless configuration parameter $Q_{ij}$ remains unchanged, whereas the momentum
parameter is continued together with its scale. 
The equivalence of the two pictures is perhaps even more transparent when considering the general solution in terms of the initial condition of the wave mode and its time derivative,
\be
h_{ij} = h_{ij}(0,\bk)\cos{(kt)} + k^{-1}\partial_t h_{ij}(0,\bk)\sin{(kt)}
=  h_{ij}(0,\bk)\cosh{(k\tau)} + k^{-1}\partial_\tau h_{ij}(0,\bk)\sinh{(k\tau)}\,.
\ee
In this way the same gravitational-wave mode
admits a real description in both pictures: the Lorentzian and the Euclidean one differ not by a
loss of reality, but by how the same phase-space data are represented. 

This point is important conceptually. Standard results on analytic Wick rotation concern the
existence of Euclidean and Lorentzian metrics as real slices of a common complexification. In that
setting, the existence of a Riemannian slice is highly restrictive, and in particular standard
Wick-rotatability implies a purely electric curvature tensor \cite{Helleland:2015wva}. By contrast, our construction does not
start from the assumption that a given Lorentzian spacetime must itself admit a Euclidean real slice in the usual sense. Rather, the
correspondence is fixed by the distinguished proper-time direction selected by the khronon and by
the accompanying continuation of the relevant scales. Therefore, the present prescription is not
meant as a generic analytic continuation through an arbitrary common complexification, but as a
more specific map between the Euclidean and Lorentzian descriptions provided by the theory
itself. 
For this reason, the known no-go theorems for generic metric Wick rotation by
themselves do not directly apply to the mode-by-mode perturbative continuation
considered here, nor, as we will show next, the Kerr correspondence.

\subsection{Rotating spacetime}
\label{kerr}

We now turn to a particularly nontrivial example of an exact solution describing a rotating spacetime - specifically, the Kerr solution - as a test case for examining the universality of the proposed Euclidean formulation. 
Since the Kerr geometry is not static, extending it to a Euclidean regime is nontrivial and challenges the applicability or at least the interpretation of conventional methods which involve complexified metrics \cite{Gibbons:1976ue,Witten:2021nzp}. The presence of rotation introduces off-diagonal metric components and prevents the existence of a global, hypersurface-orthogonal timelike Killing vector field. Although a synchronous coordinate system should, in principle, be attainable\footnote{We note three apparently independent attempts (which do not seem to agree with each other) at deriving a synchronous coordinate system for the Kerr metric \cite{Novello:2010af,Sorge:2021nqm,Khatsymovsky:2021vjq}.}, we choose to adopt a convenient ansatz informed by Doran’s metric \cite{Doran:1999gb}, which incorporates nontrivial lapse and shift functions. Thus, we will study a generic axially symmetric $Spin(4)$ connection given as follows:
%The Kerr solution, since it describes a rotating spacetime, is a perhaps yet more nontrivial example. With the khronon gauge, (co)frame fields are required to be synchronous, and it is not straightforward to describe a rotating spacetime in such frame. While there are several literatures in which synchronous formulation of the Kerr metric is presented
%\footnote{We note three apparently independent attempts (which do not seem to agree with each other) at deriving the synchronous coordinate system for the Kerr metric \cite{Novello:2010af,Sorge:2021nqm,Khatsymovsky:2021vjq}.}, the following setup corresponding to \cite{Doran:1999gb} is one of the few working examples. 
\bs \label{Kerrconnection1}
\ba
\bA^{41} & = & -\frac{1}{\tau} \lp \beta \bdiff \tau + \frac{{\sigma}}{f} \bdiff r- {\alpha}  \sin^2 \theta \bdiff \varphi \rp\,, \quad \bA^{42} = - \frac{{\sigma}}{\tau}\bdiff\theta\,, \quad \bA^{43} = -\frac{{f}\sin\theta}{\tau}\bdiff\varphi\,,
\\
\bA^{12} & = & O \bdiff \tau + P \bdiff r + Q \bdiff \theta + R \bdiff \varphi\,, \quad
\bA^{13}  =  S \bdiff \tau + T \bdiff r + U \bdiff \theta + V \bdiff \varphi\,, \quad
\bA^{23}  =  W \bdiff \tau + X \bdiff r + Y \bdiff \theta + Z \bdiff \varphi\,. \quad\quad
\ea
\es
The `electric' components of the connection involve the 4 functions $f(r)$, $\sigma(r,\theta)$, $\alpha(r,\theta)$, $\beta(r,\theta)$ which conveniently parameterise an axially symmetric metric geometry, and the completely generic `magnetic' components of the connection then involve the 3$\times$4=12 free functions $O(r,\theta), \dots , Z(r,\theta)$. Then the anti/self-dual connection is
\bs
\ba
\x\bA^1 & = &  -\lp \pm \frac{\beta}{\tau} + W \rp \bdiff\tau - \lp\pm\frac{\sigma}{\tau f} + X \rp\bdiff r - Y\bdiff\theta - \lp Z \mp \frac{ \alpha}{\tau} \sin^2\theta \rp \bdiff\varphi\,, \\
\x\bA^2 & = & S \bdiff\tau + T \bdiff r + U \bdiff\theta + \lp V \mp \frac{\sigma}{\tau} \rp \bdiff\varphi\,, \\
\x\bA^3 & = & - O \bdiff\tau - P \bdiff r - Q \bdiff\theta - \lp R \pm \frac{f}{\tau}\sin \theta \rp \bdiff\varphi\,.
\ea
\es
In the absence of matter source, the connection EoM (\ref{efe3}) reduce to %\tomi{some $\alpha \rightarrow \beta$ here:}
\bs \ba
&& - \lb \frac{\alpha  P \sin \theta + \beta f' }{\sigma} + \frac{2}{\tau} - \frac{U}{\sigma} + \frac{  \alpha }{ f \sigma^2} \lp 2 \sigma \cos \theta -\sigma\mrq  \sin \theta \rp + \frac{  \alpha\mrq \sin \theta + \sigma R \csc \theta }{f \sigma} + \frac{f \beta  \sigma' }{ \sigma^2} \rb \star \bbe^1
\nn \\&& 
- \lb \frac{f}{\sigma} (\beta  P +T ) - \frac{ \alpha'}{\sigma} \sin\theta - O \rb \star \bbe^2 
- \lb \frac{f}{\sigma} (P -\beta  T )-\sigma \mrq + S \rb \star \bbe^3 \nn \\&& 
- \lb \frac{ \alpha  T \sin\theta +Q +f'}{\sigma} + \frac{f \sigma'}{\sigma^2}  + \frac{V}{f}\csc \theta \rb \bbe^4 =0 \,,
\\&&
 \lb \frac{\alpha \sigma'}{\sigma^2}  \sin\theta - \frac{f \beta}{\sigma} P + O - \frac{\beta \mrq}{\sigma} + \frac{\beta \sigma \mrq}{\sigma^2} - \frac{Y}{\sigma} \rb \star \bbe^1
- \lb \frac{\alpha P \sin\theta + \beta f'}{\sigma} + \frac{2}{\tau} + \frac{f}{\sigma} \left(\beta' +X \right)+ \frac{R}{f} \csc \theta \rb \star \bbe^2 
\nn \\&& 
- \lb f \left(\frac{\sigma'}{\sigma} -\beta X \right)+  Q +\sigma W \rb \star \bbe^3
- \lb \frac{\alpha}{\sigma}  X \sin\theta + \frac{1}{\sigma} \cot \theta + \frac{\sigma \mrq}{\sigma^2} - \frac{f}{\sigma} P + \frac{Z}{f} \csc \theta \rb \star \bbe^4 =0\,,
\\&&
- \lb \frac{\alpha X - \alpha' \beta + \alpha \beta'}{\sigma}  \sin\theta - S + \frac{1}{\sigma} \cot\theta + \frac{f \beta}{\sigma} T + \frac{Z}{f} \csc\theta \rb \star \bbe^1
+ \lb \frac{\alpha}{\sigma}  T \sin\theta + W + \frac{f'}{\sigma} - \frac{f \beta}{\sigma} X + \frac{V}{f} \csc \theta \rb \star \bbe^2
\nn \\&& 
+ \lb \frac{f \beta  \sigma'}{\sigma^2} + \frac{ f \left(\beta' +X \right)-U }{\sigma} + \frac{2}{\tau}  \rb \star \bbe^3
+\lb \lp \frac{\alpha'}{\sigma} + \frac{\alpha \sigma'}{\sigma^2} \rp  \sin\theta  - \frac{f T + Y}{\sigma} \rb \star \bbe^4 =0\,,
\ea \es
where $'$ and $\mrq$ respectively denote $\partial_r$ and $\partial_\theta$. These 3$\times$4 equations yield us the solutions for the 3$\times$4 functions in the `magnetic' part of the connection in terms of the 4 functions that appear in the `electric' part of the connection: 
%\tomi{some $\alpha \rightarrow \beta$ here:}
\bs \ba
&&O= \frac{\beta \mrq \sigma + \beta  \sigma \mrq}{2 \sigma ^2}
+ \frac{\alpha'\beta^2 - \alpha \beta \beta'}{2\sigma}  \sin\theta\,, \quad
P = \frac{\sigma \mrq}{f \sigma} + \frac{\alpha'\beta - \alpha \beta'}{2f}  \sin\theta\,, \quad \nn
\\&&
Q = -\frac{f \sigma'}{\sigma }\,, \quad
R = - \frac{\alpha}{2\sigma} \lp \alpha' \beta - \alpha \beta' \rp \sin^3 \theta -\frac{ \alpha \mrq \sigma + \alpha  \sigma \mrq }{2 \sigma^2} \sin^2 \theta - \frac{1}{ \sigma} \lp   \alpha \cos\theta + \beta f f' \rp \sin\theta - \frac{f}{\tau} \sin\theta \,.
\\&& 
S=  \frac{ \alpha ' \beta + \alpha \beta ' }{2 \sigma}  \sin\theta + \frac{\beta \beta\mrq \sigma - \beta^2 \sigma\mrq}{2 \sigma^2}\,, \quad
T= \frac{\alpha '}{ f}  \sin\theta + \frac{\beta\mrq \sigma - \beta \sigma\mrq}{2 f \sigma}\,, \quad
\nn \\&&
U= \frac{\alpha \mrq \sigma - \alpha  \sigma \mrq}{2 f \sigma}  \sin\theta + \frac{ \alpha}{f}  \cos\theta + \frac{f \beta  \sigma' }{\sigma } + \frac{\sigma}{\tau} \,, \quad
V= -\frac{ \alpha \alpha '}{\sigma}  \sin^3\theta - \frac{ \alpha \beta\mrq \sigma - \alpha \beta \sigma\mrq}{2 \sigma^2}  \sin^2\theta - \frac{f f'}{\sigma^2} \sin\theta \,.
\\&&
W = \frac{\alpha\mrq \sigma - \alpha \sigma\mrq}{2 f \sigma^2} \beta  \sin\theta + \frac{ \alpha }{f \sigma}\beta  \cos\theta - \frac{ f \beta \beta '}{ \sigma} - \frac{\beta}{\tau}\,, \quad \nn \\&&
X= \frac{ \alpha \mrq \sigma - \alpha  \sigma \mrq }{2 f^2 \sigma }   \sin\theta + \frac{ \alpha }{f^2 } \cos\theta - \beta' - \frac{\sigma}{\tau f }\,, \quad
Y = \frac{ \alpha  \sigma'}{\sigma }  \sin\theta - \frac{\beta \mrq \sigma - \beta  \sigma \mrq}{2 \sigma }\,, \quad \nn \\&&
Z = -\frac{ \alpha \alpha\mrq \sigma- \alpha^2 \sigma\mrq}{2 f \sigma^2} \sin^3\theta - \frac{\alpha^2}{ f \sigma}  \sin^2\theta \cos\theta + \frac{ \tau f \alpha ' \beta - \tau f \alpha \beta ' - 2 \alpha \sigma }{2 \tau \sigma} \sin^2\theta - \frac{f}{\sigma} \cos\theta\,. 
\ea\es
Plugging this solution into energy-momentum equations (\ref{efe1}, \ref{efe2}), we finally obtain the solutions for the remaining 4 functions:
\be \label{kerrsolution}
f = \pm\sqrt{r^2-a_E^2}\,, \quad
\sigma = \pm\sqrt{r^2 - a_E^2\cos^2 \theta}\,, \quad
\alpha = \pm a_E\beta\,, \quad \beta = \pm\sqrt{\frac{\kappa m_S r}{4\pi \sigma^2}}\,. 
\ee
Two integration constants appear in the solution, $a_E$ and $m_S$. When considering the emergent pseudo-Riemannian geometry, it is crucial to note that the parameter $a_E \equiv J_E/m_S$ will correspond to the angular momentum $J$ per unit mass $m_S$ in the rotating spacetime, characterising the intrinsic rotation or “twist” of the geometry, conventionally encoded in off-diagonal metric components that mix space and time directions. When performing a Wick rotation, $J_E \sim m_S r^2\dot{\varphi} \rightarrow -i m_S r^2\partial_t\varphi \sim -iJ$, the angular momentum parameter  would formally become imaginary, $a_E \rightarrow -ia$. This change occurs because the rotation, fundamentally tied to temporal flow, does not straightforwardly translate into the Euclidean regime; the analytic continuation alters the nature of the time coordinate and thus affects quantities dependent on it. This situation is quite analogous to what we already encountered with gravitational waves in section \ref{flrw} above: in Lorentzian spacetime, they represent oscillatory, propagating disturbances, but in the Euclidean regime, their equations become elliptic, leading to exponentially growing or decaying solutions rather than oscillations. Our approach ensures that the metric remains real (or equivalently, the khronon field and the Bartels frame remain real) and well-defined in both Lorentzian and Euclidean descriptions, meaning that the angular momentum as well retains a consistent, physically meaningful interpretation across the analytic continuation without becoming complex. This reflects a significant advantage, as it allows for a unified treatment of rotation that avoids the complications of complexified parameters common in standard Euclideanisations. To be explicit, the solution (\ref{kerrsolution}) gives us the pseudo-Riemannian geometry described by the Cartan frame
\bs
\label{Lkerr}
\ba
\bbe^0 & = & \bdiff t\,, \\
\bbe^1 & = & \sqrt{\frac{r^2 + a^2\cos^2\theta}{r^2 + a^2}}\bdiff r
+\sqrt{\frac{m_S r}{4\pi m_P^2\lp r^2 + a^2\cos^2\theta\rp}}\lp \bdiff t - a\sin^2\theta\bdiff\varphi\rp\,, \\
\bbe^2 & = & \sqrt{r^2 + a^2\cos^2\theta}\bdiff\theta\,, \\
\bbe^3 & = & \sqrt{r^2 + a^2}\sin\theta\bdiff\varphi\,,
\ea
\es
which is the same as the one reported in Ref.\cite{Doran:1999gb} (we have chosen the first plus signs and two last minus signs at (\ref{kerrsolution}) to match the conventions of Ref.\cite{Doran:1999gb}). 
Now the pseudo-Riemannian (Lorentzian) dynamics are understood as the manifestation of a deeper, fundamentally Euclidean structure. The familiar causal and dynamical features of spacetime arise as analytic continuations - appearances - of the underlying real Riemannian geometry,
\bs
\label{Ekerr}
\ba
\bbe^4 & = & \bdiff \tau\,, \\
\bbe^1 & = & \sqrt{\frac{r^2 - a^2_E\cos^2\theta}{r^2 - a^2_E}}\bdiff r
+\sqrt{\frac{\kappa m_S r}{4\pi \lp r^2 - a^2_E\cos^2\theta\rp}}\lp \bdiff \tau + a_E\sin^2\theta\bdiff\varphi\rp\,, \\
\bbe^2 & = & \sqrt{r^2 - a_E^2\cos^2\theta}\bdiff\theta\,, \\
\bbe^3 & = & \sqrt{r^2 - a_E^2}\sin\theta\bdiff\varphi\,.
\ea
\es
Moreover, this Riemannian geometry emerges from the configuration of the khronon field $\phi$, whose magnitude, measured in conventionally assigned units of $\kappa$, sets the structure of time in ultimately arbitrary units of mass-like $m_P$.

\section{Cosmology}
\label{cosmology}

We shall now investigate the generic 6-parameter theory and begin by focusing on its cosmological solutions. Since the part $I_{(0)}$ in (\ref{TheAction}) is a surface integral, for the purposes of classical dynamics the action we consider reduces to  
\be
I = \int \bDiff\phi^I\wedge\bDiff\phi^J\wedge\lp g_+ \+\bF_{IJ}  - g_- \m\bF_{IJ}\rp - \int \star \lambda - \int \lp \bDiff\phi^I\wedge\bt_I + \bA^{IJ}\wedge\bO_{IJ}\rp\,. 
\ee
From this, we obtain the EoM
\bs
\ba
\bDiff\lb 2\lp g_+\+\bF_{IJ}- g_-\m\bF_{IJ}\rp \wedge \bDiff\phi^J - \lambda\star\bDiff\phi_I - \bt_I \rb & = & 0\,, \\
\bDiff\lb g_+\+{}\lp\bDiff\phi_I\wedge\bDiff\phi_J\rp - g_-\m{}\lp\bDiff\phi_I\wedge\bDiff\phi_J\rp\rb & - & 2\lp g_+\phi_{[I}\+\bF_{J]K}- g_-\phi_{[I}\m\bF_{J]K}\rp \wedge\bDiff\phi^K \nn \\
& = &  - \phi_{[I}\bt_{J]} - \lambda\phi_{[I}\star\bDiff\phi_{J]} + \bO_{IJ}\,,  
\ea
\es
which we can rearrange in the more convenient form,
\bs
\ba
2\g\bF_{IJ}\wedge\bDiff\phi^J & = & \bt_I + \lambda\star\bDiff\phi_I + \bM_I\,, \\
\bDiff\g{}\lp\bDiff\phi_I\wedge\bDiff\phi_J\rp & = & \phi_{[I}\bM_{J]} + \bO_{IJ}\,, \label{Ceom} \\
\bDiff \bM_I & = & 0\,, 
\ea
\es
by defining the projector
\be
\g X_{IJ} \equiv g_+ \+ X_{IJ} - g_- \m X_{IJ}\,. 
\ee
In the symmetry-broken phase, we obtain for the khronon EoM 
\bs
\label{khrononEoM}
\ba
\lp g_+ \+\bF_i + g_-\m\bF_i\rp\wedge\bDiff\phi^i & = & \bt^4 + \lambda\star\bDiff\phi + \bM^4\,, \label{Sync1} \\
\lp g_+ \+\bF^i + g_-\m\bF^i\rp\wedge\bdiff\phi - \epsilon^i{}_{jk}\lp g_+\+\bF^j - g_-\m\bF^j\rp\wedge\bDiff\phi^k & = & - \bt^i - \lambda\star\bDiff\phi^i - \bM^i\,, \label{Sync2} 
\ea
and for the connection EoM
\ba
g_+\kappa^{-1/2}\lp \bT^i\wedge\bdiff\phi + \epsilon^i{}_{jk}\bT^j\wedge\bDiff\phi^k\rp & = & - \frac{1}{2}\phi\bM^i - \+\bO^i\,, \label{Tplus}\\
g_-\kappa^{-1/2}\lp \bT^i\wedge\bdiff\phi - \epsilon^i{}_{jk}\bT^j\wedge\bDiff\phi^k\rp & = & \frac{1}{2}\phi\bM^i - \m\-\bO^i\,. \label{Tminus}  
\ea
\es
Here we have taken the anti/self-dual projections of (\ref{Ceom}), but it is sometimes more convenient to instead consider the (4-i) and the (i-j) components of (\ref{Ceom}),
\bs
\ba
\label{CEoM1}
\lp g_+ - g_-\rp\bT^i\wedge\bbe^4 + \lp g_+ + g_-\rp\epsilon^i{}_{jk}\bT^j\wedge\bbe^k & = & -\kappa\phi\bM^i - 2\kappa\bO^{4i}\,, \\
\label{CEoM2}
\lp g_+ + g_-\rp\bT^i\wedge\bbe^4 + \lp g_+ - g_-\rp\epsilon^i{}_{jk}\bT^j\wedge\bbe^k & = & \epsilon^i{}_{jk}\kappa\bO^{jk}\,, 
\ea
\es
as the effective source term from the integration form appears only for the former components. 
The two khronon field equations (\ref{Sync1}) and (\ref{Sync2}) specialised to the homogeneous and isotropic background now give, respectively,
\bs
\label{SyncF}
\ba
-\frac{3}{a^2}\lb \lp g_+ + g_-\rp\lp A^2 + B^2\rp - 2\lp g_+ - g_-\rp AB\rb\star \bdiff\tau & = & \sqrt{\kappa}\lp \bt^4 + \kappa^{-3/2}\lambda\star\bdiff\tau + \bM^4 \rp\,, \\ 
\frac{1}{a^2}\lb  \lp g_+ + g_-\rp\lp A^2 + B^2 - 2a\dot{A}\rp - 2\lp g_+ - g_-\rp\lp AB - a\dot{B}\rp\rb\star\bbe^i & = & -\sqrt{\kappa}\lp \bt^i  + \kappa^{-3/2}\lambda\star\bbe^i + \bM^i \rp\,,      
\ea
\es
whereas the two connection field equations (\ref{Tplus},\ref{Tminus}) can be lumped together as
\be \label{SyncT}
g_{\pm}\lp \pm \dot{A}\tau + B\rp\star\bbe^i = \pm \frac{a\kappa}{4}\phi\bM^i + \frac{a\kappa}{2}\prescript{\pm}{}\bO^i\,.  
\ee
It can be first instructive to check the two cases $g_\mp = 0$ in the absence of spin currents, but taking into account the perfect fluid source,
\bs
\ba
\bt^i & = & \sqrt{\kappa}p \star\bbe^i\,, \\
\bt^4 & = & -\sqrt{\kappa}{\rho} \star\bdiff\tau\,.  
\ea
\es
In these special limits, the two torsion equations $g_\pm\lp \dot{A}\tau \pm B\rp = 0$ imply that $\bM^i=0$ and that $B = \mp \dot{A}\tau$. Plugging this solution into (\ref{SyncF}) yields in both cases the Friedmann equations 
\bs
\label{SyncF2}
\ba
3g_\pm\lp\frac{\dot{a}}{a}\rp^2 & = & \kappa\lp \rho + \hat{\rho}\rp + \Lambda \quad \Leftrightarrow \quad 
3 g_\pm H^2 =  m_P^{-2}\lp \rho + \hat{\rho}\rp + \Lambda \\
g_\pm\lb 2 \frac{\ddot{a}}{a} + \lp\frac{\dot{a}}{a}\rp^2 \rb & = &  -\kappa p + \Lambda  \quad \Leftrightarrow \quad 
g_\pm\lp 2\partial_t H + 3H^2\rp = -m_P^{-2} p + \Lambda  \,,
\ea
\es
where $\Lambda \equiv \kappa^{-1}\lambda = -m_P^2\lambda$, and the effective energy density $\hat{\rho}$ again obeys the ideal dust conservation equation, implying the conservation of the matter source with the equation of state $p/\rho$. As expected, both the cases ($g_\mp =0$, $g_\pm = 1$) reproduce the system already found in \ref{flrw}. %However, now that the parameter $\lambda$ has been included, we note that the cosmological constant has the different sign in the Lorentzian regime. Let us remark in passing that this could potentially be the key to unlock a new physical relevance to the AdS/CFT correspondence.     

For generic $g_\pm$, in the absence of the integration forms $\bM^i=0$ and spin currents $\bO^{IJ}=0$, the two torsion equations  (\ref{SyncT}) together with (\ref{FRWtorsion}) imply that $\dot{A}=0$ and $B=0$. Nontrivial cosmological solutions thus in general require nonvanishing $\bM^i \neq 0$. We make an Ansatz for the dark matter form, 
\bs
\ba
%\bM^i & = & -a^{-1}\phi^{-2}M\star\bbe^i\,, \label{Mi} \\
\bM^i & = & \sqrt{\kappa}\hat{p}\star\bbe^i\,, \\ 
\bM^4 & = & -\sqrt{\kappa}\hat{\rho}\star\bdiff\tau\,.
\ea
\es
In the absence of material spin currents, two torsion equations (\ref{SyncT})  
\be \label{SyncT2}
g_\pm\lp  \dot{A}\tau \pm B\rp = \frac{a}{4}\tau\kappa\hat{p}\,, \\
\ee 
are together solved by 
\bs
\label{BMsolution}
\ba
B & = & - \frac{g_+ - g_-}{g_+ + g_-} \tau \dot{A}\,, \label{BM1} \\
\kappa\hat{p} & = &  \frac{8 g_+ g_-}{ g_+ + g_-} \frac{\dot{A}}{a}\,.  \label{BM2}
\ea
\es
This solution requires that $g_+ \neq -g_-$, and we will check the special case of equal $g = \pm g_\pm$ later.
The Euler equation $\bDiff\bM^i = \bdiff \bM^i + \bA^i{}_j\wedge\bM^j - \bA^4{}_i\wedge\bM_4 =0$ is trivially satisfied, since each of the three terms vanishes identically for the chosen cosmological Ansatz. However, the continuity equation, $\bDiff \bM_4 = \bdiff \bM_4 + \bA_{4i}\wedge\bM^i = 0$ gives now
\be \label{DMcont}
\dot{\hat{\rho}} + 3\frac{\dot{a}}{a}\hat{\rho} =  -3\frac{\hat{p}}{\tau} = \frac{24 g_+ g_-}{\kappa\lp g_+ + g_-\rp}\frac{ A \dot{A}}{a^2} =  \frac{24 g_+ g_-}{\kappa\lp g_+ + g_-\rp\tau^2}\lp \frac{\dot{a}}{a} - \frac{1}{\tau}\rp\,.
\ee
The effective pressure term can be seen as a source of the effective energy density. The homogeneous equation has a solution that contributes an ideal dust part into the energy density $\hat{\rho}_{\text{CDM}} \in \hat{\rho}$, where $ \hat{\rho}_{\text{CDM}} \sim a^{-3}$, and in this sense a cold dark matter contribution still arises as an integration constant. However, the total effective fluid cannot be ideal dust and thus potential problems with caustics could be avoided. Moreover, dark matter can be now be associated with the spin current $\hat{\bO}_{IJ}=\phi_{[I}\bM_{J]}$, an exotic possibility not often explored in the literature  \cite{Izaurieta:2020xpk,Elizalde:2022vvc,Barriga:2024hpe}.

It will be convenient to use the short-hand notations for the parameter combinations
\bs
\label{alphabeta}
\ba
\alpha & \equiv & \frac{\lp g_+ - g_-\rp^2}{g_+ + g_-}\,, \\
\beta & \equiv & -\frac{4 g_ + g_-}{g_+ + g_-}\,, \\
\gamma & \equiv & \frac{g_+ + g_-}{g_+ - g_-} =  \frac{g_+ - g_-}{\alpha}\,. \label{gammadef}
\ea
\es
Note that we have defined the parameter combination $\gamma$ such that it corresponds to the well-known Barbero-Immirzi parameter  in the context of loop quantum gravity \cite{BarberoG:1994eia,Immirzi:1996di}. We may rewrite the action (\ref{I2}) as
\be
I_{(2)} = \frac{1}{2}\lp g_+ + g_-\rp\int \bDiff\phi^I\wedge\bDiff\phi^J\lp \frac{1}{2}\epsilon_{IJKL}+\frac{1}{\gamma}\delta_{IK}\delta_{JL}\rp\wedge\bF^{KL}\,,
\ee
which upon the substitution $\bDiff\phi^I \rightarrow (g_++g_-)^{-\frac{1}{2}}\kappa^{-\frac{1}{2}}\bbe^I$ would become (Euclideanised) action for loop quantum gravity, with the Barbero-Immirzi parameter $\gamma$ controlling the relative coupling strengths of the first, ``Palatini'' term and the second,
``Holst'' term. Since now torsion plays a crucial role, both these terms are relevant to dynamics. After this remark, let us proceed with the cosmological equations. 

\subsection{The Friedmann equations}
\label{TheFriedmann}

Plugging now the solution (\ref{BM1}) into the field equations (\ref{SyncF}) we obtain
\bs
\ba
3\alpha\lp \frac{\dot{a}}{a} \rp^2 - 3\frac{\beta}{\tau^2} & = & \kappa\lp \rho + \hat{\rho}\rp - \Lambda\,, \\
2\alpha\frac{\ddot{a}}{a} +  \alpha \lp \frac{\dot{a}}{a} \rp^2 - \frac{\beta}{\tau^2} & = & -\kappa p - \Lambda\,.  
\ea
\es
These equations are consistent with
\bs
\ba
\dot{\rho} + 3\frac{\dot{a}}{a}\lp \rho + p\rp & = & 0\,, \\
\dot{\hat{\rho}} + 3\frac{\dot{a}}{a}\lp \hat{\rho} + \frac{2\beta}{\kappa\tau^2}\rp & = &  \frac{6\beta}{\kappa\tau^3}\,. 
\ea
\es
Now let $\kappa^{-\frac{1}{2}}\tau \rightarrow m_P t$. Then the above equations read
\bs
\label{FriedmannFinal}
\ba \label{FriedmannFinal1}
3\alpha H^2 - 3\beta t^{-2} & = & m_P^{-2}\lp \rho + \hat{\rho} \rp + \Lambda\,, \\ 
\alpha\lp 3H^2 + 2 H' \rp - \beta t^{-2} & = & -m_P^{-2} p + \Lambda\,, 
\ea
\es 
and
\bs
\ba
\partial_t{\rho} + 3H\lp \rho + p\rp & = &  0\,, \label{mattercont} \\ 
\partial_t{\hat{\rho}} + 3H\hat{\rho} & = & 6\beta m_P^2 \lp t^{-3} - H t^{-2}\rp\,.   
\ea
\es
It is clear that the Euclidean solutions can be mapped to Lorentzian solutions and vice versa, taking into account the effective flip of the sign of $\Lambda$ noticed earlier. 
%Note, in particular, that the terms proportional to $\beta$, which modify the standard Friedmann equations, flip their sign under the
%Wick rotation.

We will first have a look at the vacuum solutions. 
There are two sets of vacuum solutions when $\rho=p=\Lambda=0$,
\be
m_P^{-2}\hat{\rho}(t) = \frac{2}{3t^2}\lp \alpha - 3\beta \pm \sqrt{\alpha^2 + 3\alpha\beta}\rp\,, \quad
a(t) \sim t^{\frac{1}{3}\lp 1 \pm \sqrt{1+3\frac{\beta}{\alpha}}\rp}\,. \label{Vacuum1}
\ee
Only in the case $\beta=0$ do we have either the Minkowski vacuum or the ideal dust solution.
Assume $\alpha=1$ and $\beta = -\epsilon$ a very small parameter. Then, in the almost-Minkowski-branch 
$M_P^{-2}\hat{\rho} \approx 3\epsilon t^{-2}$ and $a \sim t^{\frac{\epsilon}{2}}$, whereas in the almost-dust-branch    
$m_P^{-2}\hat{\rho}   \approx(4/3 + \epsilon) t^{-2}$ and $a \sim t^{\frac{2}{3} - \frac{\epsilon}{2}}$. Depending on the sign of $g_-$, we may thus render the effective dark matter equation of state slightly negative or slightly positive.

\subsection{The special case $g_+ = -g_-$}

Let us then check the special case $g_+ = -g_- \equiv g$. For simplicity, let $\lambda=0$, since the cosmological constant can always be absorbed in the source term and interpreted as a part of $\rho$. The connection equation (\ref{SyncT2}) forces then to set $\dot{A}=0$, and we denote the constant $A$ as $A=-a_0$.
Thus, the khronon is essentially the scale factor $\phi = a/a_0$. The other condition we obtain is 
\be
\kappa\hat{p} = \frac{4g}{a\tau }B = \frac{4g\sqrt{\kappa}}{a_0\phi^2}B\,. 
\ee 
This implies the strange Friedmann equations (\ref{SyncF})
\bs
\ba
 12 g B & = & a_0\phi^2 \kappa^{3/2}\lp \rho + \hat{\rho}\rp\,, \\ 
4g\lp \tau \dot{B} + 2B\rp & = & -a_0\phi^2\kappa^{3/2}p\,.
\ea
\es
In this case, it occurs that both $a\dot{\rho} + 3\dot{a}(\rho+p) = 0$ and $a\dot{\hat{\rho}} + 3\dot{a}(\hat{\rho}+\hat{p}) = 0$, because now $\dot{a}/a=1/\tau$, which means $H=1/t$ in the Lorentzian picture. In achiral gravity,
the universe always expands like it would be curvature-dominated. The vacuum solution is given by $\hat{p} = \hat{\rho}/3$, whilst the equation of state would have the opposite sign in the standard theory in order to drive the same expansion. If there is matter with the equation of state $w=p/\rho$, implying that $\rho \sim \phi^{-3(1+w)}$, then the integration form has to be chosen such that
\be
\hat{\rho} = -\frac{\rho}{1-3w} + \frac{m_0}{\phi^4}\,,
\ee  
where the first piece compensates the effect of the matter source, and the integration constant $m_0$ determines the energy density of the ``dark radiation'' which drives the $\dot{a}/a \sim 1/\tau$ expansion. 

\subsection{The special case $g_+ = g_-$}
\label{dymaxion}

Another special case is $g_+ = g_- \equiv g$, for which $\alpha=0$, $\beta=-2g$. We note that the solution (\ref{Vacuum1}) breaks down in this limit. Therefore we go back to the Friedmann equations (\ref{SyncF}) which reduce to
\be
2g \tau^{-2} = \frac{\kappa}{3}\lp \rho + \hat{\rho}\rp = -\kappa p\,. 
\ee
We see that now $\hat{\rho}=-(1+3w)\rho$, and this solution only exists if it is sourced by matter with pressure\footnote{This may change in the presence of spatial curvature, but in this article we restrict to the flat FLRW cosmology.}. Then the expansion rate is given  
by the matter equation of state as usual, $a \sim \phi^{\frac{2}{3(1+w)}}$. Yet, the system is very unusual. To check its consistency, we 
should verify the conservation equation for the dark matter. For the RHS of the equation (\ref{DMcont}) we get
\bs
\be
\dot{\hat{\rho}} + 3\frac{\dot{a}}{a}\hat{\rho} = 3w\lp 1+3w\rp\frac{\dot{a}}{a}\rho\,,  
\ee
and for the LHS we get
\be
-3\frac{\hat{p}}{\tau} = 12g\frac{A\dot{A}}{\kappa a^2} = -\frac{4g(1+3w)}{(1+w)\kappa\tau^3} = 3w\lp 1+3w\rp\frac{\dot{a}}{a}\rho\,. 
%-\frac{6g}{a^2}(A^2)' = = -\frac{4g}{(1+w)\phi a} \frac{\partial A^2}{\partial a}\,. 
%\frac{12g}{\phi^2}\lp \frac{1}{\phi} - \frac{a'}{a}\rp 
\ee
\es
These match, confirming that the system, though weird, is consistent. The $g_+ = g_-$ theory was called
``quasi-topological'' \cite{Koivisto:2024asr} because, on the one hand, the action does not reduce to a boundary term but, on the other hand, there are no local degrees of freedom \cite{Nikjoo:2023flm}. The latter property reflects the emergence of an extra symmetry as $g_- \rightarrow g_+$ \cite{Nikjoo:2023flm} (see also appendix A of \cite{Koivisto:2024asr}). This would, in an extension of the present work that would allow varying coupling constants ( - ``the dymaxion'' - ) in terms of running parameters or dynamical scalars, suggest a scenario wherein the dynamical universe emerges from the ``quasi-topological'' phase of enhanced symmetry \cite{Koivisto:2024asr}. 

Action principles, which are essentially equivalent to our case $g_+ = g_-$, appear in the works of several different authors in the literature \cite{Aldrovandi:2004uz,Diakonov:2011im,Obukhov:2012je,Vladimirov:2014gma,Vergeles:2019xfh,Volovik:2024iiy}; however, the physical contents of the theory have not been understood in these works, since the action principles have been assumed to be dynamically equivalent to general relativity. 

\section{Large-scale structure}
\label{largescale}

We shall now proceed to investigate the cosmological physics of the $Spin(4)$ theory, focusing on the evolution of structure in the universe. 
Rather than restrict ourselves to a specific background or gauge choice from the outset, we will develop the formalism of cosmological perturbation theory in full generality (up to linear order), allowing a unified treatment of scalar, vector, and tensor modes on arbitrary (spatially flat) FLRW spacetimes. This will provide the tools necessary for analysing a wide range of physical phenomena, from generation of primordial fluctuations and their imprint on the cosmic microwave background to the propagation of gravitational waves and the dynamics of structure formation at late times.

Linear perturbation theory offers a crucial and highly nontrivial testing ground for the consistency of the theory. Unlike the homogeneous and isotropic background, the perturbed universe generically admits no symmetries, exposing potential inconsistencies in the proposed underlying structure. Thus, already the successful reproduction of standard results in cosmology a stringent check on the theory. Beyond merely reproducing the standard results of cosmological perturbation theory in the presence of ideal dust in the case $\beta=0$, the generalised $\beta \neq 0$ cases reveal qualitatively new features absent in any conventional setting. This enriches the theoretical landscape, but most importantly, it also offers new observational signatures.

We consider the generic fluctuations and provide a summary of the parameterisations below in Table \ref{tab:field_classification}. There are nice recent papers on cosmological perturbation\footnote{For an introduction to and a reference for cosmological perturbation theory we recommend Kurki-Suonio's lecture notes at https://www.mv.helsinki.fi/home/hkurkisu/.} formalism with general connections \cite{Aoki:2023sum,Castillo-Felisola:2024atv}; those should be consistent with the formulation below.   

\begin{table}[h]
    \centering
    \setlength{\tabcolsep}{10pt}
    \renewcommand{\arraystretch}{1.2}
    \begin{tabular}{|l|c|c|c|c|c|c|}
        \hline
        \textbf{Fields} & {scalars} & {pseudos.} & {vectors} & {pseudov.} & {tensors} & {pseudot.} \\
        \hline
        Cartan khronon & $\varphi$ &  &  &  &  &  \\
        \multicolumn{1}{|r|}{\textit{ total \# d.o.f.} 1} & $1\times 1$ & 0 & 0 & 0 & 0 & 0 \\
        \hline
        $Spin(4)$ connection & ${c},{r},{s},{\psi}$  & $\tilde{c},\tilde{r},\tilde{s},\tilde{\psi}$  & $v_i,u_i,w_i$ & $\tilde{u}_i,\tilde{v}_i,\tilde{w}_i$ & $h_{ij}$ & $\tilde{h}_{ij}$   \\
        \multicolumn{1}{|r|}{\textit{total \# d.o.f.} 24}  & $4\times 1$ & $4\times 1$ & $3\times 2$  & $3\times 2$  & $1\times 2$ &  $1\times 2$  \\
        \hline
        Dark matter  & $n,m,\chi,\delta\hat{\rho},\delta\hat{p}$ & $\tilde{m}$ & $m_i,n_i,{q}_i$ & $\tilde{m}_i$  & $m_{ij}$ &  \\
        \multicolumn{1}{|r|}{\textit{total \# d.o.f.} 16} & $5\times 1$ & $1\times 1$ & $3\times 2$ & $1\times 2$ & $1\times 2$  & 0 \\
        \hline
        \hline
        Bartels frame & $c,r,\Psi$ & $\tilde{s}$ & $v_i,w_i$ & $\tilde{u}_i$  & $h_{ij}$ &  \\
        \multicolumn{1}{|r|}{\textit{total \# d.o.f. } 12} & $3\times 1$ & $1\times 1$ & $2\times 2$ & $1\times 2$ & $1\times 2$  & 0 \\
        \hline
        Metric & $\partial{\varphi}/\partial\phi,\sqrt{\kappa}\varphi+a^2\dot{c},r,\Psi$ &  & $v_i,w_i$ &  & $h_{ij}$ &  \\
        \multicolumn{1}{|r|}{\textit{total \# d.o.f. } 10} & $4\times 1$ & $0$ & $2\times 2$ & $0$ & $1\times 2$  & 0 \\
        \hline
    \end{tabular}
    \caption{Classification of fields by their transformation properties under the little Lorentz group. The two top rows are the theory's fundamental fields; the dark matter field arises from integrating the equations of motion; the two bottom rows are composite fields.}
    \label{tab:field_classification}
\end{table}

\subsection{Spin-2 perturbations}

In the general theory, we should generalise the Ansatz (\ref{GWansatz}) for tensor perturbations by taking into account transverse-traceless fluctuations also in the integration 3-form $\bM_I$, besides those in the connection 1-form. The generic Ansatz can then be given as
\bs
\label{GWansatz2}
\ba
\bA^{4i} &  =  & A\lp \delta^i_j + h^i{}_j\rp \bdiff x^j\,, \quad
\bA^{ij}  =  B\lp \epsilon^{ij}{}_k + \epsilon^{ijl}\tilde{h}_{lk}\rp\bdiff x^k\,, \\
\bM^4 & = & -\sqrt{\kappa}\hat{\rho}\star\bbe^4\,, \quad
\bM^i = \lp \sqrt{\kappa}\hat{p}\delta^i_j + \kappa^{-3/2}m^i{}_j\rp\star\bbe^j\,, 
\ea
\es
where the new perturbation field $m_{ij}$ is dimensionless, symmetric, transverse and traceless. From the set of connection equations (\ref{CEoM1}), we can derive an expression for $\tilde{h}_{ij}$ in terms of the other perturbations,
\be
\tilde{h}_{ij} = h_{ij} + \frac{1}{B}\lp \gamma a\dot{h}_{ij} + \epsilon_{(i}{}^{kl}h_{j)k,l} - a\alpha\gamma\kappa^{-1}\tau m_{ij}\rp\,.
\ee
We note that this consistently generalises (\ref{hsolution}). Plugging this into the other set of connection equations (\ref{CEoM2}), we obtain a solution for the perturbation $m_{ij}$,
\be
m_{ij} =  \frac{ \beta\gamma\kappa \dot{h}_{ij}}{\tau}\,.  
\ee
As expected, this vanishes when $\beta=0$ or $\gamma =0$. Using these results in the khronon EoM (\ref{khrononEoM}), we arrive at the evolution equation for the gravitational waves,
\be
\ddot{h}_{ij} + 3(\dot{a}/a)\dot{h}_{ij} + (\gamma/ a)^{2}\nabla^2 h_{ij} = 0\,,
\ee
consistently with \cite{Nikjoo:2023flm}. As in subsection \ref{flrw}, we obtain the standard wave equation in the Lorentzian stage, wherein now, however, the gradient term is modulated by the prefactor $\gamma^2$. 
Thus, the speed of propagation of gravitational waves is strictly equal to the speed of light only in the case of strictly right-handed (or strictly left-handed) gravity.  

\subsection{Spin-1 perturbations}

In the irreducible decomposition of the connection, half of the degrees of freedom, i.e. 6$\times$2 components, are encoded into spin-1 perturbations in terms of 6 transverse (pseudo)vectors. We parameterise those as:
\bs
\label{vectorA}
\ba
\bA^{4}{}_i & = & A v_i\bdiff\tau + A\lp \delta_{ij} + 2w_{(i,j)} + \epsilon_{ijk} \tilde{u}^k\rp \bdiff x^j\,, \\
\bA^{ij} & = &  B\epsilon^{ijk}\tilde{v}_k\bdiff \tau  + B\lp \epsilon^{ij}{}_k + 2\delta^{[i}_k u^{j]} + \epsilon^{ijl} \tilde{w}_{(l,k)}\rp\bdiff x^k\,. \label{vectorAm}
\ea    
\es
Here $\tilde{u}$, $\tilde{v}$ and $\tilde{w}$ are the independent pseudovectors; $u$, $v$ and $w$ are the independent vectors. We have freedom to adjust the frame by a rotation given by a pseudovector $\tilde{r}^i$, $\delta_{\tilde{r}} \bA^{ij} = \epsilon^{ij}{}_k\bDiff \tilde{r}^k$, which allows us to set $\tilde{v}^k=0$. In addition, we have the freedom to perform arbitrary diffeomorphisms, and a convenient coordinate system could be obtained by choosing a gauge s.t. $v^i=0$, because then we would obtain the metric in the synchronous gauge. For generality, we shall proceed without 
any gauge fixing of the coordinates nor of the $SO(3)$ frame. However, because the Lorentz frame is now fixed indeed up to $SO(3)$ rotations, $\delta\phi^i = 0$, there can now be no vector perturbations in the khronon field. In the generic case, the 3-form $\bM^I$ features four independent vector modes, and can be parameterised as
\bs
\ba
\bM^4 & = & -\sqrt{\kappa}\hat{\rho}\star\bbe^4
+ q_i \star \bbe^i \,, \\
\bM^i & = & n^i\star\bbe^4 +  \lp \sqrt{\kappa}\hat{p}\delta^i_j 
+ m^i{}_{,j} + m_j{}^{,i} + \epsilon^i{}_{jk}\tilde{m}^k\rp\star\bbe^j\,.  \label{vectorMm}
\ea
\es
Using the sourced connection equations (\ref{CEoM1}), we obtain expressions for the (pseudo)vector perturbations contained in the `magnetic' connection (\ref{vectorAm}) as
\bs
\ba
 u^i  & = &  \tilde{u}^i - \frac{1}{2B}\lp \nabla^2 w^i + \epsilon^{ijk}\tilde{u}_{j,k}\rp + \frac{\sqrt{\kappa}\tau^2 A}{2\alpha\gamma^2 B} n^i\,,  \\
\tilde{v}^i & = & - v^i + \frac{1}{B}\lp \dot{\tilde{u}}^i -\frac{1}{2}\epsilon^{ijk}v_{j,k}\rp + \frac{\tau\sqrt{\kappa}}{\alpha\gamma^2 B}\lp \frac{1}{2\gamma}n^i - \tilde{m}^i\rp\,, \\
\tilde{w}^i & = & w^i - \frac{1}{2B}\lb \tilde{u}^i - \epsilon^{ijk}w_{j,k} + \gamma a \lp 2\dot{w}^i + v^i\rp - \frac{2\tau\sqrt{\kappa}}{\alpha\gamma}m^i\rb\,.  
\ea
\es
Using then the unsourced connection equations (\ref{CEoM2}) we obtain the expressions for vector perturbations contained in the spatial components of the integration 3-form $\bM^i$ (\ref{vectorMm}) as
\bs
\ba
n^i & = & 0\,, \\
m^i & = & \frac{\beta}{\tau\sqrt{\kappa}}\lp \dot{w}^i + \frac{1}{2}v^i\rp\,, \\
\tilde{m}^i & = & 0\,. 
\ea
\es
While continuity equation is trivially satisfied, the Euler equations $\bDiff \bM^i = 0$ identify the perturbation in the temporal part of integration form $\bM^4$.
\be
q^i = - \frac{\beta}{\sqrt{\kappa} \tau A} \nabla^2 \lp \dot{w}^i + \frac{1}{2} v^i \rp
\ee
Using these results in the field equations (\ref{khrononEoM}), they reduce to expressions in terms of the coordinate-invariant perturbation $V^i \equiv 2\dot{w}^i + v^i$. In the absence of vector matter sources we obtain
\bs
\ba \label{vectorpeq1}
\alpha \nabla^2 V^i & = & 0\,, \\
\alpha\lp \dot{V}^i + 3 \frac{\dot{a}}{a}V^i\rp & = & 0\,.
\ea
\es
It is clear that vector perturbations do not propagate.

\begin{comment}
First, we use the connection EoM (\ref{efe3}) to solve the pseudovectors,
\bs
\ba
\tilde{u}^i & = & w^i + \frac{1}{2B}\lp \nabla^2 u^i - \epsilon^{ijk}\tilde{u}_{j,k}\rp\,, \\ 
\tilde{v}^i & = & v^i - \frac{1}{B}\lp \dot{u}^i - \frac{1}{2}\epsilon^{ijk}v_{j,k}\rp\,, \\ 
\ea
\es
\end{comment}

\subsection{Spin-0 perturbations}

For generality, we don't fix into the synchronous coordinate system, but let the khronon $\phi$ depend on all the coordinates. At the level of cosmological background, the khronon may only be some function of $\tau$, and for simplicity we choose the linear function, but there may be some generic perturbation $\varphi$. So, 
\be
\phi = \kappa^{-1/2}\tau + \varphi\,. 
\ee
%Defining the background lapse function as $n=\sqrt{\kappa}\dot{\Phi}$, 
%\tomi{If one doesn't set $n=1$ then it has to be taken into account that $a=-\Phi A$. (It would seem in the Khronon EoM below $n$ is present but not in the connection EoM.)}
%\lucy{The $n$ in the result comes from $\diff\tau=n\diff t$. I've also noticed the missing $\sqrt{\kappa}$ and now running the code with $\kappa$ included. I think it is best to set $n=1$ here in the tetrad and also in the result ($\diff\tau=n\diff t$) what do you think? (I haven't typed in the full result because of some strange coefficients and still checking.}
We introduce 4 independent scalars $c$, $r$, $s$ and $\psi$, and 4 pseudoscalars $\tilde{c}$, $\tilde{r}$, $\tilde{s}$ and $\tilde{\psi}$ to describe the generic scalar perturbation of the connection  %\cmt{$ r_{i,j}\rightarrow r_{,ij}$ below?}
\bs
\label{scalarA}
\ba
\bA^{4}{}_i & = & A \dot{c}_{,i} \bdiff\tau + A\lp \lp 1 - \psi \rp \delta_{ij} + \Delta^i_j r + \epsilon_{ij}{}^{k} \tilde{s}_{,k} \rp \bdiff x^j\,, \\
\bA^{ij} & = &  B\epsilon^{ijk}\dot{\tilde{c}}_{,k} \bdiff \tau  + B\lp \epsilon^{ij}{}_k \lp 1 - \tilde{\psi} \rp + \epsilon^{ijl} \Delta_{lk} \tilde{r}  + 2 s_{,[i} \delta_{j]k}\rp\bdiff x^k\,, 
\ea    
\es
where we introduced the traceless $\Delta^i_i=0$ double-derivative operator $\Delta_{ij} f \equiv f_{,ij} - \frac{1}{3}\delta_{ij}\nabla^2 f$. 
It follows that the frame is  
\bs
\ba
\bbe^4 & = & \lp 1 + \sqrt{\kappa}\dot{\varphi} \rp \bdiff \tau + \sqrt{\kappa} \varphi_{,i} \bdiff x^i \,, \\ 
\bbe^i & = &  a\dot{c}^{,i}\bdiff\tau + a\lb\lp 1 + \Psi\rp\delta^i_j + \Delta^i_j r + \epsilon^i{}_{jk}\tilde{s}^k\rb\bdiff x^j\,,
\ea
\es
where we have introduced $\Psi \equiv \frac{\sqrt{\kappa}}{\tau} \varphi - \psi$.  
There are 4 additional scalar degrees of freedom in the integration form
\bs \label{scalarM}
\ba
\bM^4 & = & - \sqrt{\kappa} \lp \hat{\rho} + \delta \hat{\rho} \rp \star\bbe^4 - \chi_{,i} \star \bbe^i\,, \\
\bM_i & = & n_{,i} \star\bbe^4 +  \lp \sqrt{\kappa} \lp \hat{p} + \delta \hat{p} \rp \delta_{ij} 
+ \Delta_{ij}m + \epsilon_{ij}{}^{k}\tilde{m}_{,k}\rp\star\bbe^j\,.  
\ea
\es
As it is well-known, not all the scalar perturbations are physical, since some of them can be eliminated by a coordinate transformation. Under a diffeomorphism parameterised by $\boldsymbol{\xi} = \xi^4\bdiff\tau + a^2\xi_{,i}\bdiff x^i$, with the action $\delta_{\bxi} \equiv \{\bdiff, \sharp\bxi\lrcorner\}$ ($\lrcorner$ denoting the interior product, $\sharp$ the musical isomorphism involving the inverse metric) the time component of $\bxi$ shifts the khronon perturbation,
\bs
\label{gaugetransf}
\be
\delta_{\bxi} \varphi = \kappa^{-\frac{1}{2}}\xi^4\,. 
\ee
The perturbed $Spin(4)$ connection is transformed as
\be
\delta_{\bxi} c =  \delta_{\bxi} \tilde{c} = \delta_{\bxi} r  =  \delta_{\bxi} \tilde{r} = \xi\,, \quad 
\delta_{\bxi} \psi  =  - \frac{\dot{A}}{A}\xi^4 -\frac{1}{3}\nabla^2\xi\,, \quad
\delta_{\bxi} \tilde{\psi} = - \frac{\dot{B}}{B}\xi^4 -\frac{1}{3}\nabla^2\xi\,, \quad
\delta_{\bxi} s  =  \delta_{\bxi} \tilde{s} = 0\,.
\ee
The perturbations in the integration form $\bM^4$ and $\bM^i$ transform as, respectively,
\ba
\delta_{\bxi}\delta\hat{\rho} & = & \lp \dot{\hat{\rho}} + 3\frac{\dot{a}}{a}\hat{\rho}\rp\xi^4 + \hat{\rho}\nabla^2\xi\,, \quad
\delta_{\bxi}\chi = -a\sqrt{\kappa}\hat{\rho}\dot{\xi}\,, \\
\delta_{\bxi} n & = & -\frac{\sqrt{\kappa}\hat{p}}{a}\xi^4\,, \quad
\delta_{\bxi}\hat{p} = \frac{\partial\lp a^2\hat{p}\xi\rp}{a^2\partial \tau} - \frac{4}{3}\hat{p}\nabla^2\xi\,, \quad
\delta_{\bxi} m = -2\sqrt{\kappa}\hat{p}\xi\,, \quad
\delta_{\bxi}\tilde{m} = 0\,. 
\ea
\es
This completes the setting-up of the scalar degrees of freedom and their transformations. 

\subsubsection{Derivation of the linear EoM}

From the sourced connection EoM's (\ref{CEoM1}), we obtain the (pseudo)scalars in the `magnetic' part of the connection,
\bs \label{Cscalar}
\ba
\dot{\tilde{c}} & = & \dot{c} - \frac{\tilde{s}}{B} - \frac{\sqrt{\kappa} \tau}{2\alpha \gamma^3 B} \lp n + 2 \gamma \tilde{m} \rp\,,  \\
 \tilde{r} & = & r - \frac{\tilde{s}}{B} + \gamma \frac{\tau A}{B} \lp  \dot{c} - \dot{r} + \frac{\sqrt{\kappa}\tau}{\alpha \gamma^2} m \rp\,,\\
s & = & - \tilde{s} + \frac{1}{B} \lp \frac{\sqrt{\kappa} \dot{A}}{A} \phi + \psi + \frac{1}{3} \nabla^2 r  - \frac{ \sqrt{\kappa} \tau^2 A}{2 \alpha \gamma^2 } n \rp\,,
\\
\tilde{\psi} &=& \frac{\sqrt{\kappa} \tau^2 A}{4\alpha \gamma B} \lp \sqrt{\kappa} \delta \hat{p} + \frac{2}{3} \nabla^2 m + 2\beta \frac{\dot{A}}{\tau^2 A} \varphi \rp - \frac{1}{2} \lp \Psi - \frac{1}{3} \nabla^2 r \rp - \frac{\gamma \tau A}{2 B} \lp \sqrt{\kappa} \frac{\dot{A}}{A} \dot{\varphi} + \dot{\psi} + \frac{1}{3} \nabla^2 \dot{r} \rp 
\,.
\ea
\es
From the unsourced connection equations (\ref{CEoM2}), we obtain 3$\times$4 independent equations. Combined with (\ref{Cscalar}), these identify the scalar perturbation of the integration 3-form (\ref{scalarM}).
\bs \label{Dscalar}
\ba
n & = & 0\,, \\
m & = & \frac{\beta}{\sqrt{\kappa} \tau} \lp \dot{c} + \dot{r} \rp\,, \\
\tilde{m} & = & 0\,,
\\
\delta \hat{p} &=& - 2 \frac{\beta}{\kappa \tau} \lp \frac{\sqrt{\kappa} \dot{A}}{\tau A} \varphi + \sqrt{\kappa} \frac{\dot{A}}{A} \dot{\varphi} + \dot{\psi} + \frac{1}{3} \nabla^2\dot{c} \rp\,. \label{deltap}
\ea
\es
\begin{comment}
\bs
\ba
&& n_{,i} = 0\,, \\
&& m_{,i,j} = -\frac{\beta}{4\sqrt{\kappa} \tau} \lp \dot{c}_{,i,j}+ \dot{r}_{,i,j} \rp \quad (i\neq j) \,, \\
&& m_{,1,1} - m_{,2,2} = -\frac{\beta}{4\sqrt{\kappa} \tau} \lp \dot{c}_{,1,1} - \dot{c}_{,2,2} + \dot{r}_{,1,1} - \dot{r}_{,2,2} \rp \,,\\
&& m_{,3,3} - m_{,1,1} = -\frac{\beta}{4\sqrt{\kappa} \tau} \lp \dot{c}_{,3,3} - \dot{c}_{,1,1}+ \dot{r}_{,3,3} - \dot{r}_{,1,1} \rp \,,\\
&& \tilde{m}_{,i} = 0\,. 
\ea
\es
\end{comment}
The Euler equation $\bDiff \bM_i = 0 $ further identify
\be \label{Escalar}
\chi = \frac{2\beta}{\sqrt{\kappa} \tau A} \lp \sqrt{\kappa} \lp \frac{\ddot{A}}{A} - \frac{\dot{A}^2}{A^2} \rp \varphi + \sqrt{\kappa} \frac{\dot{A}}{A} \dot{\varphi} + \dot{\psi} + \frac{1}{3}\nabla^2 \dot{r} \rp \,,
\ee
and the continuity equation $\bDiff \bM_4 = 0 $ yields the relation between the perturbation in density $\delta \hat{\rho}$ and in pressure $\delta\hat{p}$. 
Applying the background continuity equation \eqref{DMcont} and the previous solutions for perturbations (\ref{Cscalar},\ref{Dscalar},\ref{Escalar}), we obtain
\ba \label{scalarDMeuclid}
\delta \dot{\hat{\rho}} + 3 \lp \frac{1}{\tau} + \frac{\dot{A}}{A} \rp \delta \hat{\rho} &=& 
\sqrt{\kappa} \hat{\rho}' \dot{\varphi} + 3\lp \frac{\sqrt{\kappa}}{\tau^2} \varphi + \sqrt{\kappa} \frac{\dot{A}}{A} \dot{\varphi} + \dot{\psi} + \frac{1}{3} \nabla^2 \dot{c} \rp 
+ \frac{6\beta}{\kappa \tau^2} \lp 2 \sqrt{\kappa} \frac{\dot{A}}{\tau A} \varphi + \frac{\dot{A}}{A} \dot{\varphi} + \dot{\psi} + \frac{1}{3} \nabla^2 \dot{c} \rp
\\\nn &&
- \frac{2\beta}{\kappa \tau^2 A^2} \nabla^2 \lb \sqrt{\kappa} \lp \frac{\dot{A}^2}{A^2} - \frac{\ddot{A}}{A} \rp \varphi - \sqrt{\kappa} \frac{\dot{A}}{A} \dot{\varphi} - \dot{\psi} - \frac{1}{3} \nabla^2 \dot{r} \rb\,.
\ea
We include the matter source
\bs
\label{mattersources}
\ba
\bt^4 & = & -\sqrt{\kappa}\lp \rho + \delta \rho\rp\star\bbe^4 + \lp \rho + p\rp\lp \sqrt{\kappa}u_{,i} + a^{-1}{\kappa}\varphi\rp_{,i}\star\bbe^i\,, \\
\bt^i & = & \lp \rho + p\rp\lp \sqrt{\kappa}u + a^{-1}\kappa\varphi\rp^{,i}\star\bbe^4 + \sqrt{\kappa}\lb \lp p + \delta p\rp\delta^i_j + \Delta^i{}_j\Pi\rb\star\bbe^j\,. 
\ea
\es
This form of the source term corresponds to the standard parameterisation of a fluid stress energy, as shown in the appendix \ref{mattercurrent}. 
\begin{comment}
\be
\bt^4 = - \sqrt{\kappa} \lp \rho + \delta \rho \rp \star \bbe^4 + \lp \rho + p \rp v_{,i} \star \bbe^i\,,
\quad
\bt_i = u_{,i} \star \bbe^4 + \lp \sqrt{\kappa} \lp p + \delta p \rp \delta_{ij} + \Delta_{ij} \Pi + \epsilon_{ij}{}^{k}\tilde{\Pi}_{,k} \rp \star \bbe^j\,,
\ee

$\bt^a\equiv e^a{}_\nu T^\nu{}_\mu \star \bbe^\mu$
\bs \ba
\bt^4 &=& - \sqrt{\kappa} \lp \rho + \delta \rho + \sqrt{\kappa} \rho \dot{\varphi} \rp \star \bbe^4 + \lp \rho + p \rp \lp v_{,i} + \sqrt{\kappa} \varphi_{,i} \rp \star \bbe^i\,,
\\
\bt_i &=& \lb \lp \rho + p \rp u_{,i} - \rho \dot{c}_{,i} \rb \star \bbe^4 - \lb \lp p + \delta p + p \lp \Psi + \Delta_{ij} r \rp \rp \delta_{ij} + \epsilon_{ij}{}^k \tilde{s}_{,k} + r_{,i,j} \rb \star \bbe^j\,,
\ea \es
\end{comment}
With these ingredients, the perturbation equations derived from the khronon EoM's are as below.
\bs 
\label{scalaeqs}
\ba
&& \delta \rho = - \delta \hat{\rho} - \frac{2 \alpha \gamma^2}{\kappa \tau^2 A^2} \nabla^2 \lp  \sqrt{\kappa} \frac{\dot{A}}{A} \varphi + \psi + \frac{1}{3} \nabla^2 r \rp 
- \frac{6 \alpha}{\kappa} \lp \frac{1}{\tau} + \frac{\dot{A}}{A} \rp \lp \frac{\sqrt{\kappa}}{\tau^2} \varphi + \sqrt{\kappa} \frac{\dot{A}}{A} \dot{\varphi} + \frac{\sqrt{\kappa}}{\tau^2} \varphi + \dot{\psi} + \frac{1}{3} \nabla^2 \dot{c} \rp 
\,, \qquad
\\
&& \lp \rho + p \rp u_{,i}  
= - \frac{\sqrt{\kappa}}{\tau A} \lp \rho + p \rp \varphi_{,i} - \frac{2 \alpha}{\kappa \tau A} \lb \sqrt{\kappa} \lp \frac{\dot{A}^2}{A^2} - \frac{\ddot{A}}{A} \rp \varphi_{,i} 
- \sqrt{\kappa}\frac{\dot{A}}{A} \dot{\varphi}_{,i} - \dot{\psi}_{,i} - \frac{1}{3}\nabla^2 \dot{r}_{,i} \rb   \,, \label{velocityeq}
\\
&& \delta p 
= \frac{2\alpha}{\kappa} \lb 3 \sqrt{\kappa} \frac{\dot{A}}{\tau^2 A} \varphi + \sqrt{\kappa} \lp 3 \frac{\dot{A}}{\tau A} + 2 \frac{\ddot{A}}{A} + \frac{\dot{A}^2}{A^2} \rp \dot{\varphi} + 3 \lp \frac{1}{\tau} + \frac{\dot{A}}{A} \rp \lp \dot{\psi} + \frac{1}{3} \nabla^2 \dot{c} \rp 
+ \sqrt{\kappa} \frac{\dot{A}}{A} \ddot{\varphi} + \ddot{\psi} + \frac{1}{3} \nabla^2 \ddot{c} \rb
\\ \nn && \qquad
+ \frac{2 \lp \alpha - \beta \rp}{3 \kappa} \lb \frac{3\sqrt{\kappa}}{\tau^3} \varphi + 3 \sqrt{\kappa} \frac{\dot{A}^2}{A^2} \dot{\varphi} + \frac{1}{\tau^2 A^2} \nabla^2 \lp \sqrt{\kappa} \frac{\dot{A}}{A} \varphi + \psi + \frac{1}{3} \nabla^2 \dot{r} \rp \rb \,,
\\
&& \Pi_{,i,j} = - \frac{\alpha - \beta}{\kappa\tau^2 A^2} \lp \sqrt{\kappa} \frac{\dot{A}}{A} \varphi_{,i,j} + \psi_{,i,j} + \frac{1}{3} \nabla^2 r_{,i,j} \rp - \frac{3\alpha}{\kappa} \lp \frac{1}{\tau} + \frac{\dot{A}}{A} \rp \lp \dot{c}_{,i,j} - \dot{r}_{,i,j} \rp 
- \frac{3\alpha}{\kappa} \lp \ddot{c}_{,i,j} - \ddot{r}_{,i,j} \rp \,.
\ea
\es   
There are only four independent (sets of) equations in the end, because the spatial components of the energy equation and time components of the momentum equations yield the identical velocity perturbation constraint (\ref{velocityeq}). Since $u$ is potential for the velocity of the fluid i.e. its gradient is the fluid's motion, $u$ is regarded as (proportional to) a time derivative. This should be taken into account when translating the physics into the Lorentzian ``time frame'', as we recall from the discussion of angular velocity in section \ref{kerr}. Thus, our Wick rotation implies $u \rightarrow -iv$, where $v$ is the usual spacetime velocity perturbation potential. Denoting derivatives wrt $t$ by a prime, recalling that $\Psi = \frac{\varphi}{m_P t} - \psi$ and defining $\Phi \equiv \frac{\varphi}{m_p}$ , the full set of field equations (\ref{scalaeqs}) are translated into the Lorentzian framework as
\bs 
\label{spin0eqs}
\ba \label{densityper}
&& \delta \rho + \delta \hat{\rho} = - 6 \alpha m_P^2 H \lp H \Phi' - \Psi' + \frac{1}{3} \nabla^2 c' \rp 
+ 2 \lp \alpha - \beta \rp \frac{m_P^2}{a^2} \nabla^2 \lp H \Phi - \Psi + \frac{1}{3} \nabla^2 r\rp
+ 6 \beta \frac{m_P^2}{t^3} \Phi \,,
\\
&& \lp \rho + p \rp v_{,i} 
%\lp = - 2 m_P^2 \lp \alpha H' + \beta t^{-2} \rp v_{,i} \rp
= - \frac{\Phi_{,i}}{a}\hat{\rho}  
+ 2 \alpha \frac{m_P^2}{a} \lp H \Phi'_{,i} - \Psi'_{,i} + \frac{1}{3} \nabla^2 r'_{,i} \rp 
- 2 \beta \frac{m_P^2}{a t^2} \Phi_{,i}\,, \label{velocityper}
\\
&& \delta p = 2 \alpha m_P^2 \lb \lp 3 H^2 + 2 H' \rp \Phi' + H \lp \Phi'' - 3\Psi' +  \nabla^2 c' \rp - \lp \Psi'' - \frac{1}{3} \nabla^2 c'' \rp \rb \label{pressureper}
\\\nn&& \qquad
- \frac{2}{3} \lp \alpha - \beta \rp \frac{m_P^2}{a^2} \nabla^2 \lp H \Phi - \Psi + \frac{1}{3} \nabla^2 r \rp
- 2 \beta \frac{m_P^2}{t^3} \Phi \,, 
\\
&& \Pi_{,i,j} = \lp \alpha - \beta \rp \frac{m_P^2}{a^2} \lp  H \Phi_{,i,j} - \Psi_{,i,j} + \frac{1}{3} \nabla^2 r_{,i,j} \rp - 3 \alpha m_P^2 H \lp c'_{,i,j} - r'_{,i,j} \rp - \alpha m_P^2 \lp c''_{,i,j} - r''_{,i,j} \rp\,. \label{anisotropyper}
\ea
\es 
\begin{comment}
The density perturbation \eqref{densityper} and the continuity equation \eqref{scalarDMcont} yield a continuity equation for the integration form and matter perturbation. 
\ba
\delta \rho' + 3H \delta \rho
&=& - \delta \hat{\rho}' - 3 H \delta \hat{\rho} - 2 \frac{m_P^2}{a^2} \lp \alpha - \beta \rp \nabla^2 \lb \lp H^2 - H' \rp \Phi - \Psi' + \frac{1}{3} \nabla^2 r' + H \lp \Phi' - \Psi + \frac{1}{3}\nabla^2 r \rp \rb
\\\nn && 
- 6 \alpha m_P^2 \lb H \lp 3 H^2 + 2 H' \rp \Phi' - \lp 3 H^2 + H' \rp \lp \Psi' - \frac{1}{3} \nabla^2 c \rp - H^2 \Phi'' + H \lp \Psi'' - \frac{1}{3} \nabla^2 c'' \rp \rb
\ea
With the background solution \eqref{FriedmannFinal1}, we obtain the relative density perturbation.
\ba
\hat{\delta} \equiv \frac{\delta \hat{\rho}}{\hat{\rho}} &=& - \lb \delta \rho - 6 \alpha m_P^2 H \lp H \Phi' - \Psi' - \frac{1}{3} \nabla^2 c' \rp 
+ 2 \alpha \gamma^2 \frac{m_P^2}{a^2} \nabla^2 \lp H \Phi - \Psi + \frac{1}{3} \nabla^2 r\rp \rb \lp - \rho + 3 \alpha H^2 - \beta t^{-2} \rp^{-1}\,
\\\nn
& \approx & \lb \delta - 6 \alpha m_P^2 \rho^{-1} H \lp H \Phi' - \Psi' - \frac{1}{3} \nabla^2 c' \rp 
+ 2 \alpha \gamma^2 \frac{m_P^2}{\rho a^2} \nabla^2 \lp H \Phi - \Psi + \frac{1}{3} \nabla^2 r\rp \rb \lp 1 - 3 \alpha \rho^{-1 }H^2 + \beta \rho^{-1} t^{-2} \rp
\ea
\end{comment}
Some consistency checks are pertinent at this point, and we have verified that these equations imply the two matter EoM (\ref{p_eqs}) given in Appendix \ref{mattercurrent}. First, we solve $\delta\rho$ and $\delta\rho'$ from (\ref{densityper}) and its first time derivative, use (\ref{velocityper}) as an expression for $v$ and use (\ref{pressureper}) as an expression for $\delta p$, and when we plug these into the continuity equation for matter (\ref{p_continuity}), it becomes
\ba \label{scalarDMcont}
\delta \hat{\rho}' + 3 H \delta \hat{\rho} &=& \Phi' \hat{\rho}' + 3 \lp H \Phi' - \Psi' + \frac{1}{3} \nabla^2 c' \rp \hat{\rho}
+ 6 \beta \frac{m_P^2}{t^2} \lb \lp \frac{2H}{t} - \frac{3}{t^2} \rp \Phi + H \Phi' - \Psi' + \frac{1}{3} \nabla^2 c' \rb 
\\\nn &&
- 2 \beta \frac{m_P^2}{a^2} \nabla^2 \lb H' \Phi + H \Phi' - \Psi'
+ \frac{1}{3} \nabla^2 r' \rb\,.
\ea
As it should, this is precisely the continuity equation for dark matter ({\ref{scalarDMeuclid}}) translated into the Lorentzian ``time frame''. Another check is given by the Euler equation for matter (\ref{p_euler}), wherein we now plug $v$ and $v'$ solved from (\ref{velocityper}) and its first time derivative, again use (\ref{pressureper}) as an expression for $\delta p$ and now also (\ref{anisotropyper}) as an expression for $\Pi$. We find that (\ref{p_euler}) is now identically satisfied, as it should since we have already solved the Euler equations for dark matter. 

\subsubsection{Newtonian gauge description}

Up to now, we have considered the generic perturbations without any gauge fixing. It can be useful to specialise into the conformal Newtonian gauge which is probably the most conventional gauge in cosmological perturbation theory. However, the new pregeometric approach suggests a slightly unconventional notation for the metric perturbations in the Newtonian gauge. This gauge is reached by implementing the diffeomorphism $\delta_{\boldsymbol{\xi}}$ 
with $\xi^4 = -a(\dot{c}-\dot{r})+\sqrt{\kappa}\varphi$ and $\xi=-a^{-2}r$. According to (\ref{gaugetransf}), the perturbations of the khronon and the connection then become
\bs
\ba
\Phi^N & = & -a\lp c' -r'\rp\,, \\
\Psi^N & = & \Psi - H\lp\Phi - ac' + ar'\rp - \frac{1}{3}\nabla^2 r\,, \\
c^N & = & c-r\,, \\
r^N & = & 0\,,
\ea
\es
where the superscript indicates that the variable is evaluated in the Newtonian gauge. Similarly, the dark matter and matter perturbations are transformed into $\delta\hat{\rho}^N$, $\delta\rho^N$, and so on. We omit the superscripts for convenience, and state the field equations (\ref{spin0eqs}) in the Newtonian gauge:
\bs 
\label{spin0eqsN}
\ba \label{densityperN}
m_P^{-2}\lp \delta \rho + \delta \hat{\rho}\rp  
& = & 6 \alpha  H \lp \Psi' - H \Phi' \rp 
- \frac{2}{a^2} \lp \alpha- \beta \rp \nabla^2 \Psi - \frac{2\beta}{a^2}\nabla^2\Phi +  \frac{6 \beta}{t^3} \Phi\,,
\\ \label{vN}
 m_P^{-2}a\lp \rho + p \rp v
& = & - m_P^{-2}\hat{\rho}\Phi  + 2 \alpha\lp H \Phi' - \Psi' \rp 
- 2 \beta \frac{\Phi}{ t^2}\,,
\\
 m_P^{-2}\delta p & = & 2\alpha\lb \lp 3 H^2 + 2 H' \rp \Phi' + H \lp \Phi'' - \Psi''- 3\Psi' \rp + \frac{1}{3a^2}{\nabla^2}\lp \Psi + \Phi'\rp\rb 
 +   \frac{2\beta }{3a^2} \nabla^2  \lp H \Phi - \Psi \rp
- \frac{2\beta}{t^3} \Phi \,, \quad
\\
 m_P^{-2}a^2\Pi & = & - \alpha \lp \Psi + \Phi' \rp 
+ \beta\lp \Psi - H\Phi \rp \,. \label{anisotropyperN}
\ea
\es 
It is interesting to note that the standard dimensionless gravitational perturbations that appear in the longitudinalised metric 
\be
\bdiff s^2 = -\lp 1+2\Phi'\rp\bdiff t\otimes\bdiff t + a^2\lp 1+2\Psi\rp\delta_{ij}\bdiff x^i\otimes\bdiff x^j\,,
\ee
are now interpreted in solely in terms of the khronon field plus the diagonal perturbation of the spatial connection, like
\be \label{newtonpotentials}
\Phi' = \frac{\partial\varphi}{\partial \phi}\,, \quad 
\Psi = \frac{\varphi}{\phi} - \psi\,.
\ee
For this reason we call the dimensionless lapse perturbation $\Phi'$ (instead of $\Phi$ as usual). Now also $\Phi$ appears in theory, and it plays the role of the dark matter velocity perturbation potential, ``$\Phi  \sim \hat{v}$''. In this way, no new fields enter into the system besides the dark matter density perturbation, consistently with the knowledge that the dynamical phase of the theory introduces only one scalar degree of freedom beyond general relativity. It is well known to those familiar with cosmological perturbation theory that the synchronous coordinate system leaves a remnant gauge freedom which is conventionally fixed by setting the cold dark matter velocity perturbation to zero. In our pregeometric theory, there is a fundamental relation between the temporal component of metric geometry and the dark matter velocity field; namely, the former is the derivative of the latter; fluctuations in time lapse are a manifestation of acceleration in dark matter.  

\subsubsection{Clustering of dark matter}

Adopting this interpretation, we consider an analogue of the usual comoving density perturbation $\delta \rho^C$, which we would now obtain as
\be
\delta\rho^C = \delta\rho + \delta\hat{\rho} + 3aH\lp \rho + p\rp v + 3\lp H\hat{\rho} + \frac{\hat{p}}{t}\rp\Phi\,.  
\ee
Using equations (\ref{densityperN}) and (\ref{vN}), we get
\bs
\label{poisson}
\be
-\nabla^2\lb \lp \alpha-\beta\rp \Psi + \beta H\Phi\rb = \frac{a^2\delta\rho^C}{2m_P^2}\,. 
%\quad \text{where} \quad G^{-1}_{eff} = 8\pi\lp\alpha - \beta\rp m_P^2\,.  
\ee
We see from (\ref{anisotropyperN}) that there will also generically occur effective anisotropic stress proportional to $\beta/\alpha$ i.e. the two gravitational potentials (\ref{newtonpotentials}) will not be equal to minus each other. Using (\ref{anisotropyperN}), we now obtain the Poisson equation in its standard form (the effective Newton's constant being determined directly by the parameter $\alpha$)
\be \label{poisson2}
\nabla^2\Phi' = \frac{a^2\delta\rho^C}{2\alpha m_P^2}\,. 
\ee
\es
The dark matter clustering is governed by (\ref{scalarDMcont}), which is rewritten in the Newtonian gauge as
\ba \label{contN}
\delta\hat{\rho}' + 3H\delta\hat{\rho} -
\hat{\rho}'\Phi' + \hat{\rho} \lp 3\Psi' - 3H\Phi' - \frac{\nabla^2}{a^2}\Phi\rp
& = & \frac{2\beta m_P^2}{a^2}\nabla^2\lp \Psi' - H\Phi' - H'\Phi\rp 
\nn \\
& + & \frac{6\beta m_P^2}{t^2}\lb -\Psi' + H\Phi' + \frac{1}{3}\lp \frac{\nabla^2}{a^2} + \frac{6Ht-9}{t^2}\rp\Phi\rb\,. 
\ea
At this point, we shall make some approximations to facilitate the analysis.

It is helpful to focus on scales inside the horizon, such that $\nabla^2 f \gg H^2f$. This is often called the `quasi-static' limit, because it allows one to neglect time derivatives, as they reflect the cosmological evolution, $f' \sim Hf$, $f'' \sim H^2 f$. Also, as we see from (\ref{poisson}), the gravitational potentials are negligible compared to the density perturbations, $\delta\rho, \delta\hat{\rho} \gg \rho\Phi',\rho\Psi$. In the quasi-static limit  (\ref{contN}) reduces to
\be
\delta\hat{\rho}' + 3H\delta\hat{\rho} = \frac{\nabla^2}{a^2}\lb 2\beta m_P^2\lp \Psi'-H\Phi'\rp + \lp \hat{\rho}-2\beta m_P^2 H' + \frac{2\beta m_P^2}{t^2}\rp\Phi\rb\,. 
\ee
The time derivative of this equation yields a second order equation for the dark matter density perturbation, sourced by $\Phi$, $\Psi$ and their derivatives. If we neglect the standard matter terms, which is justified in a vacuum and in regions dominated by dark matter, we can then use the field equations (\ref{spin0eqsN}) to obtain expressions for the $\Phi$, $\Psi$ and their derivatives ($\Phi'$ being already obtained at (\ref{poisson2}), since at small scales $\delta{\hat{\rho}}^C \approx \delta{\hat{\rho}}$ in the Newtonian gauge), in order to set the second order equation into an autonomous form. We report only the final result, in terms of the fractional energy density perturbation $\hat{\delta} \equiv \delta\hat{\rho}/\hat{\rho}$,
\ba \label{theresult}
\hat{\delta}'' & + & \lb 2H - \frac{2\beta m_P^2}{t^3}\lp \frac{6-6Ht}{\hat{\rho}} + \frac{t^2}{\hat{\rho}t^2 + 2\beta m_P^2}\rp\rb\hat{\delta}' \nn \\
& = & \left\{ \frac{\hat{\rho}}{2\alpha m_P^2} + 2\beta \frac{m_P^2 \hat{\rho}  t^2 \lb 3 \alpha 
    H t (5 H t-7)+9\alpha-5 \beta \rb +  6 \beta  m_P^4 \lb\alpha  H t (5 H t-8)+4\alpha-\beta \rb -\hat{\rho}^2 t^4}{\alpha  \hat{\rho}  t^4 \left(2 \beta  m_P^2+\hat{\rho} 
   t^2\right)}\right\}\hat{\delta}\,.\quad 
\ea
When $\beta=0$, we recover the standard evolution equation for density perturbations in cold dark matter. In a generic theory, both the friction term and the effective Newton's constant (as read from the curly bracket) are modulated by time-dependent background terms.
The result verifies the existence of a single scalar (extra) degree of freedom without any instabilities or other pathologies. It is remarkable that in the $\beta \neq 0$ case the effective fluid has pressure $\hat{p} \neq 0$, but nevertheless there doesn't appear any gradient term in (\ref{theresult}). In other words, the generalised dark matter is still characterised by vanishing sound speed\footnote{This property of Lorentz gauge theory was anticipated in a completely different approach of phenomenological fluid Lagrangians \cite{Iosifidis:2024ksa}. A nonzero sound speed does not imply a physical pathology (though an imaginary sound speed implies an instability); however, it compromises the phenomenological viability of a wide range of alternative dark sectors from Cardassian and Chaplygin models to interacting quintessence scalars, 3-forms and vectors to Palatini-f(R) and various other modified gravities, see \cite{Iosifidis:2024ksa} for references.}. This is only possible because the effective fluid is associated a with spin current. 

In addition to the modified growth rate of clustering, one of the various avenues for testing the theory is through the presence of an effective anisotropic stress \cite{Hu:1998kj,Koivisto:2005mm,Mota:2007sz}, which manifests as a difference between the two scalar gravitational potentials (\ref{newtonpotentials}). This difference is captured by the ``gravitational slip'' parameter $\eta \equiv -\Psi/\Phi'$ \cite{Amendola:2007rr,Daniel:2008et,CosmoVerse:2025txj}. 
Expanding to
first order in $\beta$ in the present quasi-static approximation,
\be
\eta = 1 + \lb \frac{2m_P^2 H}{\hat{\rho}} (\log{\hat{\delta}})'+ \frac{1}{\alpha}\rb\beta + \mathcal{O}(\beta^2)\,,
\ee
confirming that this effect is absent in the right-handed limit.
The gravitational slip can be observationally constrained by combining galaxy clustering (which probes the Newtonian potential $\Phi'$) with weak lensing (which probes the lensing potential $\Phi'-\Psi$).  

\begin{comment}
\ba
\delta \hat{\rho}' + 3 H \delta \hat{\rho} &=& \Phi' \hat{\rho}' + 3 \lp H \Phi' - \Psi'  -\frac{k^2}{3a^2} \Phi \rp \hat{\rho}
+ 6 \beta \frac{m_P^2}{t^2} \lb \lp \frac{2H}{t} - \frac{3}{t^2} \rp \Phi + H \Phi' - \Psi' \rb 
\\\nn &&
+ 2 \beta \frac{m_P^2}{a^2} k^2 \lb \lp H' - \frac{1}{t^2} \rp \Phi + H \Phi' - \Psi' \rb
\ea
Then we find
\bs \ba
&& \frac{2m_P^2}{a^2} \nabla^2 \lb \lp \alpha- \beta \rp \Psi + \beta H \Phi \rb = - \lp \delta \rho^N + \delta \hat{\rho}^N \rp + 3H \lp av^N + \Phi \rp \lp \hat{\rho} + 2 \alpha m_P^2 H' + 2 \beta \frac{m_P^2}{t^2} \rp + 6 \beta \frac{m_P^2}{t^3} \Phi \,,
\\
&& \delta p ^N + \frac{2}{3} \Pi^N = - 2 \alpha m_P^2 \lb \lp 3 H^2 + 2 H' \rp \Phi' + H \lp \Phi'' - \Psi''- 3\Psi' \rp \rb
- 2 \beta \frac{m_P^2}{t^2} \lb \Phi' + \lp H -\frac{1}{t}\rp \Phi \rb - 2 \beta \frac{m_P^2}{t^3} \Phi \,, 
\ea \es
%where we have introduced the relative density perturbation of the integration form $\hat{\delta} \equiv \frac{\delta \hat{\rho}}{\hat{\rho}}$.
\ba
\hat{\delta}'_N&=&-3H \hat{\delta} - \frac{\hat{\rho}'}{\hat{\rho}} \hat{\delta} + \frac{\hat{\rho}'}{\hat{\rho}} \Phi' + 3 H \Phi' - \Psi' + \frac{1}{a^2} \nabla^2 \Phi 
\\\nn &&\quad
+ 2 \beta \frac{m_P^2}{\hat{\rho}t^2} \lb H \Phi' - \Psi' - \lp 3 \frac{H}{t} + \frac{1}{t^2} \rp \Phi \rb + 2 \beta \frac{m_P^2}{\hat{\rho}a^2} \nabla^2 \lp H'\Phi + H \Phi' - \Psi' \rp
\ea
\end{comment}

\section{Spherical symmetry}
\label{spherical}

We now turn to the study of spherically symmetric solutions, which provide a crucial testing ground for the nonperturbative structure of the theory. Unlike the cosmological setting, where deviations from homogeneity are treated perturbatively, spherical symmetry allows for fully nonlinear configurations while still permitting significant analytic control. This sector is particularly well-suited for probing the theory's strong-field regime, examining the nature of horizons, and assessing the viability of black hole solutions. It also offers a sharp diagnostic for potential pathologies, such as singularities, asymptotic non-flatness, or unphysical branches that may arise in generalised gravity theories. In the present framework, grounded in a Euclidean 
$Spin(4)$ gauge structure, we will see that spherically symmetric solutions behave in a controlled and physically reasonable manner, further reinforcing the consistency of the theory beyond linear perturbations. While a full classification of solutions—especially beyond the restricted “one-handed gravity” sector—remains an open problem, this initial exploration reveals no inconsistencies and shows encouraging agreement with the cosmological sector, suggesting a coherent and unified underlying structure.

After first deriving the generic field equations in the generic spherically symmetric case, and further analysing them in the absence of material spin currents, we both reproduce the known static solutions when $\beta=0$ and show their absence when $\beta \neq 0$ in \ref{nostatic}, and then briefly check the simplest, homogeneous solutions when $\beta \neq 0$ in \ref{ongeneric}. 
  
\subsection{The general field equations}

Without loss of generality, a spherically symmetric configuration can be parameterised as\footnote{The coefficients $\bA^{4i}$ have been put into a convenient form by exhausting the freedom to choose coordinates, whilst the coefficients $\bA^{ij}$ are completely generic.}: 
\bs
\label{sphericalAnsatz}
\ba
\bA^{41} & = & -\frac{F}{\tau}\bdiff\rho\,, \quad \bA^{42} = -\frac{r}{\tau}\bdiff\theta\,, \quad \bA^{43} = -\frac{r\sin\theta}{\tau}\bdiff\varphi\,, \\
\bA^{12} & = & A\bdiff\theta + B\sin\theta\bdiff\varphi\,, \quad \bA^{13} = -B\bdiff\theta + A\sin\theta\bdiff\varphi\,, \quad 
\bA^{23} =  C\bdiff\tau + D\bdiff\rho - \cos\theta\bdiff\varphi\,. 
\ea
\es
There are now 6 functions $F$, $r$, $A$, $B$, $C$ and $D$, of 2 variables, $\tau$ and $\rho$. Note that because of the angular parameterisation, the connection functions $A$ and $B$ are now dimensionless, and also $F$ is a dimensionless function. Due to
spherical symmetry, all the 6 functions depend on the 2 variables via $r(\tau,\rho)$, and if we impose staticity, then all the functions may depend only on $r$. In the following, a dot denotes derivative wrt $\tau$ as previously, and a prime denotes derivative wrt $\rho$. 

The connection (\ref{sphericalAnsatz}) can equivalently be presented as
\bs
\ba
\x\bA^1 & = &  -C\bdiff\tau - \lp\pm\frac{F}{\tau} + D\rp\bdiff\rho + \cot\theta\bdiff\tilde{\varphi}\,, \\
\x\bA^2 & = & -\lp \pm \frac{r}{\tau} + B\rp\bdiff\theta + A\viff\,, \\
\x\bA^3 & = & -A\bdiff\theta - \lp \pm \frac{r}{\tau} + B\rp\viff\,,
\ea
\es
where we introduced the short-hand notation $\viff \equiv \sin\theta\bdiff\varphi$. 
We first compute the torsion,
\bs
\label{sphericalT}
\ba
\bT^1 & = & \lp \dot{F} - \frac{F}{\tau}\rp\bdiff\tau\wedge\bdiff\rho - 2Br\bdiff\theta\wedge\viff\,, \\
\bT^2 & = & \lp \dot{r} - \frac{r}{\tau}\rp\bdiff\tau\wedge\bdiff\theta + Cr\bdiff\tau\wedge\viff + \lp r'+FA\rp\bdiff\rho\wedge\bdiff\theta
+ \lp FB + Dr\rp \bdiff\rho\wedge\viff\,, \\
\bT^3 & = & -Cr\bdiff\tau\wedge\bdiff\theta +\lp \dot{r} - \frac{r}{\tau}\rp\bdiff\tau\wedge\viff 
- \lp FB + Dr\rp \bdiff\rho\wedge\bdiff\theta + \lp r'+FA\rp\bdiff\rho\wedge\viff\,,
\ea
\es
and then the anti/self-dual field strengths,
\bs
\label{sphericalF}
\ba
\x\bF^1 & = & \lb \mp \frac{\partial}{\partial\tau}\lp\frac{F}{\tau}\rp{} + C' - \dot{D}\rb \bdiff\tau\wedge\bdiff\rho + \lb A^2 -1 +\lp \frac{r}{\tau} \pm  B\rp^2\rb\bdiff\theta\wedge\viff\,, \\
\x\bF^2 & = & -\lb \pm\frac{\partial}{\partial\tau}\lp\frac{r}{\tau}\rp  + AC + \dot{B} \rb\bdiff\tau\wedge\bdiff\theta +\lb\dot{A} -\lp\pm \frac{r}{\tau}+B\rp C\rb\bdiff\tau\wedge\viff \nn \\ & - & \lb \pm \frac{r'}{\tau} + A\lp\pm \frac{F}{\tau} + D\rp  + B'\rb\bdiff\rho\wedge\bdiff\theta + \lb A' - \lp \frac{r}{\tau} \pm B\rp\lp\frac{F}{\tau} \pm D\rp\rb\bdiff\rho\wedge\viff\,, \\
\x\bF^3 & = & -\lb\dot{A} -\lp \pm \frac{r}{\tau}+B\rp C\rb\bdiff\tau\wedge\bdiff\theta - \lb \pm \frac{\partial}{\partial\tau}\lp\frac{r}{\tau}\rp  + AC + \dot{B} \rb\bdiff\tau\wedge\viff \nn \\
& - &  \lb A' - \lp \frac{r}{\tau} \pm B\rp\lp\frac{F}{\tau} \pm D\rp\rb\bdiff\rho\wedge\bdiff\theta
-  \lb \pm\frac{r'}{\tau} + A\lp\pm\frac{F}{\tau} + D\rp  + B'\rb\bdiff\rho\wedge\viff\,. 
\ea
\es
The third set of field equation  (\ref{Tplus},\ref{Tminus}) for the spherically symmetric torsion (\ref{sphericalT}) reduces to 
%a double set of two independent equations, both sets with two 3-form components in the RHS,
\bs
\label{Tspher}
\ba
\frac{g_\pm}{\sqrt{\kappa}r}\lb \lp \dot{r} - \frac{r}{\tau} \mp B\rp\star\bbe^1 - \lp \frac{r'}{F} + A\rp \star\bbe^4\rb  & = & \frac{1}{2}\tau \bM^1 \mp \x\bO^1\,, \\
\frac{g_\pm}{\sqrt{\kappa}Fr}\lb  \lp r\dot{F} + \dot{r}F - 2F\frac{r}{\tau} \mp FB \mp Dr\rp\star\bbe^2 + \lp \pm r' \pm FA + F Cr\rp\star\bbe^3\rb & = & 
\tau \bM^2 \mp 2\x\bO^2\,, \\
\frac{g_\pm}{\sqrt{\kappa}Fr}\lb  \lp \pm r' \pm FA + F Cr\rp\star\bbe^2 - \lp r\dot{F} + \dot{r}F - 2F\frac{r}{\tau} \mp FB \mp Dr\rp\star\bbe^3 \rb & = & 
\tau \bM^3 \pm 2\x\bO^3\,, 
\ea
\es
The spherically symmetric Ansatz suggested for the integration form $\bM^I$ by (\ref{Tspher}) is of the form
\bs
\ba
\bM^1 & = & -\frac{\sqrt{\kappa}\hat{p}_r}{2}\star\bbe^1 - \frac{\hat{v}}{2\kappa^{\frac{3}{2}}}\star \bbe^4\,, \\  
\bM^2 & = & -\frac{\sqrt{\kappa}\hat{p}_\theta}{2}\star\bbe^2 + \hat{\pi}\star \bbe^3\,, \\
\bM^3 & = &\frac{\hat{\pi}}{\kappa^{\frac{3}{2}}}\star \bbe^2 +  \frac{\sqrt{\kappa}\hat{p}_\theta}{2}\star\bbe^3\,, \\
\bM^4 & = &  \frac{\hat{w}}{2\kappa^{\frac{3}{2}}}\star\bbe^1 + \frac{\sqrt{\kappa}\hat{\rho}}{2}\star \bbe^4\,.
\ea
\es
Thus, yet 6 new functions of 2 variables appear in the system. To make progress from here, we need to make some assumptions about the matter sources.

We will first restrict to the case of spinless matter.
In the absence of spin currents, $\x\bO^i=0$, we obtain the set of 2$\times$4 equations
\bs
\label{torsionEQ}
\ba
4g_\pm\lp \dot{r}-\frac{r}{\tau}\mp B\rp & = & -\tau r\kappa\hat{p}_r\,, \\
2g_\pm\lp \frac{r'}{F} + A\rp & = & \tau \kappa^{-1/2}\hat{v}\,, \\
2g_\pm \lp \frac{\dot{F}}{F} + \frac{\dot{r}}{r} - \frac{2}{\tau} \mp \frac{B}{r} \mp \frac{D}{F}\rp & = & -\tau \kappa\hat{p}_\theta\,, \\
g_\pm  \lp \pm \frac{r' }{F}\pm A + Cr\rp & = & r\tau\kappa^{-1/2} \hat{\pi}\,. 
\ea
\es
Solving these equations, we find that two of the imperfect terms in $\bM^i$ vanish, $\hat{v}=\hat{\pi}=0$, and the effective pressure is determined as
\bs
\label{pressures}
\ba
\kappa\hat{p}_r & = & \frac{2\beta}{\tau}\lp \frac{\dot{r}}{r}-\frac{1}{\tau}\rp\,, \\
\kappa\hat{p}_\theta & = & \frac{\beta}{\tau}\lp \frac{\dot{F}}{F} + \frac{\dot{r}}{r}-\frac{2}{\tau}\rp\,. 
\ea
\es
The equations (\ref{torsionEQ}) also give solutions to the 4 of the 6 functions parameterising the connection,  
\be
A = -\frac{r'}{F}\,, \quad B = \gamma^{-1}\lp \dot{r} -\frac{r}{\tau}\rp\,, \quad C = 0\,, \quad D= \gamma^{-1}\lp \dot{F}-\frac{F}{\tau}\rp\,,
\ee
where we recall the parameter $\gamma$ defined at (\ref{gammadef}). 
%where we defined, in addition to (\ref{alphabeta}), the parameter 
The field strengths (\ref{sphericalF}) then become
\bs
\ba
\x\bF^1 & = & -\lb \gamma^{-1}\ddot{F} - \lp \gamma^{-1} \mp 1\rp\lp\frac{\dot{F}}{\tau} - \frac{F}{\tau^2}\rp \rb\bdiff\tau\wedge\bdiff\rho
+ \lb \lp\frac{r'}{F}\rp^2  + \lp \lp \gamma^{-1} \mp 1\rp\frac{r}{\tau^2} - \gamma^{-1}\frac{\dot{r}}{\tau}\rp^2 - 1 \rb\bdiff\theta\wedge\viff\,,  \\
\x\bF^2 & = & -\lb \gamma^{-1} \ddot{r} - \lp \gamma^{-1} \mp 1\rp\lp\frac{\dot{r}}{\tau} - \frac{r}{\tau^2}\rp\rb \bdiff\tau\wedge\bdiff\theta
+ \lp \frac{\dot{F} r'}{F^2}  - \frac{\dot{r}'}{F}\rp\bdiff\tau\wedge\viff \nn \\
& + & \gamma^{-1}\lp \frac{r' \dot{F}}{F} - \dot{r}'\rp  \bdiff\rho\wedge\bdiff\theta 
+ \lb \frac{F'r'}{F^2} - \frac{r''}{F} - \lp \lp \gamma^{-1} \mp 1\rp\frac{F}{\tau} - \gamma^{-1}\dot{F}\rp  \lp \lp \gamma^{-1} \mp 1\rp\frac{r}{\tau} - \gamma^{-1} r\rp\rb \bdiff\rho\wedge\viff\,, \quad \quad \\
\x\bF^3 & = & -\lp \frac{\dot{F} r'}{F^2}  - \frac{\dot{r}'}{F}\rp\bdiff\tau\wedge\bdiff\theta   -\lb \gamma^{-1} \ddot{r} - \lp \gamma^{-1} \mp 1\rp\lp\frac{\dot{r}}{\tau} - \frac{r}{\tau^2}\rp\rb \bdiff\tau\wedge\viff \nn \\
& - &  \lb \frac{F'r'}{F^2} - \frac{r''}{F} - \lp \lp \gamma^{-1} \mp 1\rp\frac{F}{\tau} - \gamma^{-1}\dot{F}\rp  \lp \lp \gamma^{-1} \mp 1\rp\frac{r}{\tau} - \gamma^{-1} r\rp\rb \bdiff\rho\wedge\bdiff\theta  + \gamma^{-1}\lp \frac{r' \dot{F}}{F} - \dot{r}'\rp  \bdiff\rho\wedge\viff\,.
\ea
\es
Thus, in the end we have arrived at the system where there are only four unknown functions, $F(\tau,\rho)$ which determines the metric of (eventually, pseudo)Riemannian geometry; $r(\tau,\rho)$ which plays the role of a radial coordinate as usual;  $\hat{\rho}(\tau,\rho)$ which represents the weight of space that can be interpreted as the energy density of effective dark matter; and $\hat{w}(\tau,\rho)$ which describes a non-standard flow in the effective dark matter field (which cannot be mimicked by any standard fluid since it would require an asymmetric energy-momentum tensor). Again, to make progress from here we have to make further assumptions about the matter sources, in particular we have to specify the material energy current $\bt^I$.

\subsection{Static solutions require $\beta=0$}
\label{nostatic}

We shall restrict to vacuum, $\bt^I=0$. Also, for simplicity we set $\lambda=0$ so that there is no effective cosmological constant term. 
The energy constraint (\ref{Sync1}) now gives 
\bs
\label{sphericalEoM}
\ba
%\frac{1}{2}\lp g_+ - g_-\rp\ddot{F}F + \lp g_+ + g_-\rp\lp \frac{F'r'r}{F^2}- \frac{r''r}{F}\rp - \alpha \dot{F}\dot{r}r + \beta F\frac{r^2}{\tau^2} & = & \frac{1}{2} F r^2\hat{\rho}\,, \label{emom1} \\
\lp g_+ + g_-\rp\lb  \lp 1-A^2\rp F + 2A'\rb -\alpha\lp F\dot{r}^2 + 2\dot{F}\dot{r}r\rp + 3\beta\frac{F r^2}{\tau^2} & = & -Fr^2\kappa\hat{\rho}\,,   \label{emom1} \\
\lp g_+ + g_-\rp \frac{\partial}{\partial\tau}\lp\frac{r'}{F}\rp & = & \frac{r}{2\kappa}\hat{w}\,. \label{emom2}
\ea
The radial momentum constraint from (\ref{Sync2}) gives
\ba
\lp g_+ + g_-\rp\lp A^2 -1\rp + \alpha\lp \dot{r}^2 + 2 r\ddot{r}\rp & = & \beta \frac{r^2}{\tau^2}\,, \label{radmom1} \\
\alpha\lp \dot{F}r' -F\dot{r}'\rp & = & 0\,. \label{radmom2}
\ea
To facilitate the analysis, we shall first look for static solutions. Then the geometrical functions should only depend of the radial function $r(\tau,\rho)$, i.e. we can consider that $F=F(r(\tau,\rho))$ and $\hat{\rho}=\hat{\rho}(r(\tau,\rho))$.
Since we exclude the special case $g_+ = g_-$ from the present analysis, (\ref{radmom2}) implies that $A$ is a constant, $\dot{A}=A'=0$ and we have that $r' = -AF$. Then we see from (\ref{emom2}) that the remaining imperfect term $\hat{w}=0$ vanishes along with the others. The final field equations, the angular parts of momentum constraints from (\ref{Sync2}) gives, after the simplifications allowed by the constancy of $A$, only one nontrivial equation,
\be
\alpha\lp \ddot{F}r + \dot{F}\dot{r} + F\ddot{r}\rp = \beta \frac{Fr}{\tau^2}\,. \label{field2}
\ee
\es
We should then check the consistency of the field equations (\ref{sphericalEoM}) by confirming their consistency with the conservation laws. 
The radial component of the equation $\bDiff \bM^i = 0$ reads
\bs
\be
\hat{p}_r'  =  \frac{2}{r}\lp \hat{p}_\theta - \hat{p}_r\rp + \frac{F \hat{w}}{\kappa^{\frac{3}{2}}\tau^2}\,. \label{field3}
\ee
With the imperfect term $\hat{w}=0$ and the pressures given by (\ref{pressures}), this equation is identically satisfied when $A$ is a constant. Also, 
the radial components of the conservation equation $\bDiff\bM^2 = \bDiff\bM^3=0$ are identically satisfied. It remains to consider the fourth component,
$\bDiff\bM^4=0$, which gives, for $\hat{w}=0$, 
\be \label{continuity}
\frac{\partial}{\partial\tau}\lp Fr^2\hat{\rho}\rp = -\frac{Fr^2}{\tau}\lp\hat{p}_r + 2\hat{p}_{\theta}\rp = \frac{ 2\beta r}{\kappa\tau^2}\lp \frac{3 Fr}{\tau}- 2F\dot{r}-\dot{F}r\rp\,,
%2\beta\frac{ r^2}{\tau^2}\lp \frac{F}{\tau} - \dot{F}\rp \,, 
\ee
\es
where we used the solution (\ref{pressures}) in the second equality.

At this stage, we identify the $\tau$ with the synchronous
 time coordinate on a pseudo-Riemannian manifold by a Wick rotation. The transformation of the Euclidean line element into a pseudo-Riemannian line element in the familiar radial Schwarzschild coordinates,
\be \label{metric}
\bbe^I\otimes\bbe_I  =  \bdiff\tau\otimes\bdiff\tau + \bbe^i\otimes\bbe_i \rightarrow 
- f^2(r) \bdiff t\otimes\bdiff t + g^2(r)\bdiff r\otimes \bdiff r + r^2 \bdiff \theta\otimes \bdiff \theta + r^2\viff \otimes \viff =  \bbe^A\otimes\bbe_A\,,  
\ee
is achieved by
\bs
\label{coordinates}
\ba
-i\bdiff\tau & \rightarrow &  A\bdiff t \pm \frac{Fg}{f}\bdiff r\,, \\
\bdiff\rho & = & \bdiff t \pm \frac{A g}{F f}\bdiff r\,.
\ea
\es
The function $f$ is given by $f^2 = A^2 - F^2$, and the function $g$ can be chosen freely in general. However, our field equations have now constrained that $r' = -AF$, which implies that $g = \mp 1/f$ in (\ref{metric},\ref{coordinates}). Thus, we already see that the Eddington parameter will retain its general-relativistic value regardless of the theory parameters $g_\pm$. Using now the coordinate transformation (\ref{coordinates}), we can re-express the derivatives wrt the khronon field $\tau$ as derivatives wrt to the radial coordinate $r$, 
\bs
\label{LBH}
\ba
\dot{r} & \rightarrow & -iF\,, \quad \ddot{r} \rightarrow -F_{,r}F\,, \\
\dot{F} & \rightarrow & -i F_{,r}F\,, \quad \ddot{F} \rightarrow -(F_{,r})^2F-F_{,rr}F^2\,.
\ea
\es
The nontrivial field equations (\ref{emom1}), (\ref{radmom1}) and (\ref{field2}) then become, respectively,
\bs
\ba
\lp g_+ + g_-\rp\lp 1 -A^2 \rp  + \alpha \lp F + 2 F_{,r} r\rp F + 3\beta\frac{r^2}{\tau^2} & = &  -r^2\kappa \hat{\rho}\,, \\
\lp g_+ + g_-\rp \lp 1 -A^2 \rp + \alpha\lp F^2 + 2F_{,r} F r\rp & = & -\beta\frac{r^2}{\tau^2}\,, \\
\alpha\lb F_{,rr} F r + (F_{,r})^2 r + 2F_{,r}F\rb  & = & - \beta\frac{r}{\tau^2}\,. 
\ea
\es
The difference of the two last equations yields a simple second-order differential equation with the solution
\be \label{BHnonsolution}
F^2 =  \gamma^2\lp A^2 -1\rp + \frac{r_S}{r} - \hat{\Lambda} r^2\,,
\ee 
where $r_S$ and $\hat{\Lambda}$ are the integration constants. On the other hand, combining the two first equations gives $\hat{\rho} = -2\beta\tau^{-2}$. 
We promptly check that this effective energy density is compatible with the continuity equation (\ref{continuity}).
However, we have only checked two linear combinations of the 
equations (\ref{LBH}). To solve the full system of equations, we find that we must set $\hat{\Lambda} = \beta \tau^{-2}/3$. But this is not a constant unless $\beta=0$: consistent static spherically symmetric solutions\footnote{We note that observational implications of a black hole metric given by (\ref{BHnonsolution}) have been considered recently \cite{Ovgun:2025stp,Umarov:2025ihy}.} exist only in the case $\beta=0$, and then we recover the Schwarzschild solution, the special limit $a \rightarrow 0$ of the Kerr solution studied in section \ref{kerr}. It is not surprising that static spherically symmetric solutions are excluded by nonzero $\beta$, since in section \ref{cosmology} we established that only the special case of maximal chiral asymmetry $g_\pm =0$ allows the exact Minkowski solution. On the other hand, we already derived exact spherically symmetric (even homogeneous) but nonstatic vacuum solutions describing the expanding universe in the $\beta \neq 0$ class of theories, recall (\ref{Vacuum1}). We should at least recover these solutions also in the present set-up, then giving up the requirement of staticity. 

\subsection{On solutions in the generic theory}
\label{ongeneric}

So, we shall take some steps back and return to the spherically symmetric Ansatz\footnote{This remains the generic spherically symmetric geometry. The familiar form of the metric (\ref{metric}) with functions $f(r) \rightarrow f(r,t)$ and  $g(r) \rightarrow g(r,t)$ can be transformed into the Lema{\^i}tre form using (\ref{coordinates}) wherein we did not make the assumption of static geometry. The resulting metric is known as the Lema{\^i}tre-Tolman-Bondi metric.}
(\ref{sphericalEoM}), keeping the connection coefficients as, instead of functions of only $r(\tau,\rho)$, generic functions of $\rho$ and $\tau$. For example, the cosmological solutions obtained at (\ref{Vacuum1}) should correspond to
\be \label{Vacuum2}
r(\tau,\rho) = \rho F(\tau,\rho)\,, 
\quad
F(\tau,\rho) = \lp\frac{\tau}{\tau_0}\rp^{\frac{1}{3}\lp 1 \pm \sqrt{1+3\frac{\beta}{\alpha}}\rp}\,,
\ee
where $\tau_0$ is some constant. 

The radial momentum constraint component (\ref{radmom2}) again sets $\dot{A}=0$, but now this does not imply that $A'=0$, so we should take $A=A(\rho)$. The energy constraint component (\ref{emom2}) still eliminates the imperfect term $\hat{w}$. The three nontrivial khronon EoM we are then left with are
\bs \label{ltb}
\ba
\lp g_+ + g_-\rp\lb  \lp 1-A^2\rp F + 2A'\rb -\alpha\lp F\dot{r}^2 + 2\dot{F}\dot{r}r\rp + 3\beta\frac{F r^2}{\tau^2} & = & -Fr^2\kappa\hat{\rho}\,, \label{ltb1} \\
\lp g_+ + g_-\rp\lp A^2 -1\rp + \alpha\lp \dot{r}^2 + 2 r\ddot{r}\rp & = & \beta \frac{r^2}{\tau^2}\,, \label{ltb2} \\
\lp g_+ + g_-\rp A' + \alpha\lp \ddot{F}r + \dot{F}\dot{r} + F\ddot{r}\rp & = & \beta \frac{Fr}{\tau^2}\,, \label{ltb3}
\ea
and we should also take into account the constraint
\be
r' = -AF\,. 
\ee
\es
We see that the $r(\tau,\rho)$ given in (\ref{Vacuum2}) solves (\ref{ltb2}) when setting $A=-1$; then the corresponding $F(\tau,\rho)$ given in (\ref{Vacuum2}) solves (\ref{ltb3}); and when plugging  (\ref{Vacuum2}) into (\ref{ltb1}) we obtain expressions for $\kappa\hat{\rho}$ which are equivalent to (\ref{Vacuum1}). Thus, the homogeneous vacuum solutions are reproduced consistently for $A=-1$, which are probably the simplest nontrivial solutions for the vacuum equations (\ref{ltb}). In fact, when $A=\mp 1$, a generic two-parameter solution for the scale factor can be obtained as 
\bs
\be
r(\tau,\rho) = g(\rho) F(\tau,\rho)\,, 
\quad
F(\tau,\rho) = \pm g'(\rho)\lp\frac{\tau}{\tau_0}\rp^{\frac{1}{3}\lp 1 - \sqrt{1+3\frac{\beta}{\alpha}}\rp}\lp \lp\frac{\tau}{\tau_0}\rp^{\sqrt{1+3\frac{\beta}{\alpha}}} + F_0\rp^\frac{2}{3}\,, \label{grhosol}
\ee
where $F_0$ is a constant, and $g(\rho)$ is an arbitrary function. This corresponds to a more complicated than  simple power-law expansion. Consequently, the expression for the effective dark matter density is also more complicated,
\be
\kappa\hat{\rho} = 
\frac{2 \alpha  \left(1+\sqrt{\frac{3 \beta }{\alpha }}+1\right) (\frac{\tau}{\tau_0})^{2 \sqrt{1+\frac{3 \beta
   }{\alpha }}}-6 \beta  \left[(\frac{\tau}{\tau_0})^{2 \sqrt{1+\frac{3 \beta }{\alpha }}}+4 F_0
   (\frac{\tau}{\tau_0})^{\sqrt{1+\frac{3 \beta }{\alpha }}}+F_0^2\right] + 2 \alpha  F_0^2
   \left(1-\sqrt{1+\frac{3 \beta }{\alpha }}\right)}{3 \tau^2 \left[(\frac{\tau}{\tau_0})^{\sqrt{1+\frac{3 \beta
   }{\alpha }}}+F_0\right]^2}\,.
\ee
\es
We have checked that this solution satisfies the continuity equation (\ref{continuity}). We see that the function $g(\rho)$ does not enter into the expression for the dark matter energy density. In the cases $A = \pm 1$, the function $g(\rho)$ in (\ref{grhosol}) only reflects the gauge freedom in choosing a radial coordinate. For some function $A(\rho) \neq \pm 1$ we would expect to find inhomogeneous geometries, and in particular, time-evolving black hole solutions. However, it is not easy to see the form of these hypothetical solutions from (\ref{ltb}).

%Black holes, the properties of their horizons and their potential singularities in the generic $\beta \neq 0$ theory is one of the important topics left open for future studies. 
The existence and the properties of black holes in the generic $\beta \neq 0$ theory is one of the important topics left open for future studies.

\section{Matter couplings}
\label{mattercouplings}

Thus far, our exploration has focused on the emergence of spacetime and gravitational dynamics from the gauge-theoretic structure based on $Spin(4)$, with the Cartan khronon playing a central role in breaking symmetry and establishing temporality. The formalism has proven to consistently reproduce both static and cosmological solutions, as well as nontrivial configurations such as the Kerr geometry. 

We now turn to an equally essential aspect of any realistic theory, its coupling to matter. The introduction of matter fields into the $Spin(4)$ gauge theory brings into focus some of the most delicate aspects of the framework - chiefly, how to formulate spinor dynamics in a setting where spacetime and even its signature are emergent rather than fundamental. While spinor fields are traditionally defined on Lorentzian spacetimes, here they must be incorporated within a manifestly real Euclidean formalism, with the Lorentzian physics understood to arise via a nontrivial analytic continuation. 

In this section, we construct the spinor representation of the Euclidean Clifford algebra, defining the associated gamma matrices in a way compatible with the chiral structure of the theory. We then proceed to define spinor fields, their transformation properties under the $Spin(4)$ gauge symmetry, and the form of their covariant derivatives. Importantly, we show how spinors source not only the composite frame field through energy-momentum, but also the gauge field through intrinsic spin currents. These spin currents may be both right- and left-handed, enriching the geometric content of the theory.
The main result of this section is the derivation of these matter currents, which provides the crucial link between matter and geometry in the full theory. 

\subsection{Gamma matrices}
\label{gammas}

The 2$\times$2 Pauli matrices,
\bs
\be \label{pauli}
  \sigma^0 = \lp
  \begin{matrix}
    1 & 0  \\
    0 & 1 
  \end{matrix}\rp\,, \,\,\,\,
  \sigma^1 = \lp
  \begin{matrix}
    0 & 1  \\
    1 & 0 
  \end{matrix}\rp\,, \,\,\,\,
   \sigma^2 = \lp
  \begin{matrix}
    0 & -i  \\
    i & 0 
  \end{matrix}\rp\,,
   \,\,\,\,
    \sigma^3 = \lp
  \begin{matrix}
    1 & 0  \\
    0 & -1
  \end{matrix}\rp\,,
\ee 
obey the algebra $\sigma^{[i}\sigma^{j]} = i\epsilon^{ij}{}_k\sigma^k$ and can be used to construct the 4$\times$4 Dirac matrices,
\be \label{dirac}
\gamma^4 = \sigma^1\otimes\sigma^0\,, \quad
\gamma^i = \sigma^2\otimes\sigma^i\,, \quad
\gamma^5 = \sigma^3\otimes\sigma^0 = \gamma^1\gamma^2\gamma^3\gamma^4\,.
\ee  
\es
The Euclidean metrics are obtained as  $\delta^{ij}\sigma^0 = \sigma^{(i}\sigma^{j)}$ and $\delta^{IJ}\mathbb{1} = \gamma^{(I}\gamma^{J)}$, where we denote the 4$\times$4 unit matrix as $\mathbb{1} \equiv\sigma^0\otimes\sigma^0$. The generators
%\bs
\be
o_{IJ} \equiv \frac{1}{2}\gamma_{[I}\gamma_{J]} = \frac{1}{4}[\gamma_I,\gamma_J] = \frac{1}{4}\lp \gamma_I\gamma_J -
\gamma_J\gamma_I\rp\,,
\ee
then realise the $\mathfrak{so}(4)$ algebra (\ref{algebra1}). 
%\be
%[\gamma_{IJ}, \gamma_{KL}] = 4\delta_{[L[I}\gamma_{J]K]} = 2\lp \delta_{L[I}\gamma_{J]K} - \delta_{K[I}\gamma_{J]L}\rp
%=  \delta_{LI}\gamma_{JK} -  \delta_{LJ}\gamma_{IK} - \delta_{KI}\gamma_{JL} + \delta_{KJ}\gamma_{IL}\,. 
%\ee
%\es
It is useful to note that
\be \label{3gamma}
\gamma^I\gamma^J\gamma^K = 2\delta^{I[J}\gamma^{K]} + \delta^{JK}\gamma^I - \epsilon^{IJKL}\gamma_L\gamma^5\,, 
\ee
which can be checked e.g. by using (\ref{dirac}) and $\sigma^i\sigma^j\sigma^k = 2\delta^{k[j}\sigma^{i]}+\delta^{ij}\sigma^k + i\epsilon^{ijk}\sigma^0$. 

\subsection{Spinors}

Let $\psi$ be a 4-component column spinor. The conjugate is defined as $\bar{\psi} \equiv \psi^\dagger\gamma$, where $\gamma$ is a Hermitian
matrix $\gamma^\dagger = \gamma$ s.t. $(\gamma^I)^\dagger = \gamma\gamma^I\gamma$. In the Lorentzian case, we'd use the $\gamma^0 \equiv -i\gamma^4$, and the appropriate choice for the $\gamma$ would be $\gamma^0$, but since working in the Euclidean signature it is possible to choose simply  
$\gamma = \mathbb{1}$. An $SO(4)$ transformation given by an orthonormal matrix with components $\Lambda^I{}_J$ is represented for spinors with the $4\times 4$ matrix $\Lambda$ such that
\be \label{spinorstrans}
\psi \rightarrow \Lambda\psi\,, \quad \bar{\psi} \rightarrow \bar{\psi}\Lambda^{-1}\,, \quad \Lambda^{-1} \gamma^I \Lambda = \Lambda^I{}_J\gamma^J\,. 
\ee
A consistency check is that $\gamma^{-1}\Lambda^\dagger\gamma = \Lambda^{-1}$, where $\Lambda = \exp{\lambda^{IJ}o_{IJ}/2}$ is the transformation parameterised by $\lambda^{IJ}$. For an infinitesimal transformation, $\Lambda \approx 1 + \lambda^{IJ}o_{IJ}/2$, one readily checks with a brief Clifford algebra that the matrix $\Lambda^I{}_J$ in (\ref{spinorstrans}) is given by  $\Lambda^I{}_J \approx \delta^I{}_J + \lambda^I{}_J$.  Thus, the transformation (\ref{spinorstrans}) requires\footnote{A consistent choice would also seem to be $\gamma=-\gamma^5$, in which case 
$(\gamma^I)^\dagger = -\gamma\gamma^I\gamma$.} that $\Lambda^{-1} = \gamma\Lambda^\dagger\gamma$, which follows from that the generators of $\Lambda$ satisfy $\gamma o_{IJ}^\dagger\gamma = -o_{IJ}$ as the direct consequence of $(\gamma^I)^\dagger = \gamma\gamma^I\gamma$. These properties hold in an arbitrary basis $\gamma^I \rightarrow U^\dagger\gamma^I U$, where $U$ is a unitary matrix. Consider the two Weyl projections
\be
\psi_\pm \equiv \frac{1}{2}\lp  \mathbb{1} \mp \gamma^5\rp\psi \quad \Rightarrow \quad
\psi = \psi_+ + \psi_-\,, \quad (\psi_\pm)_\pm = \psi_\pm\,, \quad (\psi_\pm)_\mp = 0\,.
\ee   
It is worth stressing that these projectors are nothing but (\ref{projectors}) adapted for the spinor representation. This can be verified explicitly by
noting that
\be
\gamma o^{IJ} = -\frac{1}{2}\epsilon^{IJKL}o_{KL}\,,
\ee
which follows by multiplying (\ref{3gamma}) with $\gamma_K$ and using that $\gamma_I\gamma^I = 4\mathbb{1}$ and that $\gamma_I\gamma^J\gamma^K\gamma^I = 4\delta^{JK}\mathbb{1}$. Thus, the projected generators can be expressed in two equivalent ways, 
\be
\x o^{IJ} \equiv P_\pm^{IJKL} o_{KL} = \frac{1}{2} o^{IJ} \pm \frac{1}{4}\epsilon^{IJKL}o_{KL}  
= \frac{1}{2}\lp \mathbb{1} \mp \gamma^5\rp o^{IJ} \equiv o^{IJ}_\pm\,. 
\ee
We recall from (\ref{rlgen}) that $\+o^{4i} = r^i$ are the 3 right-handed and $\m o^{41} = -l^i$ are the 3 left-handed rotations. For the projections of the conjugates, the convention
\be
\bar{\psi}_\pm \equiv \frac{1}{2}\bar{\psi}\lp \mathbb{1} \mp \gamma^5\rp\,, 
\ee
is logical in the way that $\overline{\psi_\pm}=\bar{\psi}_\pm$. However, then the property $\psi_\pm\psi_\pm=0$ familiar from the Lorentzian spinor geometry is lost. We readily check that now we have, instead, 
\bs
\label{scalarsandvectors}
\ba
\bar{\psi}\psi  & = &   \bar{\psi}_+\psi_+ +  \bar{\psi}_-\psi_- \in \mathbb{R}\,, \quad
\bar{\psi}_\pm{\psi}_\mp = 0\,, \\
\bar{\psi}\gamma^I{\psi} & = & \bar{\psi}_+\gamma^I{\psi}_-
+  \bar{\psi}_-\gamma^I{\psi}_+ \in \mathbb{R}\,, \quad
\bar{\psi}_\pm\gamma^I{\psi}_\mp  =  0\,. 
\ea
\es
There are no non-trivial real null vectors in Riemannian manifolds, but the two complex vectors $\bar{\psi}_+\gamma^I{\psi}_-$ and $\bar{\psi}_-\gamma^I{\psi}_+$ are both null. They are the complex conjugates of each other, so only their sum is real.   
Thus, we can form real scalars and vectors from spinors as usual, but the flipping of the respective signs in (\ref{scalarsandvectors}) reflects the crucial difference to the Lorentzian case:  now the conjugates of the right-handed spinors are still right-handed spinors. We remind that in the conventional $Spin(1,3)\times\mathbb{C}=SL(2,\mathbb{C})\times SL(2,\mathbb{C})$ formulation there are two Lorentz groups, one of which acts on the right-handed and one which acts on the left-handed spinors, and the conjugates belong to the inequivalent representations. One identifies the physical Lorentz group by a holomorphic elimination of the spurious half of the complex transformations, i.e. via imposing that the action of the $SL(2,\mathbb{C})$ from the left is the Hermitian conjugate of the action of the $SL(2,\mathbb{C})$ from the right.
In contrast, in present $Spin(4)$ formulation, no complexification is required. The left- and the right-handed transformations are naturally identified with the irreducible subgroups of the one and the same real group, the conjugates of the spinors being accommodated into the conjugate representations of the respective subgroup. 

Due to this crucial difference, the Euclidean Dirac action cannot be the straightforward translation of the well-known Lorentzian Dirac action. The kinetic term for a Dirac field in Minkowski background involves the Hermitian term $i\bar{\psi}\slashed{\partial}\psi = i\bar{\psi}_+\slashed{\partial}\psi_+ + i\bar{\psi}_-\slashed{\partial}\psi_-$. The operator $\slashed{\partial}$ is translated nicely into Euclidean regime, since the $\gamma^0\partial_t = \gamma^4\partial_\tau$.
However, if we naively replace here the spinors we just constructed in the Euclidean geometry, we do not obtain quadratic kinetic terms for the two Weyl components, but according to (\ref{scalarsandvectors}), we get $\bar{\psi}\slashed{\partial}\psi = \bar{\psi}_+\slashed{\partial}\psi_- + \bar{\psi}_-\slashed{\partial}\psi_+$. One way to deal with this is by   
``doubling'' the degrees of freedom by taking $\bar{\psi}$ as another, independent field, not related to $\psi$ by conjugation \cite{Osterwalder:1973dx}. Then, however, the Hermitian property is lost.

Schwinger had written down a consistent Euclidean action for Dirac spinors already in 1959 \cite{Schwinger:1959zz}. It is explained in Ref.\cite{vanNieuwenhuizen:1996tv} that the action can be understood as a Wick rotation of the Lorentzian action which not only switches to imaginary time but also rotates the spinor indices. Geometrically, the rotation matrix $S = \exp{(\gamma^4\gamma^5\theta/2)}$ can be understood as a boost along a fifth dimension. It acts on spinors like $\psi \rightarrow S\psi$ and on conjugates like $\psi^\dagger \rightarrow \psi^\dagger S$. (This action is analogous to the conventional Wick rotation of spacetime vectors, whose contra- and covariant zeroth components are rotated identically.) Let us write the Dirac Lagrangian including a mass term:
\be \label{kinetic}
i\psi^\dagger S\gamma^0\lp  \slashed{\partial} + m\rp S\psi 
= \psi^\dagger S\gamma^4\lp  \slashed{\partial} + m\rp S\psi = 
\psi^\dagger \gamma^4 S^{-1}\lp  \slashed{\partial} + m\rp S\psi = \psi^\dagger\gamma^4\lp \delta^\mu_I\gamma^I_E\partial_\mu + m\rp\psi\,.   
\ee 
The first form is manifestly the usual Dirac Lagrangian when $\theta=0$ that is, $S=1$. In the second form we just recalled that $\gamma^4=i\gamma^0$. In the third form we used the property of the $S$-matrix that $S\gamma^4 = \gamma^4 S^{-1}$ \cite{vanNieuwenhuizen:1996tv}. In the final form we wrote the $\slashed{\partial}_E$ explicitly in terms of the $S$-rotated gamma-matrices, $\gamma_E^I \equiv S\gamma^IS^{-1}$. For $\theta=\pi/2$, we obtain 
\be
\gamma^I_E = \gamma^I\,, \quad \gamma_E^4=\gamma^5\,, \quad 
\gamma_E^5 = \gamma^1_E\gamma^2_E\gamma^3_E\gamma^4_E = -\gamma^4\,.
\ee
The kinetic term (\ref{kinetic}) suggests identifying the conjugate spinor as $\bar{\psi}_E=\psi_E^\dagger\gamma^5_E = -\psi_E^\dagger\gamma^4$, i.e. $\gamma_E=\gamma_E^5$. Note that the rotation of the spinor indices rotates also the Lorentz generators. It still holds that $\gamma_E o^{IJ}_E \gamma_E = -(o^{IJ}_E)^\dagger$, and we note that now
\be
(\gamma_E^I)^\dagger = -\gamma_E\gamma_E^I\gamma_E\,, \quad (\gamma_E^5)^\dagger = \gamma_E\gamma_E^5\gamma_E\,.
\ee
The projected spinors in the rotated basis are 
\be
\psi_{E\pm} = \frac{1}{2}\lp 1 \mp \gamma^5_E\rp\psi_E =  \frac{1}{2}\lp 1 \pm \gamma^4\rp\psi_E\,.
\ee
To recapitulate, the equivalent action to Schwinger's,
\be \label{schwing}
L_E = \bar{\psi}_E\lp \slashed{\partial}_E + m\rp\psi_E =
\bar{\psi}_{E+}\slashed{\partial}_E\psi_{E-} + 
\bar{\psi}_{E-}\slashed{\partial}_E\psi_{E+}
+ m\lp \bar{\psi}_{E+}{\psi}_{E+} + 
\bar{\psi}_{E-}{\psi}_{E-}\rp\,,
\ee
is manifestly Hermitian and admits the interpretation as a Wick rotation of the Dirac action taking into account the rotation of the spinor indices. The latter part of the transformation reconciles the
apparent difference of the conjugacy property of Euclidean versus Lorentzian spinors  $\overline{\psi_{L\pm}}=\bar{\psi}_{L\mp}$. 

It is also possible to construct kinetic terms for Euclidean Majorana spinors without a ``doubling'' of the degrees of freedom. Instead of Hermitian conjugation, one should then turn to possible alternative complex structures, since reality in the Lorentzian regime may correspond to Osterwalder-Schrader positivity in the Euclidean regime \cite{Osterwalder:1973dx}; see Wetterich \cite{Wetterich:2010ni} for a systematic discussion. 

A simple interpretation of Weyl spinors was proposed recently \cite{Woit:2023idu}. In fact, this interpretation is suggested by the matter coupling we had already adopted for the right-handed ($g_-=0$) gravity model \cite{Gallagher:2023ghl} along the lines of the matter coupling proposed in self-dual loop quantum gravity \cite{Ashtekar:1989ju}. In the Lorentzian regime, one can consider the kinetic term
\be \label{kinetic1}
L_+ = \frac{i}{2}\lp \bar{\psi}\slashed{\partial}\psi_+ - \bar{\psi}\cev{\slashed{\partial}}\psi_-\rp\,, 
\ee
wherein the actions of the derivative operator on the components $\psi_-$ and $\bar{\psi}_-$ are projected out, and thus we will obtain only the couplings of the self-dual connection with matter. We perform the same spinorial Wick rotation to (\ref{kinetic1}) as earlier in the case of (\ref{kinetic}),
\be \label{kinetic2}
L_+ = \frac{i}{2}\lp \psi^\dagger S\gamma^0\slashed{\partial}S\psi_+ - \slashed{\partial}(\psi^\dagger S\gamma^0) S\psi\rp
=  \frac{1}{2}\lp \psi^\dagger S\gamma^4\slashed{\partial}S\psi_+ - (\psi^\dagger S\gamma^4)\cev{\slashed{\partial}} S\psi\rp
 =  \frac{1}{2}\lp \bar{\psi}_E\slashed{\partial}_E\psi_{E+} -
\bar{\psi}_E\cev{\slashed{\partial}}_E \psi_{E-}\rp\,.
\ee
This coupling is Hermitian, $L^\dagger_+ = L_+$, and allows the consistent minimal coupling to self-dual gravity the results of which were reported in Ref.\cite{Gallagher:2023ghl}. An improvement due to the Euclidean formulation would seem to be that the self-dual coupling prescription within the context of the complexified Lorentz group \cite{Ashtekar:1989ju,Gallagher:2023ghl} is not manifestly Hermitean. 

When restricting this prescription to pure Weyl spinors, say to the right-handed ones, an apparent problem is that the component $\psi_-$ appears in both pieces of the action \eqref{kinetic2}. The simple resolution is to regard $\psi_-$ as transforming in the conjugate representation of $\psi_+$, so that the would-be left-handed component is instead identified with the conjugate of the right-handed one \cite{Woit:2023idu}. In this way, one obtains a consistent kinetic term involving only the right-handed Weyl spinor. Thus, Weyl spinors are naturally incorporated into the Euclidean Lorentz gauge theory. In particular, one avoids both the complexification of the symmetry typically required in the standard Lorentzian approach and the doubling of the field content required in the standard Euclidean counterpart.

In this sense, the $Spin(4)$ framework does not merely accommodate spinors consistently, but appears to provide their most natural setting.

\subsection{Energy and spin currents}
\label{EandScurrents}

Our aim is to construct the gravitational generalisation of the Schwinger Lagrangian for fermions (\ref{schwing}). We drop the indices of the Euclidean basis in $\gamma^I_E, \psi_E$ and write just $\gamma^I, \psi$, since the following derivations will be basis-independent.    

The covariant derivative of a spinor and a conjugate spinor are
\bs
\label{spinorderi}
\ba
\bDiff\psi & = & \bdiff\psi + \frac{1}{2}\bA^{IJ} o_{IJ}\psi\,, \\
\bDiff\bar{\psi} & = & \bdiff\bar{\psi} - \frac{1}{2}\bA^{IJ}\bar{\psi} o_{IJ}\,. 
\ea
\es
Since the connection coefficients are real, $\bDiff\bar{\psi} = \overline{\bDiff\psi}$. Defining $\bA \equiv \+\bA + \m\bA$ where, consistently with definitions in section \ref{reducibility},
\bs
\ba
\+\bA & \equiv & \+\bA^i r_i\,, \quad \text{where} \quad 
r^i = -\frac{1}{8}\epsilon^{ijk}\gamma_j\gamma_k - \frac{1}{4}\gamma^{[4}\gamma^{i]}\,, \\ 
\m\bA & \equiv & \m\bA^i l_i\,, \quad \text{where} \quad 
l^i = -\frac{1}{8}\epsilon^{ijk}\gamma_j\gamma_k + \frac{1}{4}\gamma^{[4}\gamma^{i]}\,,
\ea
\es
we can expand (\ref{spinorderi}) as
\bs
\ba
\bDiff\psi & = & \bdiff\psi + \bA\psi =  \bDiff\psi_+ + \bDiff\psi_- = \lp\bdiff + \+\bA\rp\psi_+ +  \lp\bdiff + \m\bA\rp\psi_- \,, \\
\bDiff\bar{\psi} & = & \bdiff\bar{\psi} - \bar{\psi}\bA =  \bDiff\bar{\psi}_+ + \bDiff\bar{\psi}_- = \bdiff\bar{\psi}_+ - \bar{\psi}_+\+\bA + \bdiff\bar{\psi}_- - \bar{\psi}_-\m\bA\,. 
\ea
\es
In the construction of a kinetic Lagrangian, we consider the Hermitian 1-form,
\ba
\frac{1}{2}\lb \lp\bar{\psi}\gamma^I\bDiff\psi\rp +
\lp\bar{\psi}\gamma^I\bDiff\psi\rp^\dagger\rb & = &
\frac{1}{2}\lp \bar{\psi}\gamma^I\bDiff\psi - 
\bDiff\bar{\psi}\gamma^I\psi\rp \nn \\
& = & 
\frac{1}{2}\lp \bar{\psi}\gamma^I\bdiff\psi - 
\bdiff\bar{\psi}\gamma^I\psi\rp + \frac{1}{4}\bA_{JK}\left\{ \gamma^I, o^{JK}\right\} \nn \\
& = & \bar{\psi}\gamma^I\bdiff\psi - \frac{1}{2}\bdiff\lp \bar{\psi}\gamma^I\psi\rp - \frac{1}{2}\star\bA^{IJ}\bar{\psi}\gamma_J\gamma^5\psi\,. 
\ea
In the last step, we used partial integration and the formula (\ref{3gamma}). The Lagrangian 4-form density can then be formed in terms of the fundamental, dimensionless matter fields $\Psi \equiv \kappa^{\frac{3}{2}}\psi$ and the dimensionless parameter $M$ related to the mass $m$ as $M=\sqrt{\kappa}m$. The Lagrangian reads
\ba \label{spinorlagr}
\bL_\psi = \frac{1}{2}\star\bDiff\phi_I\wedge\lp \bar{\Psi}\gamma^I\bDiff\Psi - 
\bDiff\bar{\Psi}\gamma^I\Psi\rp + \star \lp M\bar{\Psi}\Psi\rp 
& = & 
\frac{1}{2}\star\bDiff\phi_I\wedge\lp  \bar{\Psi}\gamma^I\bdiff\Psi - 
\bdiff\bar{\Psi}\gamma^I\Psi
- \star\bA^{IJ}\bar{\psi}\gamma_J\gamma^5\Psi\rp + \star \lp m \bar{\Psi}\Psi\rp  \nn \\
\text{in the geometric phase\,:}
%\overset{\text{nondeg.}}{=}
& = & 
 \frac{1}{2} \ast\lb \lp\slashed{\partial}\bar{\psi}\rp\psi
- \bar{\psi}\slashed{\partial}\psi - \frac{1}{2}\epsilon^{IJKL}A_{IJK}\bar{\psi}\gamma_L\gamma^5\psi
+ 2m\bar{\psi}\psi\rb
\,. 
%\overset{\text{up to b.t.}}{=} \star\bbe_I\wedge\bar{\psi}\lp \gamma^I\bdiff\psi - \frac{1}{2}\star \bA^{IJ}\gamma_J\gamma^5\psi\rp \overset{\text{nondeg.}}{=}
%\star\lp \bar{\psi}\slashed{D}\psi\rp \,. 
\ea
In the second line we have assumed the existence of inverse $\bbie_I$, s.t. we can write $\slashed{\partial} = \gamma^I\bbie_I\lrcorner\bdiff$ and $A_{IJK}=\bbie_I\lrcorner\bA_{JK}$. This density generates the source 3-forms
\bs
\ba
\bt^I_\psi & = & -\frac{1}{2}\star\lp\bbe^I\wedge\bbe^J\rp\wedge\lp
\bar{\psi}\gamma_J\bDiff\psi - 
\bDiff\bar{\psi}\gamma_J\psi\rp + \star\bbe^I m \bar{\psi}\psi\,, \\
\bO^{IJ}_\psi & = & -\frac{1}{4}\epsilon^{IJ}{}_{KL}\lp\star\bbe^K\rp\bar{\psi}\gamma^L\gamma^5\psi
%= -\frac{1}{2}\star\lp\star\bbe^I\rp\bar{\psi}\gamma^J\gamma^5\psi 
\,,
\ea
\es
which contribute, respectively, an energy current and a spin current. The latter decomposes into anti/self-dual pieces according to (\ref{projectors}) as
\be
\x\bO^{IJ}_\psi = \mp\bar{\psi}\x{}\lp\star\bbe^{[I}\gamma^{J]}\rp\gamma^5\psi\,. 
\ee
In the case of the right-handed model in section \ref{righthand}, this first appears problematical: since the field equations enforce $\m \bO^{IJ}_\psi=0$, extra constraints seem to be imposed on the fermion fields which obstruct the viable dynamics for these fields. However, when taking into account the presence of dark 'matter' with the spin current $\hat{\bO}^{IJ}$, a consistent dynamical system is recovered. In generic models with $g_- \neq 0$, the antiself-dual spin current has a novel impact on the spacetime structure, whereas the self-dual spin current contributes to the gravitational dynamics in some analogy with the axial spin current familiar from the Einstein-Cartan model. 

To close this final section, let us note that one important topic left beyond the scope of the present article is the incorporation of the remaining gauge interactions. A first-order pregeometric theory for the Yang-Mills fields of the standard model, compatible with the Lorentz gauge theory of gravity, has been considered by Gallagher {\it et al.} \cite{Gallagher:2022kvv,Gallagher:2023ghl,Gallagher:2024haq}, and it would be interesting to verify whether that reformulation can also be embedded consistently into the present Euclidean framework.

At the same time, the $Spin(4)$ theory of gravity may suggest a different, more twistorial route towards unification, as we briefly comment in the outlook \ref{overview} below. The Schwinger Lagrangian \cite{Schwinger:1959zz}, following the formulation of Ref.~\cite{vanNieuwenhuizen:1996tv}, is based on a rotation involving a fifth dimension, which hints at an extension of the geometric framework by an additional imaginary dimension. In a hypothetical twistorial completion, this construction could be further extended by introducing a sixth gamma matrix, while the khronon would be promoted from a 4-vector to a biquaternion. Such a picture would place space and time on a yet more symmetric footing, and may naturally accommodate structures resembling the electromagnetic, weak, and strong interactions\footnote{For related mathematical insight, see \cite{Woit:2021bmb}.}, with the distinction between these internal gauge interactions and gravitational dynamics arising from the breaking of symmetry between the three spatial dimensions and the three extra dimensions that collapse into an external time. %From this perspective, electromagnetism in Lorentzian spacetime may be interpreted as a manifestation of the scale symmetry of the underlying Euclidean space.

\section{Conclusions and perspectives}
\label{perspectives}

%The khronon field provides a canonical notion of Euclidean time through the foliation defined by its level sets. In sectors where $\bdiff\phi \neq 0$, this makes it possible to formulate the Euclidean-to-Lorentzian correspondence intrinsically, coordinate-independently and without reference to a background metric. Our conjecture is that the two corresponding 'frames' for describing physics - the Euclidean and the Lorentzian one - are both available for all physically relevant solutions of the theory. In the present article we have not proven this statement in full generality, but we have shown that the proposed 'recipe' works consistently in a set of nontrivial examples, including homogeneous cosmology, linear perturbations around FLRW, rotating spacetimes, and spherically symmetric configurations. Since these already cover a large part of the domain in which general relativity is typically applied, they provide substantial evidence in favour of the conjecture. We have shown that the Wick rotation arises naturally and consistently as a reinterpretation of units, bridging the Euclidean and Lorentzian descriptions without modifying the underlying field content. The same field configurations admit dual interpretations, which become physically equivalent once the role of the khronon in setting time and energy scales is properly recognised.
%The structure of the theory is heuristically illustrated in figure \ref{graph}, highlighting the two important steps of our recipe, the symmetry breaking and the Wick rotation.

In this work we have constructed a real $Spin(4)$ gauge theory in which a khronon field $\phi^I$ selects a preferred foliation and thereby makes a Euclidean-to-Lorentzian correspondence dynamically meaningful. The conjecture suggested by this framework is that the two corresponding “frames” for describing physics - the Euclidean and the Lorentzian one - are both available for all physically relevant solutions of the theory. While no general proof has been given here, we have demonstrated that the prescription works consistently in a set of nontrivial cases, including homogeneous cosmology, linear perturbations around FLRW backgrounds, rotating spacetimes, and spherically symmetric configurations. These already cover a large part of the domain in which general relativity is typically applied, and therefore provide substantial support for the conjecture.

The present prescription should not be confused with a conventional analytic Wick rotation between an arbitrarily specified Euclidean metric and an arbitrarily specified Lorentzian one. Rather, the continuation is fixed by the proper-time direction selected by the khronon field, and is accompanied by the corresponding continuation of the physical scales used to interpret that direction. In this sense, the map is not a generic passage between different real slices of a complexified metric, but a specific field-dependent correspondence tied to the physical structure of the solution itself. This is precisely why the standard no-go results for generic metric Wick rotation do not directly apply in the present setting.

The structure of the theory is heuristically illustrated in Fig. 1, highlighting the two key ingredients of the construction: the symmetry breaking and the generalised Wick rotation.

\begin{center}
\begin{figure}
\begin{tikzpicture}[
  node distance=1.2cm and 1.2cm,
  box/.style={draw, rounded corners, text width=4.8cm, align=left, minimum height=1.2cm, fill=blue!5},
  arrow/.style={-{Latex}, thick}
]

\node[box, text width=9.5cm] (spin4) {\textbf{Spin(4) gauge theory} \\  %Gauge group: Spin(4) $\cong$ SU(2)$_L$ $\times$ SU(2)$_R$\\\textbf{Field content:}
\begin{itemize}
  \setlength\itemsep{2pt}
  \item[\textbullet] Cartan khronon $\phi^I$ (from the egg spinor)
  \item[\textbullet] Gauge connection $\bA^{IJ}$ (from the symmetry principle)
\end{itemize}
};

\node[box, below left=of spin4, xshift=2.0cm, yshift=-0.6cm] (action) {\textbf{Action principle}\\Topological: $\bF \wedge \bF,\,\, \bF\wedge\star\bF$\\Dynamical: $\bDiff\phi \wedge \bDiff\phi \wedge \x\bF$\\Couplings: $g_\pm$, $\lambda$};

\node[box, below right=of spin4, xshift=-2.0cm, yshift=-0.6cm] (matter) {\textbf{Matter coupling}\\Spinors: $\psi$, coupled via $\bDiff\psi$\\Energy current: $\bt^I$ \\Spin current: $\bO^{IJ}$
%\\Dark matter: $M^I \propto \phi^I$
};

\node[box, below=of spin4, yshift=-2.0cm] (geometry) {\textbf{Emergent geometry}\\
%Bartels frame and torsion structure:\\[3pt]
Bartels frame $\bbe^i = \sqrt{\kappa}\bDiff\phi^i$ \\
Torsion $\bT^i = \frac{\sqrt{\kappa}\phi}{2}(\prescript{-}{}\bF^i - \prescript{+}{}\bF^i)$
};

\node[box, below=of geometry] (euclidean) {\textbf{Euclidean description}\\ ``Time'' variable $ \tau = \sqrt{\kappa}\phi$ \\  Riemannian metric $\bDiff\phi^I\otimes\bDiff\phi_I$ \\ Chiral ``dynamics'', with $\bM^I$};

\node[box, below=of euclidean] (phenom) {\textbf{Spacetime description}\\ Time variable $t=\phi/m_P$\\ Pseudo-Riemannian metric \\ Gravitation with dark matter};

% Arrows with label on Spin(4) -> Emergent
\draw[arrow] (spin4) -- node[left, xshift=-2mm] {\(\phi^I \rightarrow \phi\, \delta^I_4\)} node[right, xshift=0.5mm] { $SO(4) \rightarrow SO(3)$ }  (geometry);
\draw[arrow] (spin4) -- (action);
\draw[arrow] (spin4) -- (matter);

\draw[arrow] (action) -- (euclidean);
\draw[arrow] (matter) -- (euclidean);'
\draw[arrow] (geometry) -- (euclidean);

\draw[arrow,<->] (euclidean) -- node[right,xshift=1mm]{ $\kappa = -m_P^{-2}$} node[left,xshift=-1mm]{$\tau^2 = -t^2$} (phenom);

\end{tikzpicture}
\caption{\label{graph} Summary of the proceedings in this article.}
\end{figure}
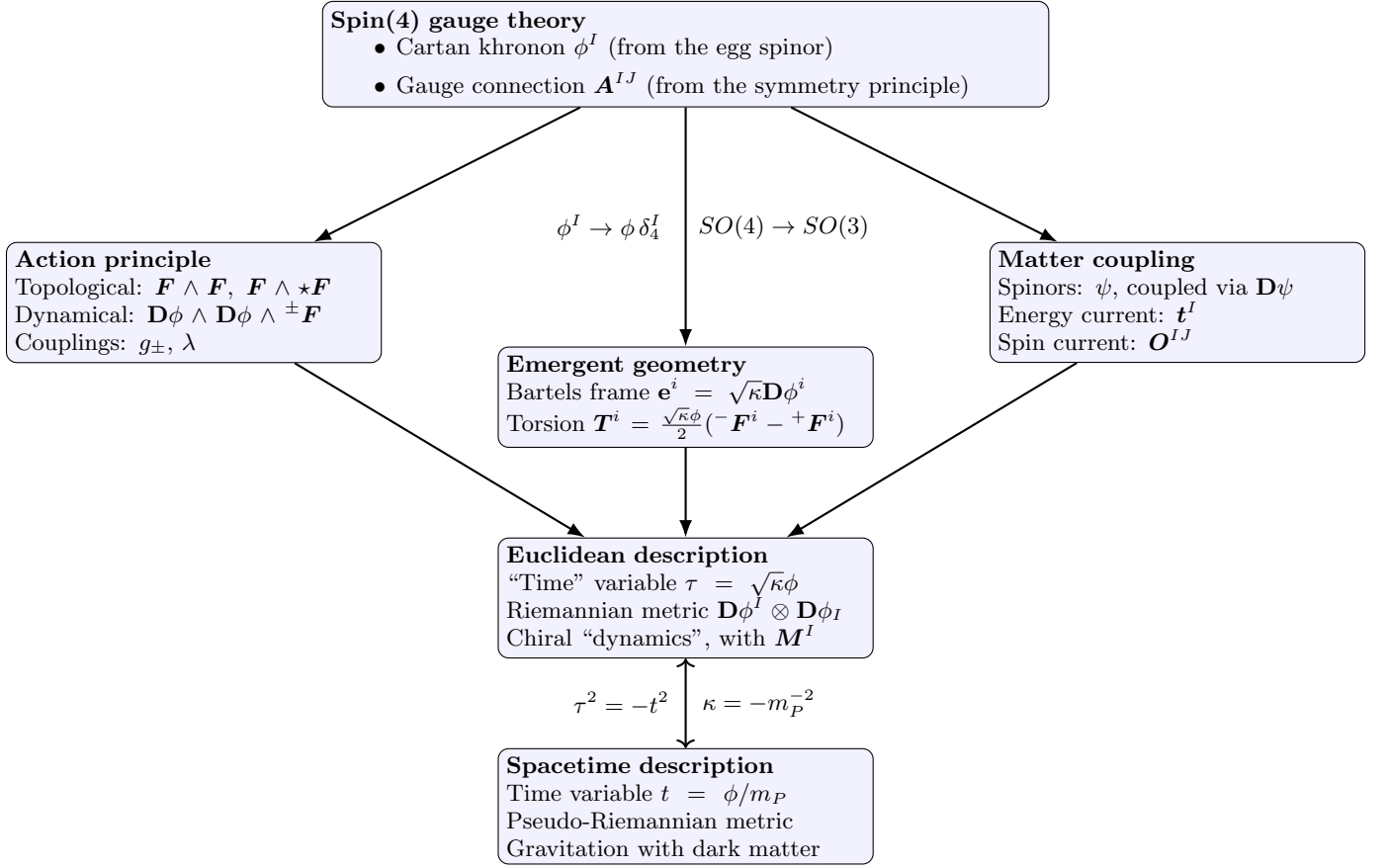
\end{center}

\subsection{Overview and open questions}
\label{overview}

%As emphasised in the introduction \ref{intro}, neither of these steps is new but rather ubiquitous in modern physics. What distinguishes our approach is the claim that the proposed implementation in $Spin(4)$ theory is unique, consistent and universal: if correct, we believe this could have far-reaching consequences. Our initial motivation arose from the path integral formulation of quantum field theory, where a compact and real-valued gauge structure offers a natural and potentially rigorous foundation. It is therefore an obvious and exciting next step to initiate a concrete study of the quantum theory and its Euclidean path integral. 

As emphasised already in the introduction, neither of those steps is new in itself. The distinctive claim of the present work is instead that both can be implemented within a single real-valued
$Spin(4)$ gauge theory, in which the khronon field renders the Euclidean-to-Lorentzian correspondence dynamical rather than externally imposed, and the correspondence is consistently implemented by the rotation of the associated physical scales. If the conjecture stated above is correct, this would furnish an economical and structurally coherent framework relating the two descriptions without enlarging the field content. Our original motivation arose from the important role of Euclidean methods in quantum field theory, particularly in the study of non-perturbative phenomena, though of course alternative approaches have been considered in the Lorentzian signature. The next task is to examine to what extent the present framework can be developed into a concrete quantum and path-integral treatment.

We reviewed the formulation of the theory in section \ref{formulation} in the language of differential forms. The structure, exploiting the reducibility of the Lorentz group, also suggests the possibility of a twistorial reformulation 
and perhaps, in that context, a route towards a unifying extension, alternative in spirit to more conventional $SO(N)$ or $Spin(N)$ (where $N>4$) attempts at unified gauge theories \cite{Krasnov:2017epi}. At present, however, such a direction remains speculative and would require a separate development. As already hinted by the Wick rotation of the spinor basis, one may further ask whether time - like space in the present formulation - could admit a quaternionic interpretation. Along related lines, de Sitter extensions of the Lorentz gauge group have been interpreted as incorporating scale invariance and introducing an additional fundamental constant through symmetry breaking  \cite{Koivisto:2021ofz}, while the conformal extension envisaged in \cite{Koivisto:2018aip,Koivisto:2019ejt} may point towards a more complete framework involving three natural scales.

%that would seem to naturally point to a unifying extension that emerges in a considerably more minimal and simpler - yet subtler - fashion than in the conventional $SO(N)$ or $Spin(N)$ (where $N>4$) attempts at unified gauge theories \cite{Krasnov:2017epi}. As already hinted by the Wick rotation of the spinor basis, time  - like space in the current formulation of the theory - would also be a purely imaginary quaternion. A possible physical interpretation of enlarging the Lorentz gauge group to de Sitter has already been established as incorporating scale invariance and introducing a second fundamental constant emerging from a symmetry breaking \cite{Koivisto:2021ofz}, while the conformal extension envisaged in \cite{Koivisto:2018aip,Koivisto:2019ejt} would add a third constant of Nature, completing the triad of natural scales that govern the Universe. 

Setting this speculation aside, the workings of the gravitational sector of the pregeometric theory were successfully demonstrated in section \ref{righthand} taking previous studies of exact solutions in Lorentz gauge theory substantially further. Future work should clarify how the Euclidean–Lorentzian duality reshuffles dynamical character - why, for instance, gravitational-wave amplitudes that are monotonic in one description emerge as oscillatory in the other, whereas a Euclidean cosmology with a steadily growing scale factor preserves its monotonicity across the dual frame. A promising clue lies in the `electric-magnetic' split of geometry: the `electric' sector (associated with tidal effects) and the `magnetic' sector (linked to frame-dragging) may map differently under this duality, selectively converting growth into oscillation. Another intriguing aspect is the Euclidean Kerr geometry, which appears to break down for $r < a$; this invites further investigation of its topological and geometrical structure.

We also derived the Friedmann equations in the most general, 6-parameter theory (\ref{TheAction}). It will be interesting to study their implications to potentially both early and late cosmology, beyond the very simplest case of vacuum that was solved analytically in section \ref{TheFriedmann}. At the level of background cosmology, it is very feasible to explore extensions of the minimal 6-parameter theory. For instance, one may allow a violation of the torsorial property of the khronon, or promote the $g_\pm$ from constants to dynamical parameters, implemented via scale-dependent running or by simply treating them as dynamical scalar fields. The new $Spin(4)$ framework might admit new modifications of gravity\footnote{In view of modifying gravity, it may seem promising that all of the 6 terms in (\ref{TheAction}) are physically viable, in contrast to the conventional framework of metric-affine gravity which allows an infinite number of terms, yet the most often-studied quadratic sectors face well-known pathologies \cite{Barker:2025xzd}.}, some of which might be well-motivated or phenomenologically relevant. 
Both these qualities may likewise characterise the particularly minimal starting point for modifications offered by the $\Lambda$CDM theory of cosmology, in which unimodularisation eliminates the effect of $\lambda$ and places the cosmological constant on a comparable footing with the cosmological cold dark matter  \cite{Gallagher:2021tgx}.
       
However, already the theory at hand invites application in cosmology, as it inherently modifies the behaviour of gravitational and matter sectors in a way that impacts structure formation. The derivation of cosmological perturbation theory in section \ref{largescale} is one of the main results of this article.
The result (\ref{theresult}) shows that our framework smoothly reproduces standard cold dark matter growth when $\beta=0$, while for $\beta \neq 0$ the growth is subject to a modified friction and a time-dependent effective Newton’s constant, yet retains a single, stable scalar degree of freedom with vanishing sound speed - made possible by the (effective) fluid’s underlying spin currents. The next step is to confront these distinctive signatures with cosmological observations: forecasts for weak lensing, cosmic microwave background temperature and polarisation spectra, matter power spectra from large‑scale structure surveys, and cross‑correlations can reveal whether the model eases the current $H_0$ and $\sigma_8$ tensions in the $\Lambda$CDM model \cite{CosmoVerse:2025txj}. With the theoretical machinery now in place, a systematic data-driven campaign can test the viability of this generalised dark matter sector.

An important question that was yet left open in the preliminary exploration of spherically symmetric exact solutions in section \ref{spherical} is the existence of black holes in the case of $\beta \neq 0$, which does not admit even an exact Minkowski solution as the pregeometric structure forces a dynamical behaviour to the emergent metric unless gravity is strictly one-handed. Of course, if the absence of solutions sufficiently resembling black holes could be proven, it would   
establish that $\beta=0$. On the other hand, were such solutions found, they would necessarily be different from those in general relativity, and thus quite interesting from both observational and theoretical perspectives. 

Finally, in section \ref{mattercouplings} we reviewed the consistent coupling of spinor fields to the Euclidean theory, drawing on (more and less) established constructions in the literature. While this ensures a geometrically coherent treatment of fermionic matter, the analysis also revealed new features: in models with $\beta \neq 0$, the anti self-dual spin current introduces novel effects on spacetime structure, while the self-dual component contributes to the gravitational dynamics in a way reminiscent of the axial spin current in Einstein–Cartan theory. With cosmological (and to some extent the more general spherically symmetric) backgrounds already under control, these provide timely and physically relevant arenas in which to explore the dynamical role of spin currents. Last but not least, features such as non-Minkowskian vacuum and novel matter couplings - even under the minimal coupling principle - open the possibility of testing the theory using the wealth of high-precision data constraining local (in this case effective) Lorentz violations from various laboratory experiments \cite{Kostelecky:2008ts}, see also \cite{Pfeifer:2024vvc,Braun:2025iqq}. This aspect of the theory calls for a more detailed scrutiny than the preliminary, back-of-the-envelope calculation given in \cite{Koivisto:2022uvd}. 

Independent of the new perspective on space and time that arises specifically from the Euclidean signature of the $Spin(4)$ framework, our study offers an intriguing extension of the Lorentz gauge theory. In particular, we have shown that
\begin{itemize}
\item[] gravitation is right-handed
\item[{\it iff}] dark matter is cold
\item[{\it iff}] the $\Lambda=0$ vacuum is Minkowski
\item[{\it iff}] the speed of graviton is the speed of light.
\end{itemize}
Even if $\beta=0$ in our Universe, these unexpected and tightly interwoven connections between its seemingly unrelated foundational aspects offer revelatory insights into its pregeometric structure.
 
\subsection{A philosophical epilogue}

\begin{displayquote}
Space, the extensive medium of the material world, is clearly the seat of the group of coordinate transformations; but the group [of physical automorphisms] seems to have its origin in the ultimate elementary particles of matter. \cite{Weyl1949-WEYPOM}
\end{displayquote}
Although scholars have acknowledged the significance of the relationship between physical and mathematical automorphisms in Hermann Weyl's thought \cite{SCHOLZ201857}, its more precise and far-reaching implications may have remained insufficiently understood, particularly in the context of spacetime and gravitation theory. The mainstream approach of ``gauging translations'' in (metric-affine, Poincar{\'e}, teleparallel, etc) gravity theories\footnote{In 1974, C.-N. Yang introduced a Lorentz gauge theory of gravity, which he described as 'conceptually superior', likely due to its alignment with Hermann Weyl’s foundational ideas \cite{Yang:1974kj}. The contrast between the mainstream approach and what may be termed the Weyl–Yang approach to gauge gravity is particularly apparent in Chapter 19 of \cite{hehl}.} conflates auxiliary mathematical structure with physical substance, a move that retains the lingering Newtonian presupposition that space, independently of any observer, furnishes an arena wherein phenomena take place \cite{Burtt1954-BURTMF}. 
 
Though Newton had published a proper theory of gravitation in the {\it Principia} (1687), the beginning of the modern science of space and time might be traced to Kant's considerations in the 18th century, on what we could nowadays call spontaneous breaking of chiral symmetry, from which he was eventually led to the view that Euclidean space and time are {\it a priori} forms of intuition \cite{VanCleve1991-VANTPO-16}.
%\cmt{Maybe a good one to cite?: https://philsciarchive.pitt.edu/713/1/parity.pdf}. 
In the 19th century, Riemann and Clifford developed a theory of gravitation and matter as manifestations of geometry and curvature, connecting physical phenomena to the structure and dynamics of space. Clifford's algebra already laid the mathematical groundwork for the unification of space and time - and, of the chiral aspects that have been 
%Riemann and Clifford developed a theory of gravitation and matter as geometry and curvature in the 19th century, and its chiral aspects have been 
elucidated in geometry (Atiyah) and gravity (Plebanski) more recently. 
%The 20th century brought quantum mechanics, which eventually suggests an
The advent of quantum mechanics in the 20th century ultimately motivates an informational perspective, with space and time as fundamental categories of cognition agreeing with Kant's %utterly compelling
unavoidable\footnote{Though of course G{\"o}del was allowed to suggest that the theory of gravity may help one to peek behind the veil of Maya.} conclusion.
%Quantum mechanics is important in the current view, which reduces space and time to the most fundamental categories of cognition according to Kant's utterly compelling
The basic units are bits, which because they are quantum-mechanical, are qubits. Moreover, because the world is only intelligible to us in space and in time, the atoms of information are {\it gauged qubits} - represented by spinors subject to Clifford's algebra. 

In the $Spin(4)$ theory, these atoms of information are counted as quaternions and thus in a perhaps less known matrix basis for spinors. The sense of causality requires an ``imaginary'' field that provides an elementary organising principle, time, born from the primordial {\it egg spinor} $\psi$. One can regard time as an illusion, but it is an illusion that definitely exists, or more to the point: nothing can exist without time. The Lorentz gauge theory gives a concrete form to an idea expressed in Weyl's characteristically unrivalled eloquence in the quotation above: external spacetime is but an intrinsic reflection of matter. 
%expressed in Weyl's unrivalled eloquence in the quote above: 

\acknowledgements{This work was supported by the Estonian Research Council grants  TARISTU24-TK10, TARISTU24-TK3, CoE TK202 “Foundations of the Universe” and PRG2608 “Space - Time  - Matter”. We would like thank an anonymous referee for helpful comments on previous versions of the manuscript.}

\bibliography{spin4refs}

%merlin.mbs apsrev4-1.bst 2010-07-25 4.21a (PWD, AO, DPC) hacked
%Control: key (0)
%Control: author (8) initials jnrlst
%Control: editor formatted (1) identically to author
%Control: production of article title (-1) disabled
%Control: page (0) single
%Control: year (1) truncated
%Control: production of eprint (0) enabled
\begin{thebibliography}{113}%
\makeatletter
\providecommand \@ifxundefined [1]{%
 \@ifx{#1\undefined}
}%
\providecommand \@ifnum [1]{%
 \ifnum #1\expandafter \@firstoftwo
 \else \expandafter \@secondoftwo
 \fi
}%
\providecommand \@ifx [1]{%
 \ifx #1\expandafter \@firstoftwo
 \else \expandafter \@secondoftwo
 \fi
}%
\providecommand \natexlab [1]{#1}%
\providecommand \enquote  [1]{``#1''}%
\providecommand \bibnamefont  [1]{#1}%
\providecommand \bibfnamefont [1]{#1}%
\providecommand \citenamefont [1]{#1}%
\providecommand \href@noop [0]{\@secondoftwo}%
\providecommand \href [0]{\begingroup \@sanitize@url \@href}%
\providecommand \@href[1]{\@@startlink{#1}\@@href}%
\providecommand \@@href[1]{\endgroup#1\@@endlink}%
\providecommand \@sanitize@url [0]{\catcode `\\12\catcode `\$12\catcode
  `\&12\catcode `\#12\catcode `\^12\catcode `\_12\catcode `\%12\relax}%
\providecommand \@@startlink[1]{}%
\providecommand \@@endlink[0]{}%
\providecommand \url  [0]{\begingroup\@sanitize@url \@url }%
\providecommand \@url [1]{\endgroup\@href {#1}{\urlprefix }}%
\providecommand \urlprefix  [0]{URL }%
\providecommand \Eprint [0]{\href }%
\providecommand \doibase [0]{http://dx.doi.org/}%
\providecommand \selectlanguage [0]{\@gobble}%
\providecommand \bibinfo  [0]{\@secondoftwo}%
\providecommand \bibfield  [0]{\@secondoftwo}%
\providecommand \translation [1]{[#1]}%
\providecommand \BibitemOpen [0]{}%
\providecommand \bibitemStop [0]{}%
\providecommand \bibitemNoStop [0]{.\EOS\space}%
\providecommand \EOS [0]{\spacefactor3000\relax}%
\providecommand \BibitemShut  [1]{\csname bibitem#1\endcsname}%
\let\auto@bib@innerbib\@empty
%</preamble>
\bibitem [{\citenamefont {Osterwalder}\ and\ \citenamefont
  {Schrader}(1973)}]{Osterwalder:1973dx}%
  \BibitemOpen
  \bibfield  {author} {\bibinfo {author} {\bibfnamefont {K.}~\bibnamefont
  {Osterwalder}}\ and\ \bibinfo {author} {\bibfnamefont {R.}~\bibnamefont
  {Schrader}},\ }\href {\doibase 10.1007/BF01645738} {\bibfield  {journal}
  {\bibinfo  {journal} {Commun. Math. Phys.}\ }\textbf {\bibinfo {volume}
  {31}},\ \bibinfo {pages} {83} (\bibinfo {year} {1973})}\BibitemShut {NoStop}%
\bibitem [{\citenamefont {Osterwalder}\ and\ \citenamefont
  {Schrader}(1975)}]{Osterwalder:1974tc}%
  \BibitemOpen
  \bibfield  {author} {\bibinfo {author} {\bibfnamefont {K.}~\bibnamefont
  {Osterwalder}}\ and\ \bibinfo {author} {\bibfnamefont {R.}~\bibnamefont
  {Schrader}},\ }\href {\doibase 10.1007/BF01608978} {\bibfield  {journal}
  {\bibinfo  {journal} {Commun. Math. Phys.}\ }\textbf {\bibinfo {volume}
  {42}},\ \bibinfo {pages} {281} (\bibinfo {year} {1975})}\BibitemShut
  {NoStop}%
\bibitem [{\citenamefont {Gibbons}\ and\ \citenamefont
  {Hawking}(1993)}]{doi:10.1142/1301}%
  \BibitemOpen
  \bibfield  {author} {\bibinfo {author} {\bibfnamefont {G.~W.}\ \bibnamefont
  {Gibbons}}\ and\ \bibinfo {author} {\bibfnamefont {S.~W.}\ \bibnamefont
  {Hawking}},\ }\href {\doibase 10.1142/1301} {\emph {\bibinfo {title}
  {Euclidean Quantum Gravity}}}\ (\bibinfo  {publisher} {WORLD SCIENTIFIC},\
  \bibinfo {year} {1993})\ \Eprint
  {http://arxiv.org/abs/https://www.worldscientific.com/doi/epdf/10.1142/1301}
  {https://www.worldscientific.com/doi/epdf/10.1142/1301} \BibitemShut
  {NoStop}%
\bibitem [{\citenamefont {Isham}(1993)}]{Isham:1992ms}%
  \BibitemOpen
  \bibfield  {author} {\bibinfo {author} {\bibfnamefont {C.~J.}\ \bibnamefont
  {Isham}},\ }\href@noop {} {\bibfield  {journal} {\bibinfo  {journal} {NATO
  Sci. Ser. C}\ }\textbf {\bibinfo {volume} {409}},\ \bibinfo {pages} {157}
  (\bibinfo {year} {1993})},\ \Eprint {http://arxiv.org/abs/gr-qc/9210011}
  {arXiv:gr-qc/9210011} \BibitemShut {NoStop}%
\bibitem [{\citenamefont {Anderson}(2012)}]{Anderson:2012vk}%
  \BibitemOpen
  \bibfield  {author} {\bibinfo {author} {\bibfnamefont {E.}~\bibnamefont
  {Anderson}},\ }\href {\doibase 10.1002/andp.201200147} {\bibfield  {journal}
  {\bibinfo  {journal} {Annalen Phys.}\ }\textbf {\bibinfo {volume} {524}},\
  \bibinfo {pages} {757} (\bibinfo {year} {2012})},\ \Eprint
  {http://arxiv.org/abs/1206.2403} {arXiv:1206.2403 [gr-qc]} \BibitemShut
  {NoStop}%
\bibitem [{\citenamefont {Wetterich}(2021)}]{Wetterich:2021ywr}%
  \BibitemOpen
  \bibfield  {author} {\bibinfo {author} {\bibfnamefont {C.}~\bibnamefont
  {Wetterich}},\ }\href {\doibase 10.1016/j.nuclphysb.2021.115526} {\bibfield
  {journal} {\bibinfo  {journal} {Nucl. Phys. B}\ }\textbf {\bibinfo {volume}
  {971}},\ \bibinfo {pages} {115526} (\bibinfo {year} {2021})},\ \Eprint
  {http://arxiv.org/abs/2101.07849} {arXiv:2101.07849 [gr-qc]} \BibitemShut
  {NoStop}%
\bibitem [{\citenamefont {Wetterich}(2022)}]{Wetterich:2021hru}%
  \BibitemOpen
  \bibfield  {author} {\bibinfo {author} {\bibfnamefont {C.}~\bibnamefont
  {Wetterich}},\ }\href {\doibase 10.1007/JHEP06(2022)069} {\bibfield
  {journal} {\bibinfo  {journal} {JHEP}\ }\textbf {\bibinfo {volume} {06}},\
  \bibinfo {pages} {069} (\bibinfo {year} {2022})},\ \Eprint
  {http://arxiv.org/abs/2101.11519} {arXiv:2101.11519 [gr-qc]} \BibitemShut
  {NoStop}%
\bibitem [{\citenamefont {Akama}(1978)}]{Akama:1978pg}%
  \BibitemOpen
  \bibfield  {author} {\bibinfo {author} {\bibfnamefont {K.}~\bibnamefont
  {Akama}},\ }\href {\doibase 10.1143/PTP.60.1900} {\bibfield  {journal}
  {\bibinfo  {journal} {Prog. Theor. Phys.}\ }\textbf {\bibinfo {volume}
  {60}},\ \bibinfo {pages} {1900} (\bibinfo {year} {1978})}\BibitemShut
  {NoStop}%
\bibitem [{\citenamefont {Wetterich}(2004)}]{Wetterich:2003wr}%
  \BibitemOpen
  \bibfield  {author} {\bibinfo {author} {\bibfnamefont {C.}~\bibnamefont
  {Wetterich}},\ }\href {\doibase 10.1103/PhysRevD.70.105004} {\bibfield
  {journal} {\bibinfo  {journal} {Phys. Rev. D}\ }\textbf {\bibinfo {volume}
  {70}},\ \bibinfo {pages} {105004} (\bibinfo {year} {2004})},\ \Eprint
  {http://arxiv.org/abs/hep-th/0307145} {arXiv:hep-th/0307145} \BibitemShut
  {NoStop}%
\bibitem [{\citenamefont {Z\l{}o\'snik}\ \emph {et~al.}(2018)\citenamefont
  {Z\l{}o\'snik}, \citenamefont {Urban}, \citenamefont {Marzola},\ and\
  \citenamefont {Koivisto}}]{Zlosnik:2018qvg}%
  \BibitemOpen
  \bibfield  {author} {\bibinfo {author} {\bibfnamefont {T.}~\bibnamefont
  {Z\l{}o\'snik}}, \bibinfo {author} {\bibfnamefont {F.}~\bibnamefont {Urban}},
  \bibinfo {author} {\bibfnamefont {L.}~\bibnamefont {Marzola}}, \ and\
  \bibinfo {author} {\bibfnamefont {T.}~\bibnamefont {Koivisto}},\ }\href
  {\doibase 10.1088/1361-6382/aaea96} {\bibfield  {journal} {\bibinfo
  {journal} {Class. Quant. Grav.}\ }\textbf {\bibinfo {volume} {35}},\ \bibinfo
  {pages} {235003} (\bibinfo {year} {2018})},\ \Eprint
  {http://arxiv.org/abs/1807.01100} {arXiv:1807.01100 [gr-qc]} \BibitemShut
  {NoStop}%
\bibitem [{\citenamefont {Gies}\ and\ \citenamefont
  {Lippoldt}(2014)}]{Gies:2013noa}%
  \BibitemOpen
  \bibfield  {author} {\bibinfo {author} {\bibfnamefont {H.}~\bibnamefont
  {Gies}}\ and\ \bibinfo {author} {\bibfnamefont {S.}~\bibnamefont
  {Lippoldt}},\ }\href {\doibase 10.1103/PhysRevD.89.064040} {\bibfield
  {journal} {\bibinfo  {journal} {Phys. Rev. D}\ }\textbf {\bibinfo {volume}
  {89}},\ \bibinfo {pages} {064040} (\bibinfo {year} {2014})},\ \Eprint
  {http://arxiv.org/abs/1310.2509} {arXiv:1310.2509 [hep-th]} \BibitemShut
  {NoStop}%
\bibitem [{\citenamefont {Gies}\ and\ \citenamefont
  {Lippoldt}(2015)}]{Gies:2015cka}%
  \BibitemOpen
  \bibfield  {author} {\bibinfo {author} {\bibfnamefont {H.}~\bibnamefont
  {Gies}}\ and\ \bibinfo {author} {\bibfnamefont {S.}~\bibnamefont
  {Lippoldt}},\ }\href {\doibase 10.1016/j.physletb.2015.03.014} {\bibfield
  {journal} {\bibinfo  {journal} {Phys. Lett. B}\ }\textbf {\bibinfo {volume}
  {743}},\ \bibinfo {pages} {415} (\bibinfo {year} {2015})},\ \Eprint
  {http://arxiv.org/abs/1502.00918} {arXiv:1502.00918 [hep-th]} \BibitemShut
  {NoStop}%
\bibitem [{\citenamefont {Lippoldt}(2016)}]{Lippoldt:2016ayw}%
  \BibitemOpen
  \bibfield  {author} {\bibinfo {author} {\bibfnamefont {S.}~\bibnamefont
  {Lippoldt}},\ }\emph {\bibinfo {title} {{Fermions in curved spacetimes}}},\
  \href@noop {} {Ph.D. thesis},\ \bibinfo  {school}
  {Friedrich-Schiller-Universit{\"a}t Jena, Physikalisch-Astronomische
  Fakult{\"a}t, Deutschland} (\bibinfo {year} {2016})\BibitemShut {NoStop}%
\bibitem [{\citenamefont {Weldon}(2001)}]{Weldon:2000fr}%
  \BibitemOpen
  \bibfield  {author} {\bibinfo {author} {\bibfnamefont {H.~A.}\ \bibnamefont
  {Weldon}},\ }\href {\doibase 10.1103/PhysRevD.63.104010} {\bibfield
  {journal} {\bibinfo  {journal} {Phys. Rev. D}\ }\textbf {\bibinfo {volume}
  {63}},\ \bibinfo {pages} {104010} (\bibinfo {year} {2001})},\ \Eprint
  {http://arxiv.org/abs/gr-qc/0009086} {arXiv:gr-qc/0009086} \BibitemShut
  {NoStop}%
\bibitem [{\citenamefont {Emmrich}(2022)}]{10.1063/5.0081140}%
  \BibitemOpen
  \bibfield  {author} {\bibinfo {author} {\bibfnamefont {C.}~\bibnamefont
  {Emmrich}},\ }\href {\doibase 10.1063/5.0081140} {\bibfield  {journal}
  {\bibinfo  {journal} {Journal of Mathematical Physics}\ }\textbf {\bibinfo
  {volume} {63}},\ \bibinfo {pages} {042302} (\bibinfo {year}
  {2022})}\BibitemShut {NoStop}%
\bibitem [{\citenamefont {Kibble}(1961)}]{Kibble:1961ba}%
  \BibitemOpen
  \bibfield  {author} {\bibinfo {author} {\bibfnamefont {T.~W.~B.}\
  \bibnamefont {Kibble}},\ }\href {\doibase 10.1063/1.1703702} {\bibfield
  {journal} {\bibinfo  {journal} {J. Math. Phys.}\ }\textbf {\bibinfo {volume}
  {2}},\ \bibinfo {pages} {212} (\bibinfo {year} {1961})}\BibitemShut {NoStop}%
\bibitem [{\citenamefont {Shaposhnikov}(2025)}]{Shaposhnikov:2025znm}%
  \BibitemOpen
  \bibfield  {author} {\bibinfo {author} {\bibfnamefont {M.}~\bibnamefont
  {Shaposhnikov}}\ }(\bibinfo {year} {2025})\ \Eprint
  {http://arxiv.org/abs/2506.11847} {arXiv:2506.11847 [hep-th]} \BibitemShut
  {NoStop}%
\bibitem [{\citenamefont {Grignani}\ and\ \citenamefont
  {Nardelli}(1992)}]{Grignani:1991nj}%
  \BibitemOpen
  \bibfield  {author} {\bibinfo {author} {\bibfnamefont {G.}~\bibnamefont
  {Grignani}}\ and\ \bibinfo {author} {\bibfnamefont {G.}~\bibnamefont
  {Nardelli}},\ }\href {\doibase 10.1103/PhysRevD.45.2719} {\bibfield
  {journal} {\bibinfo  {journal} {Phys. Rev. D}\ }\textbf {\bibinfo {volume}
  {45}},\ \bibinfo {pages} {2719} (\bibinfo {year} {1992})}\BibitemShut
  {NoStop}%
\bibitem [{\citenamefont {Koivisto}\ and\ \citenamefont
  {Zlosnik}(2023)}]{Koivisto:2022uvd}%
  \BibitemOpen
  \bibfield  {author} {\bibinfo {author} {\bibfnamefont {T.~S.}\ \bibnamefont
  {Koivisto}}\ and\ \bibinfo {author} {\bibfnamefont {T.}~\bibnamefont
  {Zlosnik}},\ }\href {\doibase 10.1103/PhysRevD.107.124013} {\bibfield
  {journal} {\bibinfo  {journal} {Phys. Rev. D}\ }\textbf {\bibinfo {volume}
  {107}},\ \bibinfo {pages} {124013} (\bibinfo {year} {2023})},\ \Eprint
  {http://arxiv.org/abs/2212.04562} {arXiv:2212.04562 [gr-qc]} \BibitemShut
  {NoStop}%
\bibitem [{\citenamefont {Westman}\ and\ \citenamefont
  {Zlosnik}(2015)}]{Westman:2014yca}%
  \BibitemOpen
  \bibfield  {author} {\bibinfo {author} {\bibfnamefont {H.~F.}\ \bibnamefont
  {Westman}}\ and\ \bibinfo {author} {\bibfnamefont {T.~G.}\ \bibnamefont
  {Zlosnik}},\ }\href {\doibase 10.1016/j.aop.2015.06.013} {\bibfield
  {journal} {\bibinfo  {journal} {Annals Phys.}\ }\textbf {\bibinfo {volume}
  {361}},\ \bibinfo {pages} {330} (\bibinfo {year} {2015})},\ \Eprint
  {http://arxiv.org/abs/1411.1679} {arXiv:1411.1679 [gr-qc]} \BibitemShut
  {NoStop}%
\bibitem [{\citenamefont {Addazi}\ \emph {et~al.}(2025)\citenamefont {Addazi},
  \citenamefont {Capozziello}, \citenamefont {Marciano},\ and\ \citenamefont
  {Meluccio}}]{Addazi:2024rzo}%
  \BibitemOpen
  \bibfield  {author} {\bibinfo {author} {\bibfnamefont {A.}~\bibnamefont
  {Addazi}}, \bibinfo {author} {\bibfnamefont {S.}~\bibnamefont {Capozziello}},
  \bibinfo {author} {\bibfnamefont {A.}~\bibnamefont {Marciano}}, \ and\
  \bibinfo {author} {\bibfnamefont {G.}~\bibnamefont {Meluccio}},\ }\href
  {\doibase 10.1088/1361-6382/ada767} {\bibfield  {journal} {\bibinfo
  {journal} {Class. Quant. Grav.}\ }\textbf {\bibinfo {volume} {42}},\ \bibinfo
  {pages} {045012} (\bibinfo {year} {2025})},\ \Eprint
  {http://arxiv.org/abs/2409.02200} {arXiv:2409.02200 [hep-th]} \BibitemShut
  {NoStop}%
\bibitem [{\citenamefont {Partanen}\ and\ \citenamefont
  {Tulkki}(2025)}]{Partanen:2023dkt}%
  \BibitemOpen
  \bibfield  {author} {\bibinfo {author} {\bibfnamefont {M.}~\bibnamefont
  {Partanen}}\ and\ \bibinfo {author} {\bibfnamefont {J.}~\bibnamefont
  {Tulkki}},\ }\href {\doibase 10.1088/1361-6633/adc82e} {\bibfield  {journal}
  {\bibinfo  {journal} {Rept. Prog. Phys.}\ }\textbf {\bibinfo {volume} {88}},\
  \bibinfo {pages} {057802} (\bibinfo {year} {2025})},\ \bibinfo {note}
  {[Erratum: Rept.Prog.Phys. 88, 069501 (2025)]},\ \Eprint
  {http://arxiv.org/abs/2310.01460} {arXiv:2310.01460 [gr-qc]} \BibitemShut
  {NoStop}%
\bibitem [{\citenamefont {Brown}\ and\ \citenamefont
  {Kuchar}(1995)}]{Brown:1994py}%
  \BibitemOpen
  \bibfield  {author} {\bibinfo {author} {\bibfnamefont {J.~D.}\ \bibnamefont
  {Brown}}\ and\ \bibinfo {author} {\bibfnamefont {K.~V.}\ \bibnamefont
  {Kuchar}},\ }\href {\doibase 10.1103/PhysRevD.51.5600} {\bibfield  {journal}
  {\bibinfo  {journal} {Phys. Rev. D}\ }\textbf {\bibinfo {volume} {51}},\
  \bibinfo {pages} {5600} (\bibinfo {year} {1995})},\ \Eprint
  {http://arxiv.org/abs/gr-qc/9409001} {arXiv:gr-qc/9409001} \BibitemShut
  {NoStop}%
\bibitem [{\citenamefont {Husain}\ and\ \citenamefont
  {Pawlowski}(2012)}]{Husain:2011tk}%
  \BibitemOpen
  \bibfield  {author} {\bibinfo {author} {\bibfnamefont {V.}~\bibnamefont
  {Husain}}\ and\ \bibinfo {author} {\bibfnamefont {T.}~\bibnamefont
  {Pawlowski}},\ }\href {\doibase 10.1103/PhysRevLett.108.141301} {\bibfield
  {journal} {\bibinfo  {journal} {Phys. Rev. Lett.}\ }\textbf {\bibinfo
  {volume} {108}},\ \bibinfo {pages} {141301} (\bibinfo {year} {2012})},\
  \Eprint {http://arxiv.org/abs/1108.1145} {arXiv:1108.1145 [gr-qc]}
  \BibitemShut {NoStop}%
\bibitem [{\citenamefont {Koivisto}(2023)}]{Koivisto:2023epd}%
  \BibitemOpen
  \bibfield  {author} {\bibinfo {author} {\bibfnamefont {T.}~\bibnamefont
  {Koivisto}},\ }\href {\doibase 10.1142/S0219887824500403} {\bibfield
  {journal} {\bibinfo  {journal} {Int. J. Geom. Meth. Mod. Phys.}\ }\textbf
  {\bibinfo {volume} {20}},\ \bibinfo {pages} {2450040} (\bibinfo {year}
  {2023})},\ \Eprint {http://arxiv.org/abs/2306.00963} {arXiv:2306.00963
  [gr-qc]} \BibitemShut {NoStop}%
\bibitem [{\citenamefont {Hebecker}\ \emph {et~al.}(2018)\citenamefont
  {Hebecker}, \citenamefont {Mikhail},\ and\ \citenamefont
  {Soler}}]{Hebecker:2018ofv}%
  \BibitemOpen
  \bibfield  {author} {\bibinfo {author} {\bibfnamefont {A.}~\bibnamefont
  {Hebecker}}, \bibinfo {author} {\bibfnamefont {T.}~\bibnamefont {Mikhail}}, \
  and\ \bibinfo {author} {\bibfnamefont {P.}~\bibnamefont {Soler}},\ }\href
  {\doibase 10.3389/fspas.2018.00035} {\bibfield  {journal} {\bibinfo
  {journal} {Front. Astron. Space Sci.}\ }\textbf {\bibinfo {volume} {5}},\
  \bibinfo {pages} {35} (\bibinfo {year} {2018})},\ \Eprint
  {http://arxiv.org/abs/1807.00824} {arXiv:1807.00824 [hep-th]} \BibitemShut
  {NoStop}%
\bibitem [{\citenamefont {Lehners}(2023)}]{Lehners:2023yrj}%
  \BibitemOpen
  \bibfield  {author} {\bibinfo {author} {\bibfnamefont {J.-L.}\ \bibnamefont
  {Lehners}},\ }\href {\doibase 10.1016/j.physrep.2023.06.002} {\bibfield
  {journal} {\bibinfo  {journal} {Phys. Rept.}\ }\textbf {\bibinfo {volume}
  {1022}},\ \bibinfo {pages} {1} (\bibinfo {year} {2023})},\ \Eprint
  {http://arxiv.org/abs/2303.08802} {arXiv:2303.08802 [hep-th]} \BibitemShut
  {NoStop}%
\bibitem [{\citenamefont {Visser}(2017)}]{Visser:2017atf}%
  \BibitemOpen
  \bibfield  {author} {\bibinfo {author} {\bibfnamefont {M.}~\bibnamefont
  {Visser}},\ }\href@noop {} {\  (\bibinfo {year} {2017})},\ \Eprint
  {http://arxiv.org/abs/1702.05572} {arXiv:1702.05572 [gr-qc]} \BibitemShut
  {NoStop}%
\bibitem [{\citenamefont {Samuel}(2016)}]{Samuel:2015oea}%
  \BibitemOpen
  \bibfield  {author} {\bibinfo {author} {\bibfnamefont {J.}~\bibnamefont
  {Samuel}},\ }\href {\doibase 10.1088/0264-9381/33/1/015006} {\bibfield
  {journal} {\bibinfo  {journal} {Class. Quant. Grav.}\ }\textbf {\bibinfo
  {volume} {33}},\ \bibinfo {pages} {015006} (\bibinfo {year} {2016})},\
  \Eprint {http://arxiv.org/abs/1510.07365} {arXiv:1510.07365 [gr-qc]}
  \BibitemShut {NoStop}%
\bibitem [{\citenamefont {Baldazzi}\ \emph {et~al.}(2019)\citenamefont
  {Baldazzi}, \citenamefont {Percacci},\ and\ \citenamefont
  {Skrinjar}}]{Baldazzi:2018mtl}%
  \BibitemOpen
  \bibfield  {author} {\bibinfo {author} {\bibfnamefont {A.}~\bibnamefont
  {Baldazzi}}, \bibinfo {author} {\bibfnamefont {R.}~\bibnamefont {Percacci}},
  \ and\ \bibinfo {author} {\bibfnamefont {V.}~\bibnamefont {Skrinjar}},\
  }\href {\doibase 10.1088/1361-6382/ab187d} {\bibfield  {journal} {\bibinfo
  {journal} {Class. Quant. Grav.}\ }\textbf {\bibinfo {volume} {36}},\ \bibinfo
  {pages} {105008} (\bibinfo {year} {2019})},\ \Eprint
  {http://arxiv.org/abs/1811.03369} {arXiv:1811.03369 [gr-qc]} \BibitemShut
  {NoStop}%
\bibitem [{\citenamefont {Singh}\ and\ \citenamefont
  {Kothawala}(2023)}]{Singh:2020hpv}%
  \BibitemOpen
  \bibfield  {author} {\bibinfo {author} {\bibfnamefont {R.}~\bibnamefont
  {Singh}}\ and\ \bibinfo {author} {\bibfnamefont {D.}~\bibnamefont
  {Kothawala}},\ }\href {\doibase 10.1140/epjc/s10052-023-11340-1} {\bibfield
  {journal} {\bibinfo  {journal} {Eur. Phys. J. C}\ }\textbf {\bibinfo {volume}
  {83}},\ \bibinfo {pages} {194} (\bibinfo {year} {2023})},\ \Eprint
  {http://arxiv.org/abs/2010.01822} {arXiv:2010.01822 [gr-qc]} \BibitemShut
  {NoStop}%
\bibitem [{\citenamefont {Kontsevich}\ and\ \citenamefont
  {Segal}(2021)}]{Kontsevich:2021dmb}%
  \BibitemOpen
  \bibfield  {author} {\bibinfo {author} {\bibfnamefont {M.}~\bibnamefont
  {Kontsevich}}\ and\ \bibinfo {author} {\bibfnamefont {G.}~\bibnamefont
  {Segal}},\ }\href {\doibase 10.1093/qmath/haab027} {\bibfield  {journal}
  {\bibinfo  {journal} {Quart. J. Math. Oxford Ser.}\ }\textbf {\bibinfo
  {volume} {72}},\ \bibinfo {pages} {673} (\bibinfo {year} {2021})},\ \Eprint
  {http://arxiv.org/abs/2105.10161} {arXiv:2105.10161 [hep-th]} \BibitemShut
  {NoStop}%
\bibitem [{\citenamefont {Witten}(2021)}]{Witten:2021nzp}%
  \BibitemOpen
  \bibfield  {author} {\bibinfo {author} {\bibfnamefont {E.}~\bibnamefont
  {Witten}},\ }\href@noop {} {\  (\bibinfo {year} {2021})},\ \Eprint
  {http://arxiv.org/abs/2111.06514} {arXiv:2111.06514 [hep-th]} \BibitemShut
  {NoStop}%
\bibitem [{\citenamefont {Valtancoli}(2023)}]{Valtancoli:2023wab}%
  \BibitemOpen
  \bibfield  {author} {\bibinfo {author} {\bibfnamefont {P.}~\bibnamefont
  {Valtancoli}},\ }\href@noop {} {\  (\bibinfo {year} {2023})},\ \Eprint
  {http://arxiv.org/abs/2306.06279} {arXiv:2306.06279 [gr-qc]} \BibitemShut
  {NoStop}%
\bibitem [{\citenamefont {Beltr{\'a}n~Jim{\'e}nez}\ and\ \citenamefont
  {Koivisto}(2024)}]{BeltranJimenez:2024ufa}%
  \BibitemOpen
  \bibfield  {author} {\bibinfo {author} {\bibfnamefont {J.}~\bibnamefont
  {Beltr{\'a}n~Jim{\'e}nez}}\ and\ \bibinfo {author} {\bibfnamefont {T.~S.}\
  \bibnamefont {Koivisto}},\ }\href@noop {} {\  (\bibinfo {year} {2024})},\
  \Eprint {http://arxiv.org/abs/2412.13946} {arXiv:2412.13946 [gr-qc]}
  \BibitemShut {NoStop}%
\bibitem [{\citenamefont {Banerjee}\ and\ \citenamefont
  {Niedermaier}(2025)}]{Banerjee:2024tap}%
  \BibitemOpen
  \bibfield  {author} {\bibinfo {author} {\bibfnamefont {R.}~\bibnamefont
  {Banerjee}}\ and\ \bibinfo {author} {\bibfnamefont {M.}~\bibnamefont
  {Niedermaier}},\ }\href {\doibase 10.1088/1361-6382/adc9ef} {\bibfield
  {journal} {\bibinfo  {journal} {Class. Quant. Grav.}\ }\textbf {\bibinfo
  {volume} {42}},\ \bibinfo {pages} {095003} (\bibinfo {year} {2025})},\
  \Eprint {http://arxiv.org/abs/2406.06047} {arXiv:2406.06047 [math-ph]}
  \BibitemShut {NoStop}%
\bibitem [{\citenamefont {Barbero~G.}(1996)}]{BarberoG:1995tgc}%
  \BibitemOpen
  \bibfield  {author} {\bibinfo {author} {\bibfnamefont {J.~F.}\ \bibnamefont
  {Barbero~G.}},\ }\href {\doibase 10.1103/PhysRevD.54.1492} {\bibfield
  {journal} {\bibinfo  {journal} {Phys. Rev. D}\ }\textbf {\bibinfo {volume}
  {54}},\ \bibinfo {pages} {1492} (\bibinfo {year} {1996})},\ \Eprint
  {http://arxiv.org/abs/gr-qc/9605066} {arXiv:gr-qc/9605066} \BibitemShut
  {NoStop}%
\bibitem [{\citenamefont {Ashtekar}(1996)}]{Ashtekar:1995qw}%
  \BibitemOpen
  \bibfield  {author} {\bibinfo {author} {\bibfnamefont {A.}~\bibnamefont
  {Ashtekar}},\ }\href {\doibase 10.1103/PhysRevD.53.R2865} {\bibfield
  {journal} {\bibinfo  {journal} {Phys. Rev. D}\ }\textbf {\bibinfo {volume}
  {53}},\ \bibinfo {pages} {2865} (\bibinfo {year} {1996})},\ \Eprint
  {http://arxiv.org/abs/gr-qc/9511083} {arXiv:gr-qc/9511083} \BibitemShut
  {NoStop}%
\bibitem [{\citenamefont {Iliev}(2000)}]{Iliev:1998su}%
  \BibitemOpen
  \bibfield  {author} {\bibinfo {author} {\bibfnamefont {B.~Z.}\ \bibnamefont
  {Iliev}},\ }\href {\doibase 10.1016/S0393-0440(99)00074-1} {\bibfield
  {journal} {\bibinfo  {journal} {J. Geom. Phys.}\ }\textbf {\bibinfo {volume}
  {34}},\ \bibinfo {pages} {321} (\bibinfo {year} {2000})},\ \Eprint
  {http://arxiv.org/abs/gr-qc/9802057} {arXiv:gr-qc/9802057} \BibitemShut
  {NoStop}%
\bibitem [{\citenamefont {Reddy}\ \emph {et~al.}(2008)\citenamefont {Reddy},
  \citenamefont {Sharma},\ and\ \citenamefont {Krishnan}}]{article}%
  \BibitemOpen
  \bibfield  {author} {\bibinfo {author} {\bibfnamefont {B.}~\bibnamefont
  {Reddy}}, \bibinfo {author} {\bibfnamefont {R.}~\bibnamefont {Sharma}}, \
  and\ \bibinfo {author} {\bibfnamefont {S.}~\bibnamefont {Krishnan}},\
  }\href@noop {} {\bibfield  {journal} {\bibinfo  {journal} {International
  Journal of Pure and Applied Mathematics}\ }\textbf {\bibinfo {volume} {47}}
  (\bibinfo {year} {2008})}\BibitemShut {NoStop}%
\bibitem [{\citenamefont {Girelli}\ \emph {et~al.}(2009)\citenamefont
  {Girelli}, \citenamefont {Liberati},\ and\ \citenamefont
  {Sindoni}}]{Girelli:2008qp}%
  \BibitemOpen
  \bibfield  {author} {\bibinfo {author} {\bibfnamefont {F.}~\bibnamefont
  {Girelli}}, \bibinfo {author} {\bibfnamefont {S.}~\bibnamefont {Liberati}}, \
  and\ \bibinfo {author} {\bibfnamefont {L.}~\bibnamefont {Sindoni}},\ }\href
  {\doibase 10.1103/PhysRevD.79.044019} {\bibfield  {journal} {\bibinfo
  {journal} {Phys. Rev. D}\ }\textbf {\bibinfo {volume} {79}},\ \bibinfo
  {pages} {044019} (\bibinfo {year} {2009})},\ \Eprint
  {http://arxiv.org/abs/0806.4239} {arXiv:0806.4239 [gr-qc]} \BibitemShut
  {NoStop}%
\bibitem [{\citenamefont {Mukohyama}\ and\ \citenamefont
  {Uzan}(2013)}]{Mukohyama:2013ew}%
  \BibitemOpen
  \bibfield  {author} {\bibinfo {author} {\bibfnamefont {S.}~\bibnamefont
  {Mukohyama}}\ and\ \bibinfo {author} {\bibfnamefont {J.-P.}\ \bibnamefont
  {Uzan}},\ }\href {\doibase 10.1103/PhysRevD.87.065020} {\bibfield  {journal}
  {\bibinfo  {journal} {Phys. Rev. D}\ }\textbf {\bibinfo {volume} {87}},\
  \bibinfo {pages} {065020} (\bibinfo {year} {2013})},\ \Eprint
  {http://arxiv.org/abs/1301.1361} {arXiv:1301.1361 [hep-th]} \BibitemShut
  {NoStop}%
\bibitem [{\citenamefont {Svidzinsky}(2017)}]{Svidzinsky:2015xbl}%
  \BibitemOpen
  \bibfield  {author} {\bibinfo {author} {\bibfnamefont {A.~A.}\ \bibnamefont
  {Svidzinsky}},\ }\href {\doibase 10.1088/1402-4896/aa93a8} {\bibfield
  {journal} {\bibinfo  {journal} {Phys. Scripta}\ }\textbf {\bibinfo {volume}
  {92}},\ \bibinfo {pages} {125001} (\bibinfo {year} {2017})},\ \Eprint
  {http://arxiv.org/abs/1511.07058} {arXiv:1511.07058 [gr-qc]} \BibitemShut
  {NoStop}%
\bibitem [{\citenamefont {Nash}(2023)}]{Nash:2023zza}%
  \BibitemOpen
  \bibfield  {author} {\bibinfo {author} {\bibfnamefont {G.}~\bibnamefont
  {Nash}},\ }\href {\doibase 10.1142/S0218271823500311} {\bibfield  {journal}
  {\bibinfo  {journal} {Int. J. Mod. Phys. D}\ }\textbf {\bibinfo {volume}
  {32}},\ \bibinfo {pages} {2350031} (\bibinfo {year} {2023})},\ \Eprint
  {http://arxiv.org/abs/2304.09671} {arXiv:2304.09671 [gr-qc]} \BibitemShut
  {NoStop}%
\bibitem [{\citenamefont {Nash}(2025)}]{Nash:2025xhe}%
  \BibitemOpen
  \bibfield  {author} {\bibinfo {author} {\bibfnamefont {G.}~\bibnamefont
  {Nash}},\ }\href {\doibase 10.1142/S0217751X25500058} {\bibfield  {journal}
  {\bibinfo  {journal} {Int. J. Mod. Phys. A}\ }\textbf {\bibinfo {volume}
  {40}},\ \bibinfo {pages} {2550005} (\bibinfo {year} {2025})},\ \Eprint
  {http://arxiv.org/abs/2503.13655} {arXiv:2503.13655 [gr-qc]} \BibitemShut
  {NoStop}%
\bibitem [{\citenamefont {Feng}\ \emph {et~al.}(2025)\citenamefont {Feng},
  \citenamefont {Mukohyama},\ and\ \citenamefont {Carloni}}]{Feng:2025xsi}%
  \BibitemOpen
  \bibfield  {author} {\bibinfo {author} {\bibfnamefont {J.~C.}\ \bibnamefont
  {Feng}}, \bibinfo {author} {\bibfnamefont {S.}~\bibnamefont {Mukohyama}}, \
  and\ \bibinfo {author} {\bibfnamefont {S.}~\bibnamefont {Carloni}},\
  }\href@noop {} {\  (\bibinfo {year} {2025})},\ \Eprint
  {http://arxiv.org/abs/2505.00112} {arXiv:2505.00112 [gr-qc]} \BibitemShut
  {NoStop}%
\bibitem [{\citenamefont {Gomes}\ \emph
  {et~al.}(2023{\natexlab{a}})\citenamefont {Gomes}, \citenamefont
  {Beltr{\'a}n~Jim{\'e}nez},\ and\ \citenamefont {Koivisto}}]{Gomes:2022vrc}%
  \BibitemOpen
  \bibfield  {author} {\bibinfo {author} {\bibfnamefont {D.~A.}\ \bibnamefont
  {Gomes}}, \bibinfo {author} {\bibfnamefont {J.}~\bibnamefont
  {Beltr{\'a}n~Jim{\'e}nez}}, \ and\ \bibinfo {author} {\bibfnamefont {T.~S.}\
  \bibnamefont {Koivisto}},\ }\href {\doibase 10.1103/PhysRevD.107.024044}
  {\bibfield  {journal} {\bibinfo  {journal} {Phys. Rev. D}\ }\textbf {\bibinfo
  {volume} {107}},\ \bibinfo {pages} {024044} (\bibinfo {year}
  {2023}{\natexlab{a}})},\ \Eprint {http://arxiv.org/abs/2205.09716}
  {arXiv:2205.09716 [gr-qc]} \BibitemShut {NoStop}%
\bibitem [{\citenamefont {Gomes}\ \emph
  {et~al.}(2023{\natexlab{b}})\citenamefont {Gomes}, \citenamefont
  {Beltr{\'a}n~Jim{\'e}nez},\ and\ \citenamefont {Koivisto}}]{Gomes:2023hyk}%
  \BibitemOpen
  \bibfield  {author} {\bibinfo {author} {\bibfnamefont {D.~A.}\ \bibnamefont
  {Gomes}}, \bibinfo {author} {\bibfnamefont {J.}~\bibnamefont
  {Beltr{\'a}n~Jim{\'e}nez}}, \ and\ \bibinfo {author} {\bibfnamefont {T.~S.}\
  \bibnamefont {Koivisto}},\ }\href {\doibase 10.1088/1475-7516/2023/12/010}
  {\bibfield  {journal} {\bibinfo  {journal} {JCAP}\ }\textbf {\bibinfo
  {volume} {12}},\ \bibinfo {pages} {010} (\bibinfo {year}
  {2023}{\natexlab{b}})},\ \Eprint {http://arxiv.org/abs/2309.08554}
  {arXiv:2309.08554 [gr-qc]} \BibitemShut {NoStop}%
\bibitem [{\citenamefont {Krasnov}(2020)}]{Krasnov:2020lku}%
  \BibitemOpen
  \bibfield  {author} {\bibinfo {author} {\bibfnamefont {K.}~\bibnamefont
  {Krasnov}},\ }\href {\doibase 10.1017/9781108674652} {\emph {\bibinfo {title}
  {{Formulations of General Relativity}}}},\ Cambridge Monographs on
  Mathematical Physics\ (\bibinfo  {publisher} {Cambridge University Press},\
  \bibinfo {year} {2020})\BibitemShut {NoStop}%
\bibitem [{\citenamefont {Nikjoo}\ and\ \citenamefont
  {Zlosnik}(2024)}]{Nikjoo:2023flm}%
  \BibitemOpen
  \bibfield  {author} {\bibinfo {author} {\bibfnamefont {M.}~\bibnamefont
  {Nikjoo}}\ and\ \bibinfo {author} {\bibfnamefont {T.}~\bibnamefont
  {Zlosnik}},\ }\href {\doibase 10.1088/1361-6382/ad1c84} {\bibfield  {journal}
  {\bibinfo  {journal} {Class. Quant. Grav.}\ }\textbf {\bibinfo {volume}
  {41}},\ \bibinfo {pages} {045005} (\bibinfo {year} {2024})},\ \Eprint
  {http://arxiv.org/abs/2308.01108} {arXiv:2308.01108 [gr-qc]} \BibitemShut
  {NoStop}%
\bibitem [{\citenamefont {Glavan}\ \emph {et~al.}(2024)\citenamefont {Glavan},
  \citenamefont {Noris},\ and\ \citenamefont {Zlosnik}}]{Glavan:2024svx}%
  \BibitemOpen
  \bibfield  {author} {\bibinfo {author} {\bibfnamefont {D.}~\bibnamefont
  {Glavan}}, \bibinfo {author} {\bibfnamefont {R.}~\bibnamefont {Noris}}, \
  and\ \bibinfo {author} {\bibfnamefont {T.}~\bibnamefont {Zlosnik}},\ }\href
  {\doibase 10.1103/PhysRevD.110.125011} {\bibfield  {journal} {\bibinfo
  {journal} {Phys. Rev. D}\ }\textbf {\bibinfo {volume} {110}},\ \bibinfo
  {pages} {125011} (\bibinfo {year} {2024})},\ \Eprint
  {http://arxiv.org/abs/2408.02763} {arXiv:2408.02763 [hep-th]} \BibitemShut
  {NoStop}%
\bibitem [{\citenamefont {Biermann}(1978)}]{keyfigure}%
  \BibitemOpen
  \bibfield  {author} {\bibinfo {author} {\bibfnamefont {K.-R.}\ \bibnamefont
  {Biermann}},\ }\href@noop {} {\bibfield  {journal} {\bibinfo  {journal}
  {Leopoldina, Mitteilungen der Deutschen Akademie der Naturforscher
  Leopoldina}\ }\textbf {\bibinfo {volume} {3}},\ \bibinfo {pages} {137}
  (\bibinfo {year} {1978})}\BibitemShut {NoStop}%
\bibitem [{\citenamefont {Ülo Lumiste}(1997)}]{LUMISTE199746}%
  \BibitemOpen
  \bibfield  {author} {\bibinfo {author} {\bibnamefont {Ülo Lumiste}},\ }\href
  {\doibase https://doi.org/10.1006/hmat.1997.2123} {\bibfield  {journal}
  {\bibinfo  {journal} {Historia Mathematica}\ }\textbf {\bibinfo {volume}
  {24}},\ \bibinfo {pages} {46} (\bibinfo {year} {1997})}\BibitemShut {NoStop}%
\bibitem [{\citenamefont {Akivis}\ and\ \citenamefont
  {Rosenfeld}(2011)}]{akivis2011elie}%
  \BibitemOpen
  \bibfield  {author} {\bibinfo {author} {\bibfnamefont {M.}~\bibnamefont
  {Akivis}}\ and\ \bibinfo {author} {\bibfnamefont {B.}~\bibnamefont
  {Rosenfeld}},\ }\href {https://books.google.ca/books?id=_8eaAwAAQBAJ} {\emph
  {\bibinfo {title} {Elie Cartan (1869-1951)}}},\ Translations of Mathematical
  Monographs\ (\bibinfo  {publisher} {American Mathematical Society},\ \bibinfo
  {year} {2011})\BibitemShut {NoStop}%
\bibitem [{\citenamefont {Volovik}(2021)}]{Volovik:2020rjz}%
  \BibitemOpen
  \bibfield  {author} {\bibinfo {author} {\bibfnamefont {G.~E.}\ \bibnamefont
  {Volovik}},\ }\href {\doibase 10.1134/S106377612104021X} {\bibfield
  {journal} {\bibinfo  {journal} {J. Exp. Theor. Phys.}\ }\textbf {\bibinfo
  {volume} {132}},\ \bibinfo {pages} {727} (\bibinfo {year} {2021})},\ \Eprint
  {http://arxiv.org/abs/2006.16821} {arXiv:2006.16821 [gr-qc]} \BibitemShut
  {NoStop}%
\bibitem [{\citenamefont {Wiesendanger}(2018)}]{Wiesendanger:2018dzw}%
  \BibitemOpen
  \bibfield  {author} {\bibinfo {author} {\bibfnamefont {C.}~\bibnamefont
  {Wiesendanger}},\ }\href {\doibase 10.1088/1361-6382/ab04e9} {\bibfield
  {journal} {\bibinfo  {journal} {Class. Quant. Grav.}\ }\textbf {\bibinfo
  {volume} {36}},\ \bibinfo {pages} {065015} (\bibinfo {year} {2018})},\
  \Eprint {http://arxiv.org/abs/1806.05037} {arXiv:1806.05037 [gr-qc]}
  \BibitemShut {NoStop}%
\bibitem [{\citenamefont {Wiesendanger}(2020)}]{Wiesendanger:2020lwa}%
  \BibitemOpen
  \bibfield  {author} {\bibinfo {author} {\bibfnamefont {C.}~\bibnamefont
  {Wiesendanger}},\ }\href {\doibase 10.1088/1361-6382/aba80b} {\bibfield
  {journal} {\bibinfo  {journal} {Class. Quant. Grav.}\ }\textbf {\bibinfo
  {volume} {37}},\ \bibinfo {pages} {195029} (\bibinfo {year}
  {2020})}\BibitemShut {NoStop}%
\bibitem [{\citenamefont {Wiesendanger}(2024)}]{Wiesendanger:2024dnp}%
  \BibitemOpen
  \bibfield  {author} {\bibinfo {author} {\bibfnamefont {C.}~\bibnamefont
  {Wiesendanger}},\ }\href@noop {} {\  (\bibinfo {year} {2024})},\ \Eprint
  {http://arxiv.org/abs/2405.03719} {arXiv:2405.03719 [gr-qc]} \BibitemShut
  {NoStop}%
\bibitem [{\citenamefont {Koivisto}\ and\ \citenamefont
  {Zheng}(2025)}]{Koivisto:2024asr}%
  \BibitemOpen
  \bibfield  {author} {\bibinfo {author} {\bibfnamefont {T.~S.}\ \bibnamefont
  {Koivisto}}\ and\ \bibinfo {author} {\bibfnamefont {L.}~\bibnamefont
  {Zheng}},\ }\href {\doibase 10.1103/PhysRevD.111.064008} {\bibfield
  {journal} {\bibinfo  {journal} {Phys. Rev. D}\ }\textbf {\bibinfo {volume}
  {111}},\ \bibinfo {pages} {064008} (\bibinfo {year} {2025})},\ \Eprint
  {http://arxiv.org/abs/2408.10100} {arXiv:2408.10100 [gr-qc]} \BibitemShut
  {NoStop}%
\bibitem [{\citenamefont {Gorji}\ \emph {et~al.}(2020)\citenamefont {Gorji},
  \citenamefont {Allahyari}, \citenamefont {Khodadi},\ and\ \citenamefont
  {Firouzjahi}}]{Gorji:2020ten}%
  \BibitemOpen
  \bibfield  {author} {\bibinfo {author} {\bibfnamefont {M.~A.}\ \bibnamefont
  {Gorji}}, \bibinfo {author} {\bibfnamefont {A.}~\bibnamefont {Allahyari}},
  \bibinfo {author} {\bibfnamefont {M.}~\bibnamefont {Khodadi}}, \ and\
  \bibinfo {author} {\bibfnamefont {H.}~\bibnamefont {Firouzjahi}},\ }\href
  {\doibase 10.1103/PhysRevD.101.124060} {\bibfield  {journal} {\bibinfo
  {journal} {Phys. Rev. D}\ }\textbf {\bibinfo {volume} {101}},\ \bibinfo
  {pages} {124060} (\bibinfo {year} {2020})},\ \Eprint
  {http://arxiv.org/abs/1912.04636} {arXiv:1912.04636 [gr-qc]} \BibitemShut
  {NoStop}%
\bibitem [{\citenamefont {Pessers}\ and\ \citenamefont {Van~der
  Veken}(2016)}]{Pessers_2016}%
  \BibitemOpen
  \bibfield  {author} {\bibinfo {author} {\bibfnamefont {V.}~\bibnamefont
  {Pessers}}\ and\ \bibinfo {author} {\bibfnamefont {J.}~\bibnamefont {Van~der
  Veken}},\ }\href {\doibase 10.1016/j.geomphys.2016.02.009} {\bibfield
  {journal} {\bibinfo  {journal} {Journal of Geometry and Physics}\ }\textbf
  {\bibinfo {volume} {104}},\ \bibinfo {pages} {163–174} (\bibinfo {year}
  {2016})}\BibitemShut {NoStop}%
\bibitem [{\citenamefont {Helleland}\ and\ \citenamefont
  {Hervik}(2018)}]{Helleland:2015wva}%
  \BibitemOpen
  \bibfield  {author} {\bibinfo {author} {\bibfnamefont {C.}~\bibnamefont
  {Helleland}}\ and\ \bibinfo {author} {\bibfnamefont {S.}~\bibnamefont
  {Hervik}},\ }\href {\doibase 10.1016/j.geomphys.2017.09.015} {\bibfield
  {journal} {\bibinfo  {journal} {J. Geom. Phys.}\ }\textbf {\bibinfo {volume}
  {123}},\ \bibinfo {pages} {424} (\bibinfo {year} {2018})},\ \Eprint
  {http://arxiv.org/abs/1504.01244} {arXiv:1504.01244 [math-ph]} \BibitemShut
  {NoStop}%
\bibitem [{\citenamefont {Gibbons}\ and\ \citenamefont
  {Hawking}(1977)}]{Gibbons:1976ue}%
  \BibitemOpen
  \bibfield  {author} {\bibinfo {author} {\bibfnamefont {G.~W.}\ \bibnamefont
  {Gibbons}}\ and\ \bibinfo {author} {\bibfnamefont {S.~W.}\ \bibnamefont
  {Hawking}},\ }\href {\doibase 10.1103/PhysRevD.15.2752} {\bibfield  {journal}
  {\bibinfo  {journal} {Phys. Rev. D}\ }\textbf {\bibinfo {volume} {15}},\
  \bibinfo {pages} {2752} (\bibinfo {year} {1977})}\BibitemShut {NoStop}%
\bibitem [{\citenamefont {Novello}\ and\ \citenamefont
  {Bittencourt}(2011)}]{Novello:2010af}%
  \BibitemOpen
  \bibfield  {author} {\bibinfo {author} {\bibfnamefont {M.}~\bibnamefont
  {Novello}}\ and\ \bibinfo {author} {\bibfnamefont {E.}~\bibnamefont
  {Bittencourt}},\ }\href {\doibase 10.1134/S0202289311030054} {\bibfield
  {journal} {\bibinfo  {journal} {Grav. Cosmol.}\ }\textbf {\bibinfo {volume}
  {17}},\ \bibinfo {pages} {230} (\bibinfo {year} {2011})},\ \Eprint
  {http://arxiv.org/abs/1004.3913} {arXiv:1004.3913 [gr-qc]} \BibitemShut
  {NoStop}%
\bibitem [{\citenamefont {Sorge}(2021)}]{Sorge:2021nqm}%
  \BibitemOpen
  \bibfield  {author} {\bibinfo {author} {\bibfnamefont {F.}~\bibnamefont
  {Sorge}},\ }\href@noop {} {\  (\bibinfo {year} {2021})},\ \Eprint
  {http://arxiv.org/abs/2112.15441} {arXiv:2112.15441 [gr-qc]} \BibitemShut
  {NoStop}%
\bibitem [{\citenamefont {Khatsymovsky}(2021)}]{Khatsymovsky:2021vjq}%
  \BibitemOpen
  \bibfield  {author} {\bibinfo {author} {\bibfnamefont {V.~M.}\ \bibnamefont
  {Khatsymovsky}},\ }\href {\doibase 10.1142/S0218271821500711} {\bibfield
  {journal} {\bibinfo  {journal} {Int. J. Mod. Phys. D}\ }\textbf {\bibinfo
  {volume} {30}},\ \bibinfo {pages} {2150071} (\bibinfo {year} {2021})},\
  \Eprint {http://arxiv.org/abs/2101.07147} {arXiv:2101.07147 [gr-qc]}
  \BibitemShut {NoStop}%
\bibitem [{\citenamefont {Doran}(2000)}]{Doran:1999gb}%
  \BibitemOpen
  \bibfield  {author} {\bibinfo {author} {\bibfnamefont {C.}~\bibnamefont
  {Doran}},\ }\href {\doibase 10.1103/PhysRevD.61.067503} {\bibfield  {journal}
  {\bibinfo  {journal} {Phys. Rev. D}\ }\textbf {\bibinfo {volume} {61}},\
  \bibinfo {pages} {067503} (\bibinfo {year} {2000})},\ \Eprint
  {http://arxiv.org/abs/gr-qc/9910099} {arXiv:gr-qc/9910099} \BibitemShut
  {NoStop}%
\bibitem [{\citenamefont {Izaurieta}\ \emph {et~al.}(2020)\citenamefont
  {Izaurieta}, \citenamefont {Lepe},\ and\ \citenamefont
  {Valdivia}}]{Izaurieta:2020xpk}%
  \BibitemOpen
  \bibfield  {author} {\bibinfo {author} {\bibfnamefont {F.}~\bibnamefont
  {Izaurieta}}, \bibinfo {author} {\bibfnamefont {S.}~\bibnamefont {Lepe}}, \
  and\ \bibinfo {author} {\bibfnamefont {O.}~\bibnamefont {Valdivia}},\ }\href
  {\doibase 10.1016/j.dark.2020.100662} {\bibfield  {journal} {\bibinfo
  {journal} {Phys. Dark Univ.}\ }\textbf {\bibinfo {volume} {30}},\ \bibinfo
  {pages} {100662} (\bibinfo {year} {2020})},\ \Eprint
  {http://arxiv.org/abs/2004.13163} {arXiv:2004.13163 [gr-qc]} \BibitemShut
  {NoStop}%
\bibitem [{\citenamefont {Elizalde}\ \emph {et~al.}(2023)\citenamefont
  {Elizalde}, \citenamefont {Izaurieta}, \citenamefont {Riveros}, \citenamefont
  {Salgado},\ and\ \citenamefont {Valdivia}}]{Elizalde:2022vvc}%
  \BibitemOpen
  \bibfield  {author} {\bibinfo {author} {\bibfnamefont {E.}~\bibnamefont
  {Elizalde}}, \bibinfo {author} {\bibfnamefont {F.}~\bibnamefont {Izaurieta}},
  \bibinfo {author} {\bibfnamefont {C.}~\bibnamefont {Riveros}}, \bibinfo
  {author} {\bibfnamefont {G.}~\bibnamefont {Salgado}}, \ and\ \bibinfo
  {author} {\bibfnamefont {O.}~\bibnamefont {Valdivia}},\ }\href {\doibase
  10.1016/j.dark.2023.101197} {\bibfield  {journal} {\bibinfo  {journal} {Phys.
  Dark Univ.}\ }\textbf {\bibinfo {volume} {40}},\ \bibinfo {pages} {101197}
  (\bibinfo {year} {2023})},\ \Eprint {http://arxiv.org/abs/2204.00090}
  {arXiv:2204.00090 [gr-qc]} \BibitemShut {NoStop}%
\bibitem [{\citenamefont {Barriga}\ \emph {et~al.}(2025)\citenamefont
  {Barriga}, \citenamefont {Izaurieta}, \citenamefont {Lepe}, \citenamefont
  {Meza}, \citenamefont {Mu{\~n}oz}, \citenamefont {Quinzacara},\ and\
  \citenamefont {Valdivia}}]{Barriga:2024hpe}%
  \BibitemOpen
  \bibfield  {author} {\bibinfo {author} {\bibfnamefont {F.}~\bibnamefont
  {Barriga}}, \bibinfo {author} {\bibfnamefont {F.}~\bibnamefont {Izaurieta}},
  \bibinfo {author} {\bibfnamefont {S.}~\bibnamefont {Lepe}}, \bibinfo {author}
  {\bibfnamefont {P.}~\bibnamefont {Meza}}, \bibinfo {author} {\bibfnamefont
  {J.}~\bibnamefont {Mu{\~n}oz}}, \bibinfo {author} {\bibfnamefont
  {C.}~\bibnamefont {Quinzacara}}, \ and\ \bibinfo {author} {\bibfnamefont
  {O.}~\bibnamefont {Valdivia}},\ }\href {\doibase
  10.1088/1475-7516/2025/02/003} {\bibfield  {journal} {\bibinfo  {journal}
  {JCAP}\ }\textbf {\bibinfo {volume} {02}},\ \bibinfo {pages} {003} (\bibinfo
  {year} {2025})},\ \Eprint {http://arxiv.org/abs/2409.15509} {arXiv:2409.15509
  [gr-qc]} \BibitemShut {NoStop}%
\bibitem [{\citenamefont {Barbero~G.}(1995)}]{BarberoG:1994eia}%
  \BibitemOpen
  \bibfield  {author} {\bibinfo {author} {\bibfnamefont {J.~F.}\ \bibnamefont
  {Barbero~G.}},\ }\href {\doibase 10.1103/PhysRevD.51.5507} {\bibfield
  {journal} {\bibinfo  {journal} {Phys. Rev. D}\ }\textbf {\bibinfo {volume}
  {51}},\ \bibinfo {pages} {5507} (\bibinfo {year} {1995})},\ \Eprint
  {http://arxiv.org/abs/gr-qc/9410014} {arXiv:gr-qc/9410014} \BibitemShut
  {NoStop}%
\bibitem [{\citenamefont {Immirzi}(1997)}]{Immirzi:1996di}%
  \BibitemOpen
  \bibfield  {author} {\bibinfo {author} {\bibfnamefont {G.}~\bibnamefont
  {Immirzi}},\ }\href {\doibase 10.1088/0264-9381/14/10/002} {\bibfield
  {journal} {\bibinfo  {journal} {Class. Quant. Grav.}\ }\textbf {\bibinfo
  {volume} {14}},\ \bibinfo {pages} {L177} (\bibinfo {year} {1997})},\ \Eprint
  {http://arxiv.org/abs/gr-qc/9612030} {arXiv:gr-qc/9612030} \BibitemShut
  {NoStop}%
\bibitem [{\citenamefont {Aldrovandi}\ \emph {et~al.}(2004)\citenamefont
  {Aldrovandi}, \citenamefont {Arcos},\ and\ \citenamefont
  {Pereira}}]{Aldrovandi:2004uz}%
  \BibitemOpen
  \bibfield  {author} {\bibinfo {author} {\bibfnamefont {R.}~\bibnamefont
  {Aldrovandi}}, \bibinfo {author} {\bibfnamefont {H.~I.}\ \bibnamefont
  {Arcos}}, \ and\ \bibinfo {author} {\bibfnamefont {J.~G.}\ \bibnamefont
  {Pereira}},\ }in\ \href@noop {} {\emph {\bibinfo {booktitle} {{Colloquium on
  Theory and Experiment in Cosmology and Gravitation in Honor of Prof. Jose
  Plinio Baptista on the Occasion of his 70th Birthday}}}}\ (\bibinfo {year}
  {2004})\ \Eprint {http://arxiv.org/abs/gr-qc/0412032} {arXiv:gr-qc/0412032}
  \BibitemShut {NoStop}%
\bibitem [{\citenamefont {Diakonov}(2011)}]{Diakonov:2011im}%
  \BibitemOpen
  \bibfield  {author} {\bibinfo {author} {\bibfnamefont {D.}~\bibnamefont
  {Diakonov}},\ }\href@noop {} {\  (\bibinfo {year} {2011})},\ \Eprint
  {http://arxiv.org/abs/1109.0091} {arXiv:1109.0091 [hep-th]} \BibitemShut
  {NoStop}%
\bibitem [{\citenamefont {Obukhov}\ and\ \citenamefont
  {Hehl}(2012)}]{Obukhov:2012je}%
  \BibitemOpen
  \bibfield  {author} {\bibinfo {author} {\bibfnamefont {Y.~N.}\ \bibnamefont
  {Obukhov}}\ and\ \bibinfo {author} {\bibfnamefont {F.~W.}\ \bibnamefont
  {Hehl}},\ }\href {\doibase 10.1016/j.physletb.2012.06.005} {\bibfield
  {journal} {\bibinfo  {journal} {Phys. Lett. B}\ }\textbf {\bibinfo {volume}
  {713}},\ \bibinfo {pages} {321} (\bibinfo {year} {2012})},\ \Eprint
  {http://arxiv.org/abs/1202.6045} {arXiv:1202.6045 [gr-qc]} \BibitemShut
  {NoStop}%
\bibitem [{\citenamefont {Vladimirov}\ and\ \citenamefont
  {Diakonov}(2014)}]{Vladimirov:2014gma}%
  \BibitemOpen
  \bibfield  {author} {\bibinfo {author} {\bibfnamefont {A.~A.}\ \bibnamefont
  {Vladimirov}}\ and\ \bibinfo {author} {\bibfnamefont {D.}~\bibnamefont
  {Diakonov}},\ }\href {\doibase 10.1134/S1063779614040145} {\bibfield
  {journal} {\bibinfo  {journal} {Phys. Part. Nucl.}\ }\textbf {\bibinfo
  {volume} {45}},\ \bibinfo {pages} {800} (\bibinfo {year} {2014})}\BibitemShut
  {NoStop}%
\bibitem [{\citenamefont {Vergeles}(2021)}]{Vergeles:2019xfh}%
  \BibitemOpen
  \bibfield  {author} {\bibinfo {author} {\bibfnamefont {S.~N.}\ \bibnamefont
  {Vergeles}},\ }\href {\doibase 10.1088/1361-6382/abebb5} {\bibfield
  {journal} {\bibinfo  {journal} {Class. Quant. Grav.}\ }\textbf {\bibinfo
  {volume} {38}},\ \bibinfo {pages} {085022} (\bibinfo {year} {2021})},\
  \Eprint {http://arxiv.org/abs/1903.09957} {arXiv:1903.09957 [hep-lat]}
  \BibitemShut {NoStop}%
\bibitem [{\citenamefont {Volovik}(2024)}]{Volovik:2024iiy}%
  \BibitemOpen
  \bibfield  {author} {\bibinfo {author} {\bibfnamefont {G.~E.}\ \bibnamefont
  {Volovik}},\ }\href {\doibase 10.3390/sym16091131} {\bibfield  {journal}
  {\bibinfo  {journal} {Symmetry}\ }\textbf {\bibinfo {volume} {16}},\ \bibinfo
  {pages} {1131} (\bibinfo {year} {2024})},\ \Eprint
  {http://arxiv.org/abs/2406.00718} {arXiv:2406.00718 [cond-mat.other]}
  \BibitemShut {NoStop}%
\bibitem [{\citenamefont {Aoki}\ \emph {et~al.}(2024)\citenamefont {Aoki},
  \citenamefont {Bahamonde}, \citenamefont {Gigante~Valcarcel},\ and\
  \citenamefont {Gorji}}]{Aoki:2023sum}%
  \BibitemOpen
  \bibfield  {author} {\bibinfo {author} {\bibfnamefont {K.}~\bibnamefont
  {Aoki}}, \bibinfo {author} {\bibfnamefont {S.}~\bibnamefont {Bahamonde}},
  \bibinfo {author} {\bibfnamefont {J.}~\bibnamefont {Gigante~Valcarcel}}, \
  and\ \bibinfo {author} {\bibfnamefont {M.~A.}\ \bibnamefont {Gorji}},\ }\href
  {\doibase 10.1103/PhysRevD.110.024017} {\bibfield  {journal} {\bibinfo
  {journal} {Phys. Rev. D}\ }\textbf {\bibinfo {volume} {110}},\ \bibinfo
  {pages} {024017} (\bibinfo {year} {2024})},\ \Eprint
  {http://arxiv.org/abs/2310.16007} {arXiv:2310.16007 [gr-qc]} \BibitemShut
  {NoStop}%
\bibitem [{\citenamefont {Castillo-Felisola}\ \emph {et~al.}(2024)\citenamefont
  {Castillo-Felisola}, \citenamefont {Gannouji}, \citenamefont
  {Morocho-L{\'o}pez},\ and\ \citenamefont
  {Rozas-Rojas}}]{Castillo-Felisola:2024atv}%
  \BibitemOpen
  \bibfield  {author} {\bibinfo {author} {\bibfnamefont {O.}~\bibnamefont
  {Castillo-Felisola}}, \bibinfo {author} {\bibfnamefont {R.}~\bibnamefont
  {Gannouji}}, \bibinfo {author} {\bibfnamefont {M.}~\bibnamefont
  {Morocho-L{\'o}pez}}, \ and\ \bibinfo {author} {\bibfnamefont
  {M.}~\bibnamefont {Rozas-Rojas}},\ }\href {\doibase
  10.1103/PhysRevD.110.023522} {\bibfield  {journal} {\bibinfo  {journal}
  {Phys. Rev. D}\ }\textbf {\bibinfo {volume} {110}},\ \bibinfo {pages}
  {023522} (\bibinfo {year} {2024})},\ \Eprint
  {http://arxiv.org/abs/2401.08198} {arXiv:2401.08198 [gr-qc]} \BibitemShut
  {NoStop}%
\bibitem [{\citenamefont {Iosifidis}\ \emph {et~al.}(2024)\citenamefont
  {Iosifidis}, \citenamefont {Jensko},\ and\ \citenamefont
  {Koivisto}}]{Iosifidis:2024ksa}%
  \BibitemOpen
  \bibfield  {author} {\bibinfo {author} {\bibfnamefont {D.}~\bibnamefont
  {Iosifidis}}, \bibinfo {author} {\bibfnamefont {E.}~\bibnamefont {Jensko}}, \
  and\ \bibinfo {author} {\bibfnamefont {T.~S.}\ \bibnamefont {Koivisto}},\
  }\href {\doibase 10.1088/1475-7516/2024/11/043} {\bibfield  {journal}
  {\bibinfo  {journal} {JCAP}\ }\textbf {\bibinfo {volume} {11}},\ \bibinfo
  {pages} {043} (\bibinfo {year} {2024})},\ \Eprint
  {http://arxiv.org/abs/2406.01412} {arXiv:2406.01412 [gr-qc]} \BibitemShut
  {NoStop}%
\bibitem [{\citenamefont {Hu}(1998)}]{Hu:1998kj}%
  \BibitemOpen
  \bibfield  {author} {\bibinfo {author} {\bibfnamefont {W.}~\bibnamefont
  {Hu}},\ }\href {\doibase 10.1086/306274} {\bibfield  {journal} {\bibinfo
  {journal} {Astrophys. J.}\ }\textbf {\bibinfo {volume} {506}},\ \bibinfo
  {pages} {485} (\bibinfo {year} {1998})},\ \Eprint
  {http://arxiv.org/abs/astro-ph/9801234} {arXiv:astro-ph/9801234} \BibitemShut
  {NoStop}%
\bibitem [{\citenamefont {Koivisto}\ and\ \citenamefont
  {Mota}(2006)}]{Koivisto:2005mm}%
  \BibitemOpen
  \bibfield  {author} {\bibinfo {author} {\bibfnamefont {T.}~\bibnamefont
  {Koivisto}}\ and\ \bibinfo {author} {\bibfnamefont {D.~F.}\ \bibnamefont
  {Mota}},\ }\href {\doibase 10.1103/PhysRevD.73.083502} {\bibfield  {journal}
  {\bibinfo  {journal} {Phys. Rev. D}\ }\textbf {\bibinfo {volume} {73}},\
  \bibinfo {pages} {083502} (\bibinfo {year} {2006})},\ \Eprint
  {http://arxiv.org/abs/astro-ph/0512135} {arXiv:astro-ph/0512135} \BibitemShut
  {NoStop}%
\bibitem [{\citenamefont {Mota}\ \emph {et~al.}(2007)\citenamefont {Mota},
  \citenamefont {Kristiansen}, \citenamefont {Koivisto},\ and\ \citenamefont
  {Groeneboom}}]{Mota:2007sz}%
  \BibitemOpen
  \bibfield  {author} {\bibinfo {author} {\bibfnamefont {D.~F.}\ \bibnamefont
  {Mota}}, \bibinfo {author} {\bibfnamefont {J.~R.}\ \bibnamefont
  {Kristiansen}}, \bibinfo {author} {\bibfnamefont {T.}~\bibnamefont
  {Koivisto}}, \ and\ \bibinfo {author} {\bibfnamefont {N.~E.}\ \bibnamefont
  {Groeneboom}},\ }\href {\doibase 10.1111/j.1365-2966.2007.12413.x} {\bibfield
   {journal} {\bibinfo  {journal} {Mon. Not. Roy. Astron. Soc.}\ }\textbf
  {\bibinfo {volume} {382}},\ \bibinfo {pages} {793} (\bibinfo {year}
  {2007})},\ \Eprint {http://arxiv.org/abs/0708.0830} {arXiv:0708.0830
  [astro-ph]} \BibitemShut {NoStop}%
\bibitem [{\citenamefont {Amendola}\ \emph {et~al.}(2008)\citenamefont
  {Amendola}, \citenamefont {Kunz},\ and\ \citenamefont
  {Sapone}}]{Amendola:2007rr}%
  \BibitemOpen
  \bibfield  {author} {\bibinfo {author} {\bibfnamefont {L.}~\bibnamefont
  {Amendola}}, \bibinfo {author} {\bibfnamefont {M.}~\bibnamefont {Kunz}}, \
  and\ \bibinfo {author} {\bibfnamefont {D.}~\bibnamefont {Sapone}},\ }\href
  {\doibase 10.1088/1475-7516/2008/04/013} {\bibfield  {journal} {\bibinfo
  {journal} {JCAP}\ }\textbf {\bibinfo {volume} {04}},\ \bibinfo {pages} {013}
  (\bibinfo {year} {2008})},\ \Eprint {http://arxiv.org/abs/0704.2421}
  {arXiv:0704.2421 [astro-ph]} \BibitemShut {NoStop}%
\bibitem [{\citenamefont {Daniel}\ \emph {et~al.}(2008)\citenamefont {Daniel},
  \citenamefont {Caldwell}, \citenamefont {Cooray},\ and\ \citenamefont
  {Melchiorri}}]{Daniel:2008et}%
  \BibitemOpen
  \bibfield  {author} {\bibinfo {author} {\bibfnamefont {S.~F.}\ \bibnamefont
  {Daniel}}, \bibinfo {author} {\bibfnamefont {R.~R.}\ \bibnamefont
  {Caldwell}}, \bibinfo {author} {\bibfnamefont {A.}~\bibnamefont {Cooray}}, \
  and\ \bibinfo {author} {\bibfnamefont {A.}~\bibnamefont {Melchiorri}},\
  }\href {\doibase 10.1103/PhysRevD.77.103513} {\bibfield  {journal} {\bibinfo
  {journal} {Phys. Rev. D}\ }\textbf {\bibinfo {volume} {77}},\ \bibinfo
  {pages} {103513} (\bibinfo {year} {2008})},\ \Eprint
  {http://arxiv.org/abs/0802.1068} {arXiv:0802.1068 [astro-ph]} \BibitemShut
  {NoStop}%
\bibitem [{\citenamefont {Di~Valentino}\ \emph {et~al.}(2025)\citenamefont
  {Di~Valentino} \emph {et~al.}}]{CosmoVerse:2025txj}%
  \BibitemOpen
  \bibfield  {author} {\bibinfo {author} {\bibfnamefont {E.}~\bibnamefont
  {Di~Valentino}} \emph {et~al.} (\bibinfo {collaboration} {CosmoVerse}),\
  }\href@noop {} {\  (\bibinfo {year} {2025})},\ \Eprint
  {http://arxiv.org/abs/2504.01669} {arXiv:2504.01669 [astro-ph.CO]}
  \BibitemShut {NoStop}%
\bibitem [{\citenamefont {{\"O}vg{\"u}n}\ and\ \citenamefont
  {Fathi}(2025)}]{Ovgun:2025stp}%
  \BibitemOpen
  \bibfield  {author} {\bibinfo {author} {\bibfnamefont {A.}~\bibnamefont
  {{\"O}vg{\"u}n}}\ and\ \bibinfo {author} {\bibfnamefont {M.}~\bibnamefont
  {Fathi}},\ }\href@noop {} {\  (\bibinfo {year} {2025})},\ \Eprint
  {http://arxiv.org/abs/2504.04331} {arXiv:2504.04331 [gr-qc]} \BibitemShut
  {NoStop}%
\bibitem [{\citenamefont {Umarov}\ \emph {et~al.}(2025)\citenamefont {Umarov},
  \citenamefont {Atamurotov}, \citenamefont {Abdujabbarov},\ and\ \citenamefont
  {{\"O}vg{\"u}n}}]{Umarov:2025ihy}%
  \BibitemOpen
  \bibfield  {author} {\bibinfo {author} {\bibfnamefont {D.}~\bibnamefont
  {Umarov}}, \bibinfo {author} {\bibfnamefont {F.}~\bibnamefont {Atamurotov}},
  \bibinfo {author} {\bibfnamefont {A.}~\bibnamefont {Abdujabbarov}}, \ and\
  \bibinfo {author} {\bibfnamefont {A.}~\bibnamefont {{\"O}vg{\"u}n}},\ }\href
  {\doibase 10.1016/j.dark.2025.101945} {\bibfield  {journal} {\bibinfo
  {journal} {Phys. Dark Univ.}\ }\textbf {\bibinfo {volume} {48}},\ \bibinfo
  {pages} {101945} (\bibinfo {year} {2025})}\BibitemShut {NoStop}%
\bibitem [{\citenamefont {Schwinger}(1959)}]{Schwinger:1959zz}%
  \BibitemOpen
  \bibfield  {author} {\bibinfo {author} {\bibfnamefont {J.}~\bibnamefont
  {Schwinger}},\ }\href {\doibase 10.1103/PhysRev.115.721} {\bibfield
  {journal} {\bibinfo  {journal} {Phys. Rev.}\ }\textbf {\bibinfo {volume}
  {115}},\ \bibinfo {pages} {721} (\bibinfo {year} {1959})}\BibitemShut
  {NoStop}%
\bibitem [{\citenamefont {van Nieuwenhuizen}\ and\ \citenamefont
  {Waldron}(1996)}]{vanNieuwenhuizen:1996tv}%
  \BibitemOpen
  \bibfield  {author} {\bibinfo {author} {\bibfnamefont {P.}~\bibnamefont {van
  Nieuwenhuizen}}\ and\ \bibinfo {author} {\bibfnamefont {A.}~\bibnamefont
  {Waldron}},\ }\href {\doibase 10.1016/S0370-2693(96)01251-8} {\bibfield
  {journal} {\bibinfo  {journal} {Phys. Lett. B}\ }\textbf {\bibinfo {volume}
  {389}},\ \bibinfo {pages} {29} (\bibinfo {year} {1996})},\ \Eprint
  {http://arxiv.org/abs/hep-th/9608174} {arXiv:hep-th/9608174} \BibitemShut
  {NoStop}%
\bibitem [{\citenamefont {Wetterich}(2011)}]{Wetterich:2010ni}%
  \BibitemOpen
  \bibfield  {author} {\bibinfo {author} {\bibfnamefont {C.}~\bibnamefont
  {Wetterich}},\ }\href {\doibase 10.1016/j.nuclphysb.2011.06.013} {\bibfield
  {journal} {\bibinfo  {journal} {Nucl. Phys. B}\ }\textbf {\bibinfo {volume}
  {852}},\ \bibinfo {pages} {174} (\bibinfo {year} {2011})},\ \Eprint
  {http://arxiv.org/abs/1002.3556} {arXiv:1002.3556 [hep-th]} \BibitemShut
  {NoStop}%
\bibitem [{\citenamefont {Woit}(2023)}]{Woit:2023idu}%
  \BibitemOpen
  \bibfield  {author} {\bibinfo {author} {\bibfnamefont {P.}~\bibnamefont
  {Woit}},\ }\href@noop {} {\  (\bibinfo {year} {2023})},\ \Eprint
  {http://arxiv.org/abs/2311.00608} {arXiv:2311.00608 [hep-th]} \BibitemShut
  {NoStop}%
\bibitem [{\citenamefont {Gallagher}\ \emph {et~al.}(2024)\citenamefont
  {Gallagher}, \citenamefont {Koivisto}, \citenamefont {Marzola}, \citenamefont
  {Varrin},\ and\ \citenamefont {Zlosnik}}]{Gallagher:2023ghl}%
  \BibitemOpen
  \bibfield  {author} {\bibinfo {author} {\bibfnamefont {P.}~\bibnamefont
  {Gallagher}}, \bibinfo {author} {\bibfnamefont {T.~S.}\ \bibnamefont
  {Koivisto}}, \bibinfo {author} {\bibfnamefont {L.}~\bibnamefont {Marzola}},
  \bibinfo {author} {\bibfnamefont {L.}~\bibnamefont {Varrin}}, \ and\ \bibinfo
  {author} {\bibfnamefont {T.}~\bibnamefont {Zlosnik}},\ }\href {\doibase
  10.1103/PhysRevD.109.L061503} {\bibfield  {journal} {\bibinfo  {journal}
  {Phys. Rev. D}\ }\textbf {\bibinfo {volume} {109}},\ \bibinfo {pages}
  {L061503} (\bibinfo {year} {2024})},\ \Eprint
  {http://arxiv.org/abs/2311.07464} {arXiv:2311.07464 [hep-th]} \BibitemShut
  {NoStop}%
\bibitem [{\citenamefont {Ashtekar}\ \emph {et~al.}(1989)\citenamefont
  {Ashtekar}, \citenamefont {Romano},\ and\ \citenamefont
  {Tate}}]{Ashtekar:1989ju}%
  \BibitemOpen
  \bibfield  {author} {\bibinfo {author} {\bibfnamefont {A.}~\bibnamefont
  {Ashtekar}}, \bibinfo {author} {\bibfnamefont {J.~D.}\ \bibnamefont
  {Romano}}, \ and\ \bibinfo {author} {\bibfnamefont {R.~S.}\ \bibnamefont
  {Tate}},\ }\href {\doibase 10.1103/PhysRevD.40.2572} {\bibfield  {journal}
  {\bibinfo  {journal} {Phys. Rev. D}\ }\textbf {\bibinfo {volume} {40}},\
  \bibinfo {pages} {2572} (\bibinfo {year} {1989})}\BibitemShut {NoStop}%
\bibitem [{\citenamefont {Gallagher}\ \emph {et~al.}(2022)\citenamefont
  {Gallagher}, \citenamefont {Koivisto},\ and\ \citenamefont
  {Marzola}}]{Gallagher:2022kvv}%
  \BibitemOpen
  \bibfield  {author} {\bibinfo {author} {\bibfnamefont {P.}~\bibnamefont
  {Gallagher}}, \bibinfo {author} {\bibfnamefont {T.}~\bibnamefont {Koivisto}},
  \ and\ \bibinfo {author} {\bibfnamefont {L.}~\bibnamefont {Marzola}},\ }\href
  {\doibase 10.1103/PhysRevD.105.125010} {\bibfield  {journal} {\bibinfo
  {journal} {Phys. Rev. D}\ }\textbf {\bibinfo {volume} {105}},\ \bibinfo
  {pages} {125010} (\bibinfo {year} {2022})},\ \Eprint
  {http://arxiv.org/abs/2202.05657} {arXiv:2202.05657 [hep-th]} \BibitemShut
  {NoStop}%
\bibitem [{\citenamefont {Gallagher}(2024)}]{Gallagher:2024haq}%
  \BibitemOpen
  \bibfield  {author} {\bibinfo {author} {\bibfnamefont {P.}~\bibnamefont
  {Gallagher}},\ }\href {\doibase 10.1103/PhysRevD.110.085010} {\bibfield
  {journal} {\bibinfo  {journal} {Phys. Rev. D}\ }\textbf {\bibinfo {volume}
  {110}},\ \bibinfo {pages} {085010} (\bibinfo {year} {2024})},\ \Eprint
  {http://arxiv.org/abs/2403.02578} {arXiv:2403.02578 [hep-th]} \BibitemShut
  {NoStop}%
\bibitem [{\citenamefont {Woit}(2021)}]{Woit:2021bmb}%
  \BibitemOpen
  \bibfield  {author} {\bibinfo {author} {\bibfnamefont {P.}~\bibnamefont
  {Woit}},\ }\href@noop {} {\  (\bibinfo {year} {2021})},\ \Eprint
  {http://arxiv.org/abs/2104.05099} {arXiv:2104.05099 [hep-th]} \BibitemShut
  {NoStop}%
\bibitem [{\citenamefont {Krasnov}\ and\ \citenamefont
  {Percacci}(2018)}]{Krasnov:2017epi}%
  \BibitemOpen
  \bibfield  {author} {\bibinfo {author} {\bibfnamefont {K.}~\bibnamefont
  {Krasnov}}\ and\ \bibinfo {author} {\bibfnamefont {R.}~\bibnamefont
  {Percacci}},\ }\href {\doibase 10.1088/1361-6382/aac58d} {\bibfield
  {journal} {\bibinfo  {journal} {Class. Quant. Grav.}\ }\textbf {\bibinfo
  {volume} {35}},\ \bibinfo {pages} {143001} (\bibinfo {year} {2018})},\
  \Eprint {http://arxiv.org/abs/1712.03061} {arXiv:1712.03061 [hep-th]}
  \BibitemShut {NoStop}%
\bibitem [{\citenamefont {Koivisto}\ and\ \citenamefont
  {Zheng}(2021)}]{Koivisto:2021ofz}%
  \BibitemOpen
  \bibfield  {author} {\bibinfo {author} {\bibfnamefont {T.~S.}\ \bibnamefont
  {Koivisto}}\ and\ \bibinfo {author} {\bibfnamefont {L.}~\bibnamefont
  {Zheng}},\ }\href {\doibase 10.1103/PhysRevD.103.124063} {\bibfield
  {journal} {\bibinfo  {journal} {Phys. Rev. D}\ }\textbf {\bibinfo {volume}
  {103}},\ \bibinfo {pages} {124063} (\bibinfo {year} {2021})},\ \Eprint
  {http://arxiv.org/abs/2101.07638} {arXiv:2101.07638 [gr-qc]} \BibitemShut
  {NoStop}%
\bibitem [{\citenamefont {Koivisto}(2018)}]{Koivisto:2018aip}%
  \BibitemOpen
  \bibfield  {author} {\bibinfo {author} {\bibfnamefont {T.}~\bibnamefont
  {Koivisto}},\ }\href {\doibase 10.1142/S0219887818400066} {\bibfield
  {journal} {\bibinfo  {journal} {Int. J. Geom. Meth. Mod. Phys.}\ }\textbf
  {\bibinfo {volume} {15}},\ \bibinfo {pages} {1840006} (\bibinfo {year}
  {2018})},\ \Eprint {http://arxiv.org/abs/1802.00650} {arXiv:1802.00650
  [gr-qc]} \BibitemShut {NoStop}%
\bibitem [{\citenamefont {Koivisto}\ \emph {et~al.}(2019)\citenamefont
  {Koivisto}, \citenamefont {Hohmann},\ and\ \citenamefont
  {Z{\l}o{\'s}nik}}]{Koivisto:2019ejt}%
  \BibitemOpen
  \bibfield  {author} {\bibinfo {author} {\bibfnamefont {T.}~\bibnamefont
  {Koivisto}}, \bibinfo {author} {\bibfnamefont {M.}~\bibnamefont {Hohmann}}, \
  and\ \bibinfo {author} {\bibfnamefont {T.}~\bibnamefont {Z{\l}o{\'s}nik}},\
  }\href {\doibase 10.3390/universe5070168} {\bibfield  {journal} {\bibinfo
  {journal} {Universe}\ }\textbf {\bibinfo {volume} {5}},\ \bibinfo {pages}
  {168} (\bibinfo {year} {2019})},\ \Eprint {http://arxiv.org/abs/1905.02967}
  {arXiv:1905.02967 [gr-qc]} \BibitemShut {NoStop}%
\bibitem [{\citenamefont {Barker}\ \emph {et~al.}(2025)\citenamefont {Barker},
  \citenamefont {Marzo},\ and\ \citenamefont {Santoni}}]{Barker:2025xzd}%
  \BibitemOpen
  \bibfield  {author} {\bibinfo {author} {\bibfnamefont {W.}~\bibnamefont
  {Barker}}, \bibinfo {author} {\bibfnamefont {C.}~\bibnamefont {Marzo}}, \
  and\ \bibinfo {author} {\bibfnamefont {A.}~\bibnamefont {Santoni}},\
  }\href@noop {} {\  (\bibinfo {year} {2025})},\ \Eprint
  {http://arxiv.org/abs/2505.23894} {arXiv:2505.23894 [hep-th]} \BibitemShut
  {NoStop}%
\bibitem [{\citenamefont {Gallagher}\ and\ \citenamefont
  {Koivisto}(2021)}]{Gallagher:2021tgx}%
  \BibitemOpen
  \bibfield  {author} {\bibinfo {author} {\bibfnamefont {P.}~\bibnamefont
  {Gallagher}}\ and\ \bibinfo {author} {\bibfnamefont {T.}~\bibnamefont
  {Koivisto}},\ }\href {\doibase 10.3390/sym13112076} {\bibfield  {journal}
  {\bibinfo  {journal} {Symmetry}\ }\textbf {\bibinfo {volume} {13}},\ \bibinfo
  {pages} {2076} (\bibinfo {year} {2021})},\ \Eprint
  {http://arxiv.org/abs/2103.05435} {arXiv:2103.05435 [gr-qc]} \BibitemShut
  {NoStop}%
\bibitem [{\citenamefont {Kostelecky}\ and\ \citenamefont
  {Russell}(2011)}]{Kostelecky:2008ts}%
  \BibitemOpen
  \bibfield  {author} {\bibinfo {author} {\bibfnamefont {V.~A.}\ \bibnamefont
  {Kostelecky}}\ and\ \bibinfo {author} {\bibfnamefont {N.}~\bibnamefont
  {Russell}},\ }\href {\doibase 10.1103/RevModPhys.83.11} {\bibfield  {journal}
  {\bibinfo  {journal} {Rev. Mod. Phys.}\ }\textbf {\bibinfo {volume} {83}},\
  \bibinfo {pages} {11} (\bibinfo {year} {2011})},\ \Eprint
  {http://arxiv.org/abs/0801.0287} {arXiv:0801.0287 [hep-ph]} \BibitemShut
  {NoStop}%
\bibitem [{\citenamefont {Pfeifer}\ \emph {et~al.}(2025)\citenamefont
  {Pfeifer}, \citenamefont {R{\"a}tzel},\ and\ \citenamefont
  {Braun}}]{Pfeifer:2024vvc}%
  \BibitemOpen
  \bibfield  {author} {\bibinfo {author} {\bibfnamefont {C.}~\bibnamefont
  {Pfeifer}}, \bibinfo {author} {\bibfnamefont {D.}~\bibnamefont {R{\"a}tzel}},
  \ and\ \bibinfo {author} {\bibfnamefont {D.}~\bibnamefont {Braun}},\ }\href
  {\doibase 10.1103/PhysRevD.111.084073} {\bibfield  {journal} {\bibinfo
  {journal} {Phys. Rev. D}\ }\textbf {\bibinfo {volume} {111}},\ \bibinfo
  {pages} {084073} (\bibinfo {year} {2025})},\ \Eprint
  {http://arxiv.org/abs/2406.12314} {arXiv:2406.12314 [gr-qc]} \BibitemShut
  {NoStop}%
\bibitem [{\citenamefont {Braun}\ \emph {et~al.}(2025)\citenamefont {Braun},
  \citenamefont {Cai}, \citenamefont {Hermes}, \citenamefont {Marchese},
  \citenamefont {Nimmrichter}, \citenamefont {Pfeifer}, \citenamefont
  {R{\"a}tzel}, \citenamefont {Redaelli},\ and\ \citenamefont
  {Ulbricht}}]{Braun:2025iqq}%
  \BibitemOpen
  \bibfield  {author} {\bibinfo {author} {\bibfnamefont {D.}~\bibnamefont
  {Braun}}, \bibinfo {author} {\bibfnamefont {R.}~\bibnamefont {Cai}}, \bibinfo
  {author} {\bibfnamefont {P.}~\bibnamefont {Hermes}}, \bibinfo {author}
  {\bibfnamefont {M.~M.}\ \bibnamefont {Marchese}}, \bibinfo {author}
  {\bibfnamefont {S.}~\bibnamefont {Nimmrichter}}, \bibinfo {author}
  {\bibfnamefont {C.}~\bibnamefont {Pfeifer}}, \bibinfo {author} {\bibfnamefont
  {D.}~\bibnamefont {R{\"a}tzel}}, \bibinfo {author} {\bibfnamefont
  {S.}~\bibnamefont {Redaelli}}, \ and\ \bibinfo {author} {\bibfnamefont
  {H.}~\bibnamefont {Ulbricht}},\ }\href@noop {} {\  (\bibinfo {year}
  {2025})},\ \Eprint {http://arxiv.org/abs/2504.10942} {arXiv:2504.10942
  [hep-ex]} \BibitemShut {NoStop}%
\bibitem [{\citenamefont {Weyl}(1949)}]{Weyl1949-WEYPOM}%
  \BibitemOpen
  \bibfield  {author} {\bibinfo {author} {\bibfnamefont {H.}~\bibnamefont
  {Weyl}},\ }\href@noop {} {\emph {\bibinfo {title} {Philosophy of Mathematics
  and Natural Science}}},\ edited by\ \bibinfo {editor} {\bibfnamefont
  {O.}~\bibnamefont {Helmer{-}Hirschberg}}\ and\ \bibinfo {editor}
  {\bibfnamefont {F.}~\bibnamefont {Wilczek}}\ (\bibinfo  {publisher}
  {Princeton University Press},\ \bibinfo {address} {Princeton, N.J.},\
  \bibinfo {year} {1949})\BibitemShut {NoStop}%
\bibitem [{\citenamefont {Scholz}(2018)}]{SCHOLZ201857}%
  \BibitemOpen
  \bibfield  {author} {\bibinfo {author} {\bibfnamefont {E.}~\bibnamefont
  {Scholz}},\ }\href {\doibase https://doi.org/10.1016/j.shpsb.2017.04.003}
  {\bibfield  {journal} {\bibinfo  {journal} {Studies in History and Philosophy
  of Science Part B: Studies in History and Philosophy of Modern Physics}\
  }\textbf {\bibinfo {volume} {61}},\ \bibinfo {pages} {57} (\bibinfo {year}
  {2018})},\ \bibinfo {note} {hermann Weyl and the Philosophy of the `New
  Physics'}\BibitemShut {NoStop}%
\bibitem [{\citenamefont {Yang}(1974)}]{Yang:1974kj}%
  \BibitemOpen
  \bibfield  {author} {\bibinfo {author} {\bibfnamefont {C.-N.}\ \bibnamefont
  {Yang}},\ }\href {\doibase 10.1103/PhysRevLett.33.445} {\bibfield  {journal}
  {\bibinfo  {journal} {Phys. Rev. Lett.}\ }\textbf {\bibinfo {volume} {33}},\
  \bibinfo {pages} {445} (\bibinfo {year} {1974})}\BibitemShut {NoStop}%
\bibitem [{\citenamefont {Blagojevi{\'c}}\ \emph {et~al.}(2013)\citenamefont
  {Blagojevi{\'c}}, \citenamefont {Hehl},\ and\ \citenamefont {Kibble}}]{hehl}%
  \BibitemOpen
  \bibfield  {author} {\bibinfo {author} {\bibfnamefont {M.}~\bibnamefont
  {Blagojevi{\'c}}}, \bibinfo {author} {\bibfnamefont {F.~W.}\ \bibnamefont
  {Hehl}}, \ and\ \bibinfo {author} {\bibfnamefont {T.~W.~B.}\ \bibnamefont
  {Kibble}},\ }\href {\doibase 10.1142/p781} {\emph {\bibinfo {title} {Gauge
  Theories of Gravitation}}}\ (\bibinfo  {publisher} {IMPERIAL COLLEGE PRESS},\
  \bibinfo {year} {2013})\ \Eprint
  {http://arxiv.org/abs/https://www.worldscientific.com/doi/pdf/10.1142/p781}
  {https://www.worldscientific.com/doi/pdf/10.1142/p781} \BibitemShut {NoStop}%
\bibitem [{\citenamefont {Burtt}(1954)}]{Burtt1954-BURTMF}%
  \BibitemOpen
  \bibfield  {author} {\bibinfo {author} {\bibfnamefont {E.~A.}\ \bibnamefont
  {Burtt}},\ }\href@noop {} {\emph {\bibinfo {title} {The Metaphysical
  Foundations of Modern Science}}}\ (\bibinfo  {publisher} {Dover
  Publications},\ \bibinfo {address} {Mineola, N.Y.},\ \bibinfo {year}
  {1954})\BibitemShut {NoStop}%
\bibitem [{\citenamefont {Van\texttt{\char126}Cleve}\ and\ \citenamefont
  {Frederick}(1991)}]{VanCleve1991-VANTPO-16}%
  \BibitemOpen
  \bibinfo {editor} {\bibfnamefont {J.}~\bibnamefont
  {Van\texttt{\char126}Cleve}}\ and\ \bibinfo {editor} {\bibfnamefont {R.~E.}\
  \bibnamefont {Frederick}},\ eds.,\ \href@noop {} {\emph {\bibinfo {title}
  {The Philosophy of Right and Left: Incongruent Counterparts and the Nature of
  Space}}}\ (\bibinfo  {publisher} {Kluwer Academic Publishers},\ \bibinfo
  {year} {1991})\BibitemShut {NoStop}%
\end{thebibliography}%

\appendix 

\section{SU(2)$\times$SU(2) formulation}
\label{quaternion}

%Having dealt with the connection, we turn to the field $\phi^I$. 
A quaternion representation of the khronon is
\be
\phi = i\lp \begin{matrix}
    i\phi^4 + \phi^3 & \phi^1 -i\phi^2  \\
    \phi^1 + i\phi^2 & i\phi^4 - \phi^3
  \end{matrix}\rp = \lp\begin{matrix}  z_1 & - z_2^* \\ 
                                   z_2 & z_1^* \end{matrix} \rp\,,
\ee
so 4 real numbers are rewritten as the 2 complex numbers  $z_1=  i\phi^4 + \phi^3$ and $\phi^1 + i\phi^2$ and they're packed into 1 quaternion. 
A property of a quaternions is that $\det{\phi} = \delta_{IJ}\phi^I \phi^J$, and that $\phi\phi^\dagger = \det{\phi}\mathbb{1}$. 
The action of $(G_-, G_+) \in SU(2) \times SU(2)$ on $\phi$ is
\be \label{trans}
\phi \rightarrow G_-\phi G_+^\dagger\,.
\ee
Let's see if this works, with the standard representation $l^i=r^i=i-\sigma^i/2$ for the su(2) algebras. If we perform an infinitesimal Lorentz transformation, $\phi^I \rightarrow \phi^I + g^I{}_J\phi^J$, the corresponding left-right transformations should be given by, according to (\ref{LRcomponents}),
%\bs
\ba
g^i_\pm & = & \pm g^{4i} - \frac{1}{2}\epsilon^i{}_{jk}g^{jk}\,.
%g^a & = & g^{4a} - \frac{1}{2}\epsilon^a{}_{bc}g^{bc}\,, \\
%g^{\bar{a}} & = & -g^{4\bar{a}} - \frac{1}{2}\epsilon^{\bar{a}}{}_{\bar{b}\bar{c}}g^{\bar{b}\bar{c}}\,. 
\ea
%\es 
Plugging into (\ref{trans}),
\be
%\phi \rightarrow \phi + \frac{i}{2}\lp - g_{\bar{a}}\sigma^{\bar{a}}\phi  + g_a\phi\sigma^a\rp  
\phi \rightarrow \phi + \frac{i}{2}\lp - g^{i}_-\sigma_{i}\phi  + g^i_+\phi\sigma_i\rp  
= \phi + i\lp \begin{matrix} 
(i g^4{}_I + g^3{}_I)\phi^I & ( g^1{}_I - i g^2{}_I ) \phi^I \\ 
(g^1{}_I + ig^2{}_I )\phi^I & (ig^4{}_I - g^3{}_I)\phi^I  
\end{matrix}\rp\,, 
\ee
verifies that our definitions are consistent. 
The transformation law (\ref{trans}) suggests that the covariant derivative of the quaternion field is given as
\be
\bDiff\phi = \bdiff\phi + \m\bA\phi + \phi\+\bA^\dagger = \bdiff\phi + \frac{i}{2}\lp  \+\bA_i\phi\sigma^i - \m\bA_i\sigma^i\phi\rp\,, 
\ee
since the connections transform as
\be
\x \bA \rightarrow G_\pm \bDiff G^\dagger_\pm = -\lp \bdiff G_\pm \rp G^\dagger_\pm + G_\pm\x\bA G^\dagger_\pm\,. 
\ee
In the second equality we took into account that in a unitary matrix representation, $\bdiff G^\dagger = -G^\dagger\lp \bdiff G\rp G^\dagger$. %Secondly, we should take into account the non-commutativity. So the above means that
%\bs
%\ba
%\+ \bA & \rightarrow & -G_+^\dagger\lp \bdiff G\rp + 
%\ea
%\es
Check:
\ba
\bDiff\phi & \rightarrow & \bdiff\lp G_- \phi G^\dagger_+\rp + \lp G_-\bDiff G^\dagger_-\rp\lp G_-\phi G_+^\dagger\rp + \lp G_-\phi G_+^\dagger\rp\lp G_+\bDiff G_+^\dagger\rp^\dagger \nn \\
& = & \lp\bdiff G_-\rp\phi G_+^\dagger+ G_-\lp\bdiff\phi\rp G_+^\dagger + G_-\phi\bdiff G_+^\dagger -  \lp\bdiff G_-\rp\phi G_+^\dagger + 
G_-\m\bA \phi G^\dagger_+ + G_-\phi G^\dagger_+\lp -G^+\bdiff G^\dagger_+ G_+ \+\bA^\dagger G^\dagger_+\rp  \nn \\
& = & G_-\lp \bDiff\phi\rp G^\dagger_+\,. 
\ea
When taking derivatives of odd forms, we have to notice that the right-handed connection is wedged from the RHS and thus with a minus sign,
\bs
\ba
\bDiff(\bDiff\phi) & = & \bdiff(\bDiff\phi) + \m\bA\wedge(\bDiff\phi) - (\bDiff\phi)\wedge\+\bA^\dagger \nn \\
 & = & \m\bF\phi + \phi\+\bF^\dagger\,, \quad \text{where} \quad \x\bF = \bdiff\x\bA + \x\bA\wedge\x\bA\,. 
\ea
As a consistency check,
\ba
\x\bF %= \bdiff\+\bA^i r_i + \+\bA^i\wedge\+\bA^j ir_j 
& = &  -\frac{i}{2}\bdiff\x\bA^i \sigma_i - \frac{1}{4}\x\bA^i\wedge\x\bA^j\sigma_i\sigma_j 
=  -\frac{i}{2}\bdiff\x\bA^i \sigma_i - \frac{1}{8}\x\bA^i\wedge\x\bA^j[\sigma_i,\sigma_j] \nn \\
& = & -\frac{i}{2}\lp \x\bA^k + \frac{1}{2}\epsilon_{ij}{}^k\x\bA^i\wedge\x\bA^j\rp \sigma_k =  -\frac{i}{2}\sigma_i \x\bF^i\,, 
\ea
we confirm that the components match with (\ref{sosocurv}). 
\es

\section{Matter current}
\label{mattercurrent}

We don't derive the cosmological source term from first principles, but use a phenomenological fluid parameterisation. The usual fluid energy-momentum (1,1)-tensor $\bT$ can be given componentwise as
\ba
\bT & = & \lb -\lp \rho + \delta\rho\rp\partial_\tau + \lp\rho+p\rp\lp  a^{-1}v + a^{-2}\sqrt{\kappa}\varphi + \dot{c}\rp^{,i}\partial_i\rb\otimes\bdiff\tau \nn \\ & + & a\lp \rho + p\rp v_{,i}\partial_\tau\otimes\bdiff x^i + \lb \lp p + \delta p\rp\delta^i_j + \Delta^i{}_j \Pi\rb\partial_i\otimes\bdiff x^j\,,
\ea
where $\rho$ and $\delta\rho$ are the energy density and its perturbation; $p$ and $\delta p$ are the pressure and its perturbation; $v$ is the scalar velocity potential; and $\Pi$ is the scalar anisotropic stress potential. The khronon perturbation $\varphi$ and the connection perturbation $\dot{c}$ enter into the off-diagonal components by raising and lowering an index
with the metric read from the line element
\be
\bdiff s^2 = \lp 1 + 2\varphi'\rp \bdiff\tau\otimes\bdiff\tau 
+ \lp \sqrt{\kappa}\varphi + a^2\dot{c}\rp_{,i}\lp  \bdiff\tau\otimes\bdiff x^i + \bdiff x^i\otimes\bdiff\tau\rp + a^2\lb \lp 1 + 2\Psi\rp\delta_{ij} + 2\Delta_{ij} r\rb \bdiff x^i\otimes\bdiff x^j\,. 
\ee
Since $\bT$ is a vector-valued 1-form, we can construct a 1-form $\bT^I = \bT\lrcorner\bbe^I$ valued in the fundamental representation of $SO(4)$ as
\bs
\label{Ta}
\ba
\bT^4 & = & -\lp\rho + \delta\rho -\rho\varphi'\rp\bdiff\tau + \lb a\lp\rho+p\rp v + p\sqrt{\kappa}\varphi\rb_{,i}\bdiff x^i\,, \\
\bT^i & = & \lb \lp \rho + p\rp \lp v + a^{-1}\sqrt{\kappa}\varphi\rp + p\dot{c}\rb^{,i}\bdiff\tau + a\lb \lp p + \delta p + \Psi\rp\delta^i_j + \delta^i_j\lp \Pi + r\rp\rb\bdiff x^j\,. 
\ea
\es
Finally, we can form the 3-form current $\bt^I = \sqrt{\kappa}\ast\bT^I = \sqrt{\kappa}T^I{}_J\star\bbe^J$, wherein the scalar components $T^I{}_J = \ie_J\lrcorner\bT^I$ are obtained by using the frame field (i.e. the inverse of the Cartan frame),
\bs
\ba
\ie_4 & = & \lp 1-\varphi'\rp \partial_\tau - \dot{c}^{,i}\partial_i\,, \\
\ie_i & = & - a^{-1}\sqrt{\kappa}\varphi_{,i}\partial_\tau + a^{-1}\lb \lp 1-\Psi\rp\delta^j_i - \Delta^j_i + \epsilon_i{}^{jk}\tilde{s}_{,k}\rb\partial_j\,. 
\ea
\es
The result 
\be \label{symmetricEMT}
T^4{}_4 = -\rho-\delta\rho\,, \quad T^4{}_i = \delta_{ij}T^j{}_4 = \lp \rho + p\rp\lp v + a^{-1}\sqrt{\kappa}\varphi\rp_{,i}\,, \quad  
T^i{}_j = \lp p + \delta p\rp\delta^i_j + \Delta^i{}_j\Pi\,,
\ee
agrees with (\ref{mattersources}). Note that there the $\star\bbe^I$ contain the linear perturbations; were replaced by only the background 3-forms $\star\bbe^I$, the extra perturbations terms appearing in $\bT^I$ at (\ref{Ta}) had to be included in the formula.

The material energy current $\bt^I$ satisfies conservation laws which are dictated by symmetry. In section 5.1. of Ref.\cite{Koivisto:2023epd} the Lorentz invariance and the coordinate invariance were shown to yield the two identities, respectively:
\bs
\ba
\bDiff\bO^{IJ} & = & \bDiff\phi^{[I}\wedge\bt^{J]}\,, \\
m_P\bDiff\bt_I & = & \bbie_I\lrcorner\bF^{JK}\wedge\lp \bO_{JK}- \phi_{[J}\bt_{K]}\rp\,. 
\ea
\es
The first of these identities is now trivial, since we do not include spin currents $\bO^{IJ}=0$, and the energy-momentum tensor is by construction symmetric. The second identity yields important information, though it is redundant with the field equations. The background part yields (\ref{mattercont}), which is extensively used in our cosmological investigations. The linear part gives two nontrivial equations for the matter source, called the continuity equation,
\bs
\label{p_eqs}
\be \label{p_continuity}
\delta\rho' + 3H\lp \delta\rho + \delta p\rp =
\lp \rho + p\rp \lb \nabla^2\lp \frac{v}{a}+c'-\frac{\Phi}{a^2}\rp - 3\Psi'\rb\,, 
\ee
and the Euler equation,
\be \label{p_euler}
a\lp\rho + p\rp \lp  v' + 4aHv\rp + \lp \rho + p\rp' v = \delta p - \frac{2}{3}\nabla^2\Pi\,. 
\ee
\es

\end{document}